\documentclass[11pt,a4paper,nofootinbib,superscriptaddress]{revtex4-1}
\pdfoutput=1
\usepackage[normalem]{ulem}
\usepackage[utf8]{inputenc}
\usepackage{graphicx,bm}
\usepackage{slashed,verbatim}
\usepackage{amssymb,graphicx,epstopdf}

 \usepackage[colorlinks=true,linktocpage=true,linkcolor=blue,citecolor=blue]{hyperref}
\usepackage{slashed,subfigure}
\usepackage{caption}
\usepackage{amsmath,amssymb,graphicx,xcolor,mathtools}
\usepackage{epstopdf,dcolumn}
\usepackage{soul}

\allowdisplaybreaks
\usepackage[normalem]{ulem}
\usepackage{cancel}
\usepackage{xcolor}
\setstcolor{red}

\def\be {\begin{equation}}
\def\ee {\end{equation}}
\def\bea {\begin{eqnarray}}
\def\eea {\end{eqnarray}}
\def\bc {\begin{center}}
\def\ec {\end{center}}
\def\nn {\nonumber}
\def\eps {\epsilon}
\def\({\left(}
\def\){\right)}
\def\mn {\mu\nu}

\newcommand \Tr{\operatorname{\text{Tr}}}
\newcommand\mycom[2]{\genfrac{}{}{0pt}{}{#1}{#2}}

\def\lrarrow{\leftrightarrow}

\def\ranglec{\rangle_{\!\!c}}
\def\rangleci{\rangle_{\!\!c_1,c_2}}

\def\sumintb{\sum\!\!\!\!\!\!\!\!\!\int\limits}
\def\sumintf{\sum\!\!\!\!\!\!\!\!\!\!\int\limits}

\DeclareGraphicsExtensions{.jpg,.pdf,.eps}

\begin{document}
\title{Pressure of a weakly magnetized hot and dense  deconfined QCD matter 
in one-loop hard-thermal-loop perturbation theory}

 \author{Aritra Bandyopadhyay}
  \email{aritrabanerjee.444@gmail.com}
  \affiliation{
 	Theory Division, Saha Institute of Nuclear Physics, HBNI, \\
 	1/AF, Bidhannagar, Kolkata 700064, India. }
 \affiliation{Departamento de F\'{\i}sica, Universidade Federal de Santa Maria, 
 	Santa Maria, RS 97105-900, Brazil}
	 \author{Bithika Karmakar}
\email{bithika.karmakar@saha.ac.in}
  \affiliation{
 	Theory Division, Saha Institute of Nuclear Physics, HBNI, \\
 	1/AF, Bidhannagar, Kolkata 700064, India. }
 \author{Najmul Haque}
 \email{nhaque@niser.ac.in}
 \affiliation{School of Physical Sciences, National Institute of Science Education and Research, HBNI,\\  Jatni, Khurda 752050, India}
 \affiliation{Institut f\"ur Theoretische Physik, Justus-Liebig--Universit\"at Giessen, 35392 Giessen, Germany}
 \author{Munshi G Mustafa}
 \email{munshigolam.mustafa@saha.ac.in}
  \affiliation{
 	Theory Division, Saha Institute of Nuclear Physics, HBNI, \\
 	1/AF, Bidhannagar, Kolkata 700064, India. }

\begin{abstract}{
We consider our recently obtained  general structure of two point (self-energy and propagator) functions of quarks and gluons in a nontrivial background like a heat bath and an external magnetic field. Based on this, here we have computed  free energy and pressure of quarks and gluons for a magnetized  hot and dense deconfined QCD matter in weak field approximation. For heat bath we have used hard thermal loop perturbation theory (HTLpt)  in presence of finite  chemical potential. For  weak field approximations we have obtained the pressure of QCD matter, both with and without the high temperature expansion. The results with high $T$ expansions are completely analytic and gauge 
independent but depends on the renormalization scale in addition to the temperature, chemical potential and  the external magnetic field. We also discuss the modification of QCD Debye mass of such  matter for an arbitrary magnetic field. Analytic  expressions for Debye mass are also obtained for both strong and weak field approximation. It is found to exhibit some interesting features depending upon the three different scales, {\it {i.e}}, the quark mass, temperature and the strength of the magnetic field. The various divergences appearing in the quark and gluon free energies are 
regulated through appropriate counter terms. In weak field approximation, the low temperature behavior of the pressure is 
found to strongly depend on the magnetic field than that at high temperature. We also discuss the specific problem with one-loop HTLpt associated with the over-counting of certain orders in coupling.}
\end{abstract}
\maketitle

\section{Introduction}
\label{intro}
Quark gluon plasma (QGP) is a thermalized color deconfined state of nuclear matter in the 
regime of quantum chromodynamics (QCD) under extreme conditions such as 
very high temperature and/or density. For the past couple of decades, different high 
energy heavy-ion-collisions (HIC) experiments are under way, e.g., RHIC at BNL, 
LHC at CERN and the  upcoming  FAIR at GSI,  to study this novel state of QCD matter within 
the largely unknown QCD phase diagram. In recent years the focus has also 
shifted toward the noncentral HIC, where a very strong magnetic field is created in 
the direction perpendicular to the reaction plane due to the spectator particles 
that are not participating in the 
collisions~\cite{Shovkovy:2012zn,D'Elia:2012tr,Fukushima:2012vr,Mueller:2014tea,
Miransky:2015ava}. Recent experimental evidences of photon anisotropy 
provided by the PHENIX Collaboration~\cite{Adare:2011zr} has also challenged the present 
theoretical tools. By assuming a presence of large anisotropic magnetic field generated 
in HIC, eventually some explanations were made~\cite{Basar:2012bp} in support of those 
experimental findings. In fact this has prompted that a theoretical study is 
much needed by considering the effects of an intense background magnetic field on various 
aspects and observables of noncentral HIC. Also some of these studies have 
subsequently revealed that the strong magnetic field generated during the 
noncentral HIC is also time dependent. More specifically, it rapidly decreases with 
time~\cite{Bzdak:2012fr,McLerran:2013hla}. Nevertheless, the inclusion of 
an external magnetic field in QGP  introduces also an extra 
energy scale in the system. At the time of the noncentral HIC, the value of the 
created magnetic field $B$ is very high compared to the temperature $T$ 
($T^2<q_fB$ where $q_f$ is the absolute charge of the quark with flavor $f$)
associated with the system. It is estimated up to the order of $q_fB \sim 15 m_\pi^2$ in the LHC at 
CERN~\cite{Skokov:2009qp}. On the other hand, neutron stars (NS), or more specifically 
magnetars are also known to possess a strong enough magnetic 
field~\cite{Duncan:1992hi,Chakrabarty:1997ef,Bandyopadhyay:1997kh}.
In this regime of study one usually works in the strong 
magnetic field approximation. 

The presence of an external anisotropic field in the medium calls for the 
appropriate modification of the present theoretical tools to investigate 
various properties of QGP and a lot of activities are in progress. Over the last 
few years, several novel phenomena came into light, e.g., chiral 
magnetic effect~\cite{Kharzeev:2007jp,Fukushima:2008xe,Kharzeev:2009fn}, magnetic catalysis 
\cite{Alexandre:2000yf,Gusynin:1997kj,Lee:1997zj} and inverse magnetic catalysis 
\cite{Bali:2011qj,Bornyakov:2013eya,Mueller:2015fka,Ayala:2014iba,Farias:2014eca,Ayala:2014gwa,Ayala:2016sln,
Ayala:2015bgv} at finite temperature; 
chiral- and color-symmetry broken/restoration 
phase~\cite{Avancini:2016fgq,Fayazbakhsh:2010bh,Fayazbakhsh:2010gc,Andersen:2012zc,Andersen:2014xxa}; 
thermodynamic 
properties~\cite{Andersen:2012zc,Andersen:2014xxa,Strickland:2012vu,Farias:2016gmy,Rath:2017fdv},
refractive indices and decay 
constant of hadrons~\cite{Fayazbakhsh:2012vr,Fayazbakhsh:2013cha, Bandyopadhyay:2016cpf,Bandyopadhyay:2018gbw,Chakraborty:2017vvg} 
and the equation of state (EOS) in holographic models~\cite{Rougemont:2015oea,Finazzo:2016mhm} in a hot magnetized medium; 
soft photon production from conformal anomaly~\cite{Basar:2012bp,Ayala:2016lvs} in 
HIC; modification of dispersion properties in a magnetized hot QED~\cite{Sadooghi:2015hha} 
and QCD~\cite{Karmakar:2018aig,Ayala:2018ina,Das:2017vfh,Hattori:2017xoo} medium; and 
various transport coefficients~\cite{Kurian:2018qwb,Kurian:2018dbn,Kurian:2017yxj},
properties of quarkonia~\cite{Singh:2017nfa,Hasan:2018kvx},
synchroton radiation~\cite{Tuchin:2013prc2}, and dilepton 
production from a hot magnetized QCD 
plasma~\cite{Bandyopadhyay:2016fyd,Tuchin:2013prc,Tuchin:2013prc2,Tuchin:2013ie,
Sadooghi:2016jyf,Bandyopadhyay:2017raf} and in strongly coupled plasma in a strong magnetic 
field~\cite{Mamo:2013efa}.
 
Thermodynamic properties of low lying hadrons in the presence of magnetic field 
have been studied in recent years within the various hadronic 
models~\cite{Strickland:2012vu,Andersen:2012zc,Andersen:2014xxa, Farias:2016gmy} .
Nevertheless, the EOS is a generic quantity and of phenomenological
importance for studying the hot and dense QCD matter, QGP, created in HIC. At zero chemical 
potential and finite temperature, lattice QCD (LQCD) established itself as the most reliable method to calculate 
thermodynamic functions. Unfortunately at finite chemical potential LQCD faces 
the infamous sign problem. Information about the thermodynamic functions in 
LQCD can still be extracted by making a Taylor expansion of the partition function 
around zero baryonic chemical potential and extrapolating the result~\cite{Allton:2005gk}. 
But due to the finite number of Taylor coefficients such extrapolation has its own 
limitations. On the other hand,  naively, the 
asymptotic freedom of QCD leads us to expect that bare perturbation theory
should be a reliable guide to calculate these properties of matter at high temperature 
and/or high density~\cite{Zhai:1995ac,Arnold:1994eb,Arnold:1994ps,Toimela:1982hv,Kapusta:1979fh,
Chin:1978gj,Shuryak:1977ut}. Although, it has been recognized early on that this is not so.
Technically, infrared divergences plague the calculation of observables at finite 
temperature, preventing the determination of high order corrections. In order to cope 
with this difficulty, whose origin is the presence of massless particles, it has been 
suggested to reorganize perturbation theory, by performing the expansion around of a 
system of massive quasiparticles. The motivation for doing so is that thermal 
fluctuations can generate a mass. It amounts to a resummation of a class of loop 
diagrams, where the loop momenta are of the order of the temperature. Such diagrams are 
those which contribute to give the excitations a thermal mass. The hard-thermal-loop  
perturbation theory (HTLpt) is one such state-of-the-art resummed perturbation 
theory~\cite{andersen1}. In HTLpt the EOS  of QCD \textit{in absence of a  magnetic field} has systematically been 
computed  within one-loop [leading order 
(LO)]~\cite{najmul13,najmul12,najmul11,sylvain2,sylvain1,andersen3,andersen2,andersen1,Haque:2018eph}, 
two loop [next-to-leading order (NLO)]~\cite{Andersen:2002ey,Andersen:2003zk,Haque:2012my,najmul2qns} 
and three loop [next-to-next-to-leading order 
(NNLO)]~\cite{3loopglue1,3loopglue2,3loopqcd1,3loopqcd2,3loopqcd3,najmul3,Haque:2014rua} 
at finite temperature and chemical potential. Although the all-loop order calculations are 
gauge invariant, the three-loop results are complete in $g^5$ and fully analytic which do 
not require any free fit parameters besides a renormalization scale. The thermomagnetic 
correction to the quark-gluon vertex in the presence of a weak magnetic field within the 
HTL approximation has recently been computed~\cite{Haque:2017nxq,Ayala:2014uua}.
Also recently, the general structure of gluon~\cite{Karmakar:2018aig,Ayala:2018ina,Hattori:2017xoo}  
and quark~\cite{Das:2017vfh} self-energy and propagator and their spectra have been obtained 
in a thermomagnetic medium within HTL approximation. Further, the thermodynamic quantities in lowest Landau level (LLL) within the strong field approximation has been calculated in Ref.~\cite{Rath:2017fdv} using HTL approximation. But in this calculation the change in the general structure of two point functions of a gluon has not been considered. It assumes for the gluonic case  without any justification that the shift in the Debye mass is the only effect of the magnetic field. However, this is not the case as it has explicitly been shown in paper-I~\cite{Karmakar:2018aig} that the presence of an external magnetic  field breaks the rotational symmetry.

In view of this, presently, a systematic determination of EOS for magnetized hot QCD 
medium is of great importance. In this article (say paper-II)\footnote{Paper-I~\cite{Karmakar:2018aig}.} we make an effort to derive the pressure of a magnetized hot and dense deconfined QCD medium created in high energy HIC.
Usually two kinds of approaches were taken in all of the previous studies of EOS in the 
presence of a magnetic field. In the first kind, the pressure remains isotropic and the system can be easily 
described in terms of standard thermodynamic 
relations~\cite{Chakrabarty:1997ef,Bandyopadhyay:1997kh}. In the second one, the 
breaking of the spherical symmetry due to the anisotropic background 
magnetic field in a preferred direction~\cite{Canuto:1969ct,Chaichian:1999gd,Broderick:2000pe,Martinez:2003dz,
PerezMartinez:2007kw} is taken into consideration. 
Subsequently, this results in an anisotropic 
pressure arising from the difference between pressure components that are transverse 
and longitudinal to the background magnetic field direction. Eventually, the difference 
in stress causes the deformation of the fireball produced in heavy ion collisions 
or the NS.  There are also some recent LQCD calculations, which incorporate 
both of these schemes~\cite{Bali:2014kia}. However, it is also shown in 
\cite{Strickland:2012vu}, that pressure 
anisotropy decreases with the increase in temperature. 
Moreover, the magnetic field created in noncentral HIC is a fast decreasing 
function of time~\cite{Bzdak:2012fr,McLerran:2013hla}, it is expected that by the time 
the quarks and gluons thermalize in a QGP medium, the magnetic field strength becomes 
sufficiently weak. By virtue of which,  temperature at that time  becomes the 
largest energy scale of the system. In this regime, in principle,  one can work within the weak magnetic field approximation in which the pressure can be considered isotropic\footnote{However, one should consider
anisotropic pressure, i.e., different pressures along the longitudinal and the transverse directions in the case of a strong magnetic field approximation~\cite{Karmakar:2019tdp}.}.
In this paper II we work on the
weak ($q_fB < m_f^2 \sim m^2_{\textrm{th}}\sim g^2T^2 < T^2$) field limit where  $m_f$ is the mass of fermion, $m_{\textrm{th}}$ is the thermal mass of a fermion,  $T$ is the
temperature, $g$ is the strong coupling and $B$ is the strength of an external magnetic field.
For the strong field case, the system is  considered to be
confined in the lowest Landau Level and the  transverse pressure vanishes in the LLL 
or in the Landau ground state due to dimensional reduction.  We note that  the pressure along the field direction, 
i.e., longitudinal pressure in strong  field approximation ($m_f^2\sim m^2_{\textrm{th}} \sim g^2T^2 < T^2 < q_fB$) 
has already been computed  in Ref.~\cite{Rath:2017fdv}. This is because the average transverse momentum of fermion vanishes as the strength of $B$ increases.  On the other hand,  a weak magnetic field case is different than that of a strong field one, as we shall see below. 

In this  paper-II, for the first time, by considering the general two point function of a  quark~\cite{Das:2017vfh} and a  gluon~\cite{Karmakar:2018aig} in a hot but weakly magnetized deconfined QCD medium, we shall compute the pressure of quark-gluon plasma within HTL approximation. As we would see that the calculation is very involved, nevertheless  the expression of free energy vis-a-vis pressure is completely analytic and gauge independent. We have used strong coupling that runs through renormalization scale and strength of the magnetic field in weak field domain. Sensitivity of the two scales on the pressure has also been discussed in details.

This paper has been organized as follows: In Sec.~\ref{setup} the basic computation of the paper is briefly outlined. The scale hierarchies in weak field approximation are discussed in Sec.~\ref{scale}. In Sec.~\ref{QF} we discuss the one-loop quark free energy for a magnetized  hot QCD medium within weak field and HTL approximations by considering the general structure of the quark propagator~\cite{Das:2017vfh}. Section~\ref{GF} discusses gluon free energy for a magnetized hot and dense QCD medium in terms of the general structure of two point functions of gauge boson as obtained in paper-I~\cite{Karmakar:2018aig}. In Sec.~\ref{mD_modified} we discuss the Debye screening mass both in strong and weak field approximations that clearly separates the two domains. In Sec.~\ref{RFPW} we obtain the finite contribution of weak field free energy in Sec.~\ref{RF} and pressure in Sec.~\ref{RP} of a magnetized hot medium.  The magnetic field dependent strong coupling is discussed in Sec.~\ref{alpha_ayala} and its validity in weak field domain is justified. In Sec.~\ref{full} we compare approximate and full results thereby justifying the use of the former. The main results with high $T$ expansion are discussed in Sec.~\ref{RES} and finally, we conclude in Sec.~\ref{conclusion}. The detailed calculations associated with various sections  are given in Appendices~\ref{a_sumint},~\ref{BS}, ~\ref{ang_av} and \ref{wohighT} and various subsections therein.

\section{Setup}
\label{setup}
The total thermodynamic free energy up to one-loop order in HTLpt in the presence of a background magnetic field, $B$, can be written as
\bea
F &=& F_q+F_g+F_0 + \Delta {\mathcal E}_0,
\label{total_fe}
\eea
where $F_q$ and $F_g$ are, respectively, the quark and gluon part of the free energy which will be computed in the presence of a magnetic field with HTL approximation. $F_0=\frac{1}{2} B^2$ is the tree-level contribution due to the constant magnetic field and the $\Delta {\mathcal E}_0$ is the HTL counterterm  given~\cite{Haque:2012my} as
\bea
\Delta {\mathcal E}_0=\frac{d_A}{128\pi^2\epsilon}m_D^4, \label{htl_count}
\eea
with $d_A=N_c^2-1$, $N_c$ is the number of color in fundamental representation and $m_D$ is the Debye screening mass in HTL approximation. 

The pressure of a system is defined as 
\be
P=-F. \label{pressure}
\ee

%\section{Strong Field Approximation}
\section{Scale hierarchies in weak field approximation}
\label{scale}
The magnetic field generated in heavy-ion collisions decreases
rapidly with time. This  provides us a simplified situation  where one can work in weak field approximation because temperature is the highest scale in the system. The presence of magnetic field $q_fB$ introduces another scale in addition to the thermal scales $gT$ and $T$. In weak field approximation one can have two hierarchies of scales:

(i) When $\sqrt{q_fB}$ is the smallest scale compared to temperature and quark mass, one can work with a hierarchy of scales $\sqrt{q_fB}< m_f < T$ and $\sqrt{q_fB}$ can be treated as perturbation. This allows one to expand, for example, the Schwinger propagator for a fermion in weak field approximation~\cite{chyi,ayala2005} up to an $\mathcal{O}[(q_fB)^2]$ as
\bea
iS^w_m(k) &=&  i\frac{\slashed{K}+m_f}{K^2-m_f^2}  (q_f B)^0+i ~(q_f B)\frac{\left(\gamma_5
\left\{(K\cdot n)\slashed{u}-(K\cdot u)\slashed{n}\right\}+i\gamma_1\gamma_2m_f\right)}{(K^2-m_f^2)^2} \nn \\
&+& i \ 2(q_fB)^2  \left[\frac{\left\{(K\cdot u)\slashed{u}-(K\cdot n)\slashed{n}\right\} 
-\slashed{K}}{(K^2-m_f^2)^3}-\frac{k_\perp^2(\slashed{K}+m_f)}{(K^2-m_f^2)^4}\right]  
+ \mathcal{O}\left[(q_f B)^3\right] ,
\eea 
which is a perturbative series of $q_fB$.  In $q_fB\rightarrow 0$, the thermomagnetic correction vanishes. 
Alternatively, the thermomagnetic effects are obtained as higher order perturbative corrections to the nonmagnetized part [i.e., HTL part as $(q_f B)^0$]. This means that for each given order in $q_fB$ in a perturbative series, one can use HTL approximation within the scale hierarchy $\sqrt {q_f B}< gT<T$ to obtain the  desired order of coupling.

(ii) When quark mass $m_f$ is the smallest scale compared to temperature and magnetic field, one may work with a hierarchy $m_f<\sqrt{q_fB}< T$ by considering $m_f$ as perturbation for a given order of $q_fB$. In this hierarchy $m_f$ in fermion propagator is either set to be zero or expanded in $m_f$ for a given order of $q_fB$.  

As discussed above, we will be working only with the hierarchy\footnote{Calculation with the other hierarchy is itself an independent problem.} $\sqrt{q_fB}<m_f<T$ in this paper.

\section{Quark free energy in the presence of weak magnetic field}
\label{QF}

\subsection{General structure of two-point fermionic function}
\label{tpff}
\begin{figure}[h!]
\centering
\includegraphics[scale=0.9]{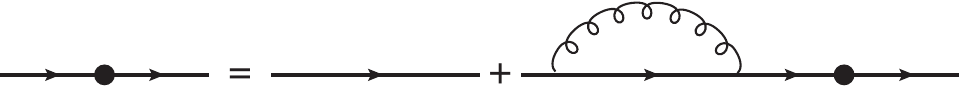}
\caption{\small Diagrammatic representation of the Dyson-Schwinger equation for one-loop effective fermion propagator.}
\label{fig:dyson_schwinger}
 \end{figure}

The inverse of the effective fermion propagator following the Dyson-Schwinger equation, as given in Fig.~\ref{fig:dyson_schwinger}, can now be written as
 \be
 S_{\textrm{eff}}^{-1}(P)  =\slashed{P}-\Sigma(P)\, . \label{inv_prop}
 \ee
 The general structure of a fermionic two-point function and its  dispersion spectrum in a hot magnetized medium has recently been discussed in detail in Ref.~\cite{Das:2017vfh}. The most general form of the fermion self-energy for CPT~\footnote{Charge conjugation, parity and time reversal.} and chirally invariant 
 theory in a hot magnetized medium becomes 
\begin{align}
\Sigma(P) &= - \mathcal{A} \,\slashed{P} - \mathcal{B}\slashed{u} - \mathcal{B}^{\prime}\gamma_{5}\,\slashed{u} 
- \mathcal{C}^{\prime}\gamma_{5}\,\slashed{n}. \label{genstructselfenergy} 
\end{align}
where $u_\mu$ is the four velocity of the heat bath and  the direction of the magnetic field~\cite{Karmakar:2018aig,Das:2017vfh}, $n_\mu$  is given as
\be
n_\mu = \frac{1}{2B} \epsilon_{\mu\nu\rho\lambda}\, u^\nu F^{\rho\lambda} =  \frac{1}{B}u^\nu {\tilde F}_{\mu\nu} \, . \label{nmu}
\ee
The background dual field tensor ${\tilde F}_{\mu\nu}$ can be written in terms   
 of field tensor ${ F}_{\mu\nu}$ as
\be
{\tilde F}_{\mu\nu}= \frac{1}{2} \epsilon_{\mu\nu\rho\lambda} F^{\rho\lambda}. \label{dual} 
\ee
Without any loss of generality we have considered the four velocity in the rest frame of the heat bath and the direction of the magnetic 
field $B$ along the $z$ direction. So, 
\begin{subequations}
 \begin{align}
 u^{\mu} &= (1,0,0,0), \label{fv4} \\
 n_{\mu} &= (0,0,0,1) . \label{mgdir}
 \end{align}
\end{subequations}
The general form of the various structure functions can be obtained from Eq.~\eqref{genstructselfenergy} as
\begin{subequations}
\begin{align}
\mathcal{A} &= \frac{1}{4}\,\,\frac{\Tr\left[\Sigma(P)\slashed{P}\right]-(P\cdot u)\,\Tr\left[\Sigma(P)\slashed{u}\right]}{(P\cdot u)^{2}-P^{2}} , \label{sta}\\
\mathcal{B} &= \frac{1}{4}\,\,\frac{-(P\cdot u)\,\Tr\left[\Sigma(P)\slashed{P}\right]+P^{2}\,\Tr\left[\Sigma(P)\slashed{u}\right]}{(P\cdot u)^{2}-P^{2}} , 
\label{stb} \\
\mathcal{B}^{\prime} &= - \frac{1}{4}\,\Tr\left[\slashed{u}\Sigma(P)\gamma_{5}\right] , \label{stbp} \\
\mathcal{C}^{\prime} &=  \frac{1}{4}\,\Tr\left[\slashed{n}\Sigma(P)\gamma_{5}\right] , \label{stcp}
\end{align}
\end{subequations}
which are also Lorentz scalars. Beside $T$, $\mu$  and $B$, these structure functions would also depend on three Lorentz scalars due to the breaking 
of both Lorentz(boost) and rotational invariance defined by
\begin{subequations}
\begin{align}
p_0=\omega &\equiv P^{\mu}u_{\mu}, \label{ome} \\
p{^3}&\equiv  -P^{\mu}n_{\mu} =p_z \, , \label{p3} \\
p_{\perp} &\equiv  \left[(P^{\mu}u_{\mu})^{2}-(P^{\mu}n_{\mu})^{2}-(P^{\mu}P_{\mu})\right] ^{1/2} \nn \\
&=\left[p_0^2 -p_3^2-P^2\right]^{1/2} = \left[p_1^2+p_2^2\right ]^{1/2} .  \label{pperp}
\end{align}
\end{subequations}

\begin{center}
\begin{figure}[tbh]
 \begin{center}
 \includegraphics[scale=.7]{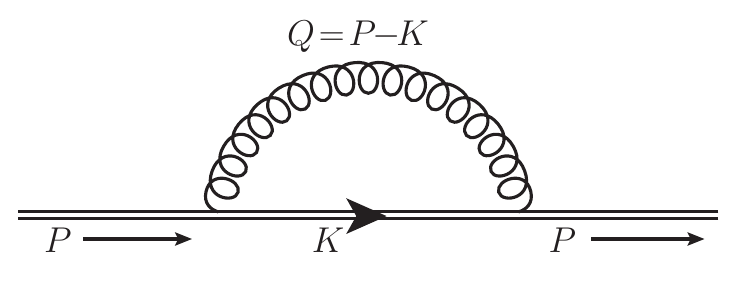} 
 \caption{Self-energy diagram for a quark in weak magnetic field approximation. 
The double line indicates the modified quark propagator in the presence of a magnetic 
field.}
  \label{quark_se}
 \end{center}
\end{figure}
\end{center}

All these structure functions in Eq.~\eqref{sta} to \eqref{stcp}  in one-loop order as shown in Fig.~\ref{quark_se}  within weak field and HTL approximations have been computed~\footnote{In Ref.~\cite{Das:2017vfh} those structure functions were computed for $\mu=0$ but we have modified it for $\mu\ne 0$} in Ref.~\cite{Das:2017vfh} as
 \begin{subequations}
 \begin{align}
\mathcal{A}(p_0,p_{\perp}, p_3) &
\, = \, -\frac{m_{\textrm{th}}^2}{p^2} \int\frac{d\Omega}{4\pi}\frac{\bm{p\cdot \hat{k}}}{P\cdot\hat{K}} \, ,
 \label{at}\\ 
% % % %  
\mathcal{B}(p_0,p_{\perp}, p_3)& \, =\, \frac{m_{\textrm{th}}^2}{p^2} \int\frac{d\Omega}{4\pi}\frac{(P\cdot u)(\bm{p\cdot\hat{k}})-p^2}{P\cdot\hat{K}} \, ,
\label{bt} \\
% % % % 
\mathcal{B}^{\prime} (p_0,p_{\perp}, p_3)
&= -m_{\textrm{eff}}^2\,\int\frac{d\Omega}{4\pi}\frac{\hat{K}\cdot n}{P\cdot\hat{K}} \, ,  \label{bprime}\\
\mathcal{C}^{\prime} (p_0,p_{\perp}, p_3)&= m_{\textrm{eff}}^2\,\int\frac{d\Omega}{4\pi}\frac{\hat{K}\cdot u}{P\cdot\hat{K}} \, .\label{cprime}
\end{align}
\end{subequations}
We emphasize that the structure functions in Eq.~\eqref{bprime} and~\eqref{cprime}  have become anisotropic in 
nature due to the breaking of the rotational invariance
in the presence of magnetic field in a given direction. Also note that
\begin{subequations}
\begin{align}
m_{\textrm{th}}^2 &= \frac{g^2 C_F T^2}{8}\left(1+4\hat{\mu}^2\right), \label{mth} \\
m_{\textrm{eff}}^2 &= 4 g^2 C_F M_{B,f}^2(T,\mu,m_f,q_{f}B)~;~~~ M_B^2= \sum\limits_f M_{B,f}^2(T,\mu,m_f,q_{f}B).
\end{align}
\end{subequations}
The magnetic mass\footnote{In  case of finite chemical potential the expression for $f_1$, the well known 
fermionic function also given in Eq.(34) of Ref~\cite{Ayala:2014uua}, gets modified as
\bea
f_1(y) =-\frac{1}{2}\ln\left(\frac{y}{4\pi}\right) + \frac{1}{4}\aleph(z) + \cdots , 
\eea 
which gets reflected in the expression of the magnetic mass in \eqref{mgmass}.
For zero chemical potential the expression for magnetic mass 
becomes~\cite{Ayala:2014uua} $M_{B,f}^2 =\frac{q_fB}{16\pi^2}\left[\ln 2-\frac{\pi 
T}{2m_f}\right]$.} for a given flavor $f$ is given as
\bea
M_{B,f}^2 &=& \frac{q_fB}{16\pi^2}\left[-\frac{1}{4}\aleph(z)-\frac{\pi 
T}{2m_f}-\frac{\gamma_E}{2}\right] , \label{mgmass}
\eea
where the function $\aleph(z)$ is defined in Eq.~(\ref{aleph}). 
Now we note that in the limit of small current quark mass ($m_f\rightarrow 0$), the magnetic mass in \eqref{mgmass} diverges. 
It has been regulated by using the prescription given in Refs.~\cite{Dolan:1973qd,Kapusta:1989tk} with a mass cutoff which is the  thermal mass $m_{\textrm{th}}$ of
the fermion.  

Combining Eq.~\eqref{genstructselfenergy} with Eqs.~\eqref{at}-\eqref{cprime} the total quark self-energy contribution of $\mathcal O[q_fB]$ in the 
presence of a weak magnetic field within HTL approximation can be written as
\bea
\Sigma(P)&=& 
m_{\textrm{th}}^2\int\frac{d\Omega}{4\pi}
\frac{\hat{\slashed{K}}}{P\cdot \hat{K}} +m_{\textrm{eff}}^2 \int\frac{d\Omega}{4\pi}\frac{\left[(\hat{K}\cdot 
n)\slashed{u}-(\hat{K}\cdot u)\slashed{n}\right]}{P\cdot \hat{K}}.
\label{qse_wfa_htl}
\eea

Using the general structure of the quark self-energy and considering the external anisotropic weak magnetic field along the $z$ (or $3$) direction~\cite{Das:2017vfh}  one can now write
\bea
S_{\textrm{eff}}^{-1}&=&\slashed{P}-\Sigma(P) = \left[\mathcal{C}(p_0,p_{\perp}, p_3)p_0\gamma_0-\mathcal{D}(p_0,p_{\perp}, p_3) 
p_i\gamma_i+\mathcal{B}' (p_0,p_{\perp}, p_3)\gamma_5\gamma_0\right. \nn \\
&& \left. +\mathcal{C}' (p_0,p_{\perp}, p_3)\gamma_5\gamma_3\right]. \label{pslash_sigma}
\eea
% % % % % % % % % % % % % % % % % 
Here
\bea
\mathcal{C}(p_0,p_{\perp}, p_3) &=& 1+\mathcal{A}(p_0,p_{\perp}, p_3)+\frac{\mathcal{B}(p_0,p_{\perp}, p_3)}{p_0}= 1-\mathcal{A}'(p_0,p_{\perp}, p_3),\\ 
\mathcal{D}(p_0,p_{\perp}, p_3) &=&  1+\mathcal{A}(p_0,p_{\perp}, p_3),
\eea
with
\bea
\mathcal{A}(p_0,p_{\perp}, p_3) &=&  -\frac{m_{\textrm{th}}^2}{p^2}\int\frac{d\Omega}{4\pi}\frac{\bm{p\cdot 
\hat{k}}}{p_0-\bm{p\cdot\hat{k}}} = ~ \frac{m_{\textrm{th}}^2}{p^2}\left[1-{\cal T}_P\right],\\
% % % % % % % % % % % % % 
\mathcal{A}'(p_0,p_{\perp}, p_3) &=&  
\frac{m_{\textrm{th}}^2}{p_0}\int\frac{d\Omega}{4\pi}\frac{1}{p_0-\bm{p\cdot \hat{k}}} = ~ 
\frac{m_{\textrm{th}}^2}{p_0^2}{\cal T}_P,\\ 
\mathcal{B}' (p_0,p_{\perp}, p_3) &=& 
-m_{\textrm{eff}}^2\int\frac{d\Omega}{4\pi}\frac{\hat{k_3}}{p_0-\bm{p
\cdot\hat{k}}} = ~ \frac{m_{\textrm{eff}}^2\,\,p_3 }{p^2}\left[1-{\cal T}_P\right], \label{af}\\ 
\mathcal{C}' (p_0,p_{\perp}, p_3) &=&  
m_{\textrm{eff}}^2\int\frac{d\Omega}{4\pi}\frac{1}{p_0-\bm{p\cdot 
\hat{k}}} = ~ \frac{m_{\textrm{eff}}^2}{p_0}{\cal T}_P,
\eea 

where we have written the coefficients in terms of
\bea
{\cal T}_P = \int\frac{d\Omega}{4\pi}\frac{p_0}{p_0-\bm{p\cdot\hat{k}}}, 
\label{ang_avg}
\eea
which is  an integral defined as the angular average over $\angle \bm{p}, \bm{\hat k}$. For convenience, the arguments
in all those structure functions will be omitted henceforth.

\subsection{One-loop quark free energy}
\label{qfw}

 In statistical field theory the partition function $Z$ can be represented as a 
functional determinant and by which  the quark  part of 
the free energy in one-loop order can be written as
\bea
F_q&=& 
-N_c\sum_f\int\frac{d^4P}{(2\pi)^4}~\ln\left({\mbox{det}}\left[S_{\textrm{eff}}^{-1}(P)\right]
\right), 
%&=& 
%-N_cN_f\int\frac{d^4P}{(2\pi)^4}~\ln\left({\mbox{det}}\left[\slashed{P}-\Sigma(P)\right]
%\right),
\label{quark_free_energy_primary}
\eea
where  $P\equiv\(p_0,p=|\bm{p}|\)$ is the four momentum of the external fermion with $N_f$ flavor. For ideal gas of quarks the free energy reads as 
\bea
F_q^\textrm{ideal} 
&=& -2N_c\sum_f\int\frac{d^4P}{(2\pi)^4}~\ln\left(-P^2\right) \nn\\
&=& 
-\frac{7\pi^2T^4}{180}N_cN_f\left(1+\frac{120}{7}\hat{\mu}^2+\frac{240}{7}
\hat{\mu}^4\right) ,
\label{F_qieal}
\eea
where $\hat \mu=\mu/2\pi T$. 

The quark free energy in terms of the inverse of general quark propagator is already defined in Eq.~\eqref{quark_free_energy_primary}. In terms of the notations introduced in Sec.~\ref{tpff}, we evaluate the determinant of Eq.~(\ref{pslash_sigma}) as
\bea
{\mbox{det}}\left[S_{\textrm{eff}}^{-1}\right] &=& (\mathcal{C}^2p_0^2-\mathcal{D}^2p^2+\mathcal{B}'^2-\mathcal{C}'^2)^2-4(p_0 
\mathcal{B}'\mathcal{C}+p_3 \mathcal{C}'\mathcal{D})^2\nn\\
&=& A_0^2-A_s^2.
\label{A0As}
\eea
% % % % % % % % % 
Combining Eqs.~(\ref{quark_free_energy_primary}) and (\ref{A0As}), the one-loop quark  free energy in 
 a weak magnetic field and HTL approximation  can be written as
\bea
F_q&=& 
% -N_cN_f\int\frac{d^4P}{(2\pi)^4}~\ln\left(det\left[\slashed{P}-\Sigma(P)\right]
% \right),\nn\\
%%%
% &=& 
-N_c\sum_f\int\frac{d^4P}{(2\pi)^4}~\ln\left(A_0^2-A_s^2\right)\nn\\
&=& -2N_cN_f\int\frac{d^4P}{(2\pi)^4}~\ln\left(P^2\right) 
-N_c\sum_f\int\frac{d^4P}{(2\pi)^4}~\ln\left(\frac{A_0^2-A_s^2}{P^4}\right)\nn\\
&=& 
-\frac{7\pi^2T^4N_cN_f}{180}\left(1+\frac{120}{7}\hat{\mu}^2+\frac{240}{7}
\hat{\mu}^4\right) \nn\\
&&\hspace{1cm}- N_c\sum_f\int\frac{d^4P}{(2\pi)^4}~\ln\left[\frac{(A_0+A_s)(A_0-A_s)}{P^4}\right].
\label{F_qHTL}
\eea
Now the argument of the logarithm in 
Eq.~(\ref{F_qHTL}) can be simplified using Eq.~(\ref{A0As}) as
\bea
&&\frac{(A_0+A_s)(A_0-A_s)}{P^4} = 1 + 
2\left(\frac{\mathcal{A}'(\mathcal{A}'-2)p_0^2-\mathcal{A}(\mathcal{A}+2)p^2+\mathcal{B}'^2-\mathcal{C}'^2}{P^2}\right) \nn\\
&&+ \frac{\left(\mathcal{A}'(\mathcal{A}'-2)p_0^2-\mathcal{A}(\mathcal{A}+2)p^2+\mathcal{B}'^2-\mathcal{C}'^2\right)^2-4(\mathcal{B}'\mathcal{C}~p_0 
+\mathcal{C}'\mathcal{D}~p_3)^2}{P^4}.
\label{log_arg}
\eea
%%%%%%%%%%%%%%%%%%%%
In the high temperature limit, the logarithmic term in Eq.~\eqref{F_qHTL} can be expanded\footnote{In Sec.~\ref{full} we will obtain the free energy without high $T$ expansion 
in the spirit of HTL perturbation theory and compare the result with the high $T$ expansion.}
in a series of coupling constants $g$ keeping terms up to $\mathcal{O}(g^4)$ as
\bea
\ln\left[\frac{(A_0+A_s)(A_0-A_s)}{P^4}\right] &=& 
2\left(\frac{\mathcal{A}'^2p_0^2-\mathcal{A}^2p^2+\mathcal{B}'^2-\mathcal{C}'^2-2\mathcal{A}'p_0^2-2\mathcal{A}p^2}{P^2}\right) \nn\\
&-& 4\left(\frac{\left(\mathcal{A}'p_0^2+\mathcal{A}p^2\right)^2+(\mathcal{B}'p_0 
+\mathcal{C}'p_3)^2}{P^4}\right)+\mathcal{O}(g^6),
\eea
with
\bea
(\mathcal{B}'^2-\mathcal{C}'^2) &=&  m_{\textrm{eff}}^4\left[\frac{p_3^2}{p^4}+\frac{{\cal 
T}_P^2 p_3^2}{p^4}-\frac{{\cal T}_P^2}{p_0^2}-\frac{2{\cal T}_P p_3^2}{p^4}\right],\\
(\mathcal{B}'p_0 +\mathcal{C}'p_3)^2 &=& m_{\textrm{eff}}^4\left[\frac{p_0^2p_3^2}{p^4}\left(1+{\cal 
T}_P^2-2{\cal T}_P\right)+\frac{{\cal T}_P^2}{p_0^2}~p_3^2+\frac{2p_3^2}{p^2}\big({\cal T}_P-{\cal T}_P^2\big)\right],\\
(\mathcal{A}'p_0^2 +\mathcal{A}p^2) &=& m_{\textrm{th}}^2,\\
(\mathcal{A}'^2p_0^2 -\mathcal{A}^2p^2) &=& m_{\textrm{th}}^4\left[\frac{{\cal 
T}_P^2}{p_0^2}-\frac{\left(1-{\cal T}_P\right)^2}{p^2}\right].
\eea

So, up to  $\mathcal{O}(g^4)$ the one-loop free energy can be written as, 
\bea
F_q &=&  N_cN_f\Bigg[ 
-\frac{7\pi^2T^4}{180}\left(1+\frac{120}{7}\hat{\mu}^2+\frac{240}{7}\hat{\mu}
^4\right)+4m_{\textrm{th}}^2\int\frac{d^4P}{(2\pi)^4}\frac{1}{P^2}\nn\\
&-&m_{\textrm{th}}^4\int\frac{d^4P}{(2\pi)^4}\left[\frac{2{\cal 
T}_P^2}{p_0^2P^2}-\frac{4}{P^4}-\frac{2}{p^2P^2}-\frac{2{\cal 
T}_P^2}{p^2P^2}+\frac{4{\cal T}_P}{p^2P^2}\right]\Bigg]\nn\\
%%%%%%%%%%%%%%%%%%%%%%%%%%%%%%%%%%%%%%%%%%%%%%%%%%%%
&-&N_c\sum_f m_{\textrm{eff}}^4 \int\frac{d^4P}{(2\pi)^4}\bigg[\frac{2}{P^2
}\left(\frac{p_3^2}{p^4}+\frac{{\cal T}_P^2 p_3^2}{p^4}-\frac{{\cal 
T}_P^2}{p_0^2}-\frac{2{\cal 
T}_P p_3^2}{p^4}\right)-\frac{4}{P^4}\bigg(\frac{p_0^2p_3^2}{p^4}\left(1+{\cal 
T}_P^2-2{\cal T}_P\right)\nn\\
&& +\ \frac{{\cal 
T}_P^2}{p_0^2}~p_3^2+\frac{2p_3^2}{p^2}\big({\cal T}_P-{\cal T}_P^2\big)\bigg)\bigg],\nn\\
%%%%%%%%%%%%%%%%%%%%%%%%%%%%%%%%%%%%%
&=& N_cN_f\Bigg[ 
-\frac{7\pi^2T^4}{180}\left(1+\frac{120}{7}\hat{\mu}^2+\frac{240}{7}\hat{\mu}
^4\right)+4m_{\textrm{th}}^2\sumintf_{\{ P\} }\frac{1}{P^2}\nn\\
&&-\ m_{\textrm{th}}^4\left(\sumintf_{\{ P\} }\frac{2{\cal T}_P^2}{p_0^2P^2}-\sumintf_{\{ P\} 
}\frac{4}{P^4}-\sumintf_{\{ P\} }\frac{2}{p^2P^2}-\sumintf_{\{ P\} }\frac{2{\cal 
T}_P^2}{p^2P^2}+\sumintf_{\{ P\} }\frac{4{\cal 
T}_P}{p^2P^2}\right)\Biggr]\nn\\
&-&N_c\sum_f 2m_{\textrm{eff}}^4\left(\sumintf_{\{ P\} 
}\frac{p_3^2}{p^4P^2}+\sumintf_{\{ P\} 
}\frac{{\cal T}_P^2 p_3^2}{p^4P^2}-\sumintf_{\{ P\} }\frac{{\cal 
T}_P^2}{p_0^2P^2}-\sumintf_{\{ P\} }\frac{2{\cal T}_P p_3^2}{p^4P^2}\right.\nn\\
&&+\  
2\left.\bigg(\sumintf_{\{ P\} }\frac{p_3^2}{p^2P^4}+\sumintf_{\{ P\} }\frac{p_3^2}{p^4P^2}-\sumintf_{\{ P\} }\frac{p_3^2 \,{\cal 
T}_P^2}{p^2P^4}+\sumintf_{\{ P\} }\frac{p_3^2 \,{\cal 
T}_P^2}{p^4P^2}-\sumintf_{\{ P\} }\frac{2p_3^2\,{\cal T}_P}{p^4P^2}+\sumintf_{\{ P\} }\frac{{p_3^2\,\cal 
T}_P^2}{p_0^2P^4}\bigg)\right),\nn\\
%%%%%%%%%%%%%%%%
&=&  
N_cN_f\Bigg[-\frac{7\pi^2T^4}{180}\left(1+\frac{120}{7}\hat{\mu}^2+\frac{240}{7}
\hat{\mu}^4\right)+\frac{m_{\textrm{th}}^2T^2}{6}\left(1+12\hat{\mu}^2\right)\nn\\
&+&4m_{\textrm{th}}^4\left[1+\frac{1- 2\Delta_3+ \Delta_4''- \Delta_3''}{2-d}\right]\sumintf_{\{P\}}\frac{1}{P^4}\Bigg]- N_c\sum_f\frac{4m_{\textrm{eff}}^4}{2-d} \bigg(\Delta_0'+\Delta_4''\Delta_0'-2\Delta_3 \Delta_0'-\Delta_3''\bigg)\nn\\
&&\times\sumintf_{\{ P\}}\frac{1}{P^4}+N_c\sum_f4m_{\textrm{eff}}^4\Bigg(\big(1-\Delta_0''\big)\Delta_0'+\big(1-\Delta_0''-\Delta_{10}\big)\Delta_0'\frac{2}{2-d}+\frac{2\big(\Delta_4''\Delta_0'-2\Delta_3\Delta_0'\big)}{2-d}\nn\\
&&+\frac{\Delta_0''}{d}-\frac{2}{d(d-2)}\Delta_{11}\Bigg)\sumintf_{\{ P\} }\frac{1}{P^4},\hspace{1cm}
\eea

%%%%5
where all the necessary sum-integrals are provided in Appendix~\ref{a_sumint}. $\Delta_i$'s are the $c$ integrations arising due to angular integral ${\cal T}_P$ which are computed in Appendix~\ref{ang_av}. Using the expressions for various sum-integrals obtained in Appendix \ref{a_sumint} we can write
%%%%%%%%%%%%%%%%%%%%%%%

\bea
F_q&=& 
N_cN_f\Bigg[-\frac{7\pi^2T^4}{180}\left(1+\frac{120}{7}\hat{\mu}^2+\frac{240}{7}
\hat{\mu}^4\right) \nn\\
&+&\frac{m_{\textrm{th}}^2T^2}{6}\left(1+12\hat{\mu}^2\right)+4m_{\textrm{th}}^4\left[
\left(\frac{\pi^2}{3}-2\right)\eps\right]\sumintf_{\{P\}}\frac{1}{P^4}\Bigg]\nn\\
&+& N_c\sum_f m_{\textrm{eff}}^4\left[\frac{2\pi^2}{9} +\left(\frac{4 \zeta (3)}{3}-\frac{8}{3}-\frac{2 \pi ^2}{27}\right) 
\epsilon\right]\sumintf_{\{P\}}\frac{1}{P^4},\nn\\
&=&  
N_cN_f\Bigg[-\frac{7\pi^2T^4}{180}\left(1+\frac{120}{7}\hat{\mu}^2+\frac{240}{7}
\hat{\mu}^4\right)+\frac{m_{\textrm{th}}^2T^2}{6}\left(1+12\hat{\mu}^2\right)+\frac{m_{\textrm{th}}^4}{
12\pi^2}\left(\pi^2-6\right)\Bigg]\nn\\
 &+& 
N_c\sum_f\frac{m_{\textrm{eff}}^4}{16}\bigg(\frac{2}{9\eps}+\frac{1}{27} \Big(12 \ln \frac{\hat\Lambda}{2}-6\aleph(z)+\frac{36 \zeta (3)}{\pi^2}-2 -\frac{72}{\pi^2}\Big)\bigg)\Bigg],\nn\\
 &=& N_c N_f\Bigg[ 
-\frac{7\pi^2T^4}{180}\left(1+\frac{120\hat\mu^2}{7}+\frac{240\hat\mu^4}{7}
\right)+\frac{g^2C_FT^4}{48}\left(1+4\hat{\mu}^2\right)\left(1+12\hat{\mu}^2\right)\nn\\
%%%%%%%%%%%%%%
&&+\, \frac{g^4C_F^2T^4}{
768\pi^2}\left(1+4\hat{\mu}^2\right)^2\left(\pi^2-6\right)+\frac{g^4C_F^2}{27N_f}
M_B^4 \bigg(12 \ln \frac{ \hat\Lambda}{2}-6\aleph(z)+\frac{36 \zeta (3)}{\pi^2}\nn\\
&&-2 -\frac{72}{\pi^2}\bigg)\Bigg]+\frac{2N_cg^4C_F^2}{9\epsilon}M_B^4
. \label{free_quark}
\eea
We note that there is no magnetic correction in the ${\mathcal O}[g^2]$ term in \eqref{free_quark}. 
Magnetic correction only appears in ${\mathcal O}[g^4]$  which is  ${\mathcal O}[ (q_fB)^2]$ .
We further note that the themomagnetic correction to the quark part of the free energy in a weak field has ${\cal O}(1/\epsilon)$ divergence, 
originating due to HTL approximation. To obtain a finite contribution  one needs an appropriate
counterterm  which will be discussed later. 

%%%%%%%%%%%%%%%%%%%%%%%%%%%%%%%%%%%%%%%%%%%%%%%%%%%%%%%%%%%%%%%%%%%%%%%%%%%%%%%%%%%%%%%%%%%
%%%%%%%%%%%%%%%%%%%%%%%%%%%%%%%%%%%%%%%%%%%%%%%%%%%%%%%%%%%
\section{Gluon Free energy  in the presence of a  magnetic field}
\label{GF}
It is convenient to calculate the gluon partition function in Euclidean space. In general the QCD partition function for a gluon can be written in Euclidean space as
\be 
{\cal Z}_g = {\cal Z}  {\cal Z}^{\textrm{ghost}},~~
{\cal Z} = N_\xi\prod_{n,\bm{p}} \sqrt{\frac{(2\pi)^D}{\det D_{\mn, E}^{-1}}},~~
{\cal Z}^{\textrm{ghost}} = \prod_{n,\bm{p}} P_E^2,
\ee
where the product over $n$ is for the discrete bosonic Matsubara frequencies 
($\omega_n=2\pi n\beta;\, \, n=0,1,2,\cdots $) due to
Euclidean time whereas  $\bm{p}$ is for the spatial momentum, $D$ is the space-time 
dimension of the theory,  $P_E^2=\omega_n^2+p^2$ is the square of the four-momentum while $D_{\mn,E}^{-1}$ is the inverse gauge boson propagator in Euclidean space.  
$N_\xi=1/(2\pi\xi)^{D/2}$ is the normalization that originates from the introduction of the  Gaussian integral at each location of position while averaging over the gauge condition function 
with a width $\xi$, the gauge fixing parameter.  

Gluon free energy can now be written as
\be
F_g = -(N_c^2-1)\frac{T}{V} \ln {\cal Z}_g = (N_c^2-1)\left[\frac{1}{2}\sumintb_{P_E}~\ln\Big[\textsf{det}
\left(D_{\mn, E}^{-1}(P_E)\right)\Big] -\sumintb_{P}~\ln P_E^2\right], \label{fe_qed}
\ee
where the gauge dependence explicitly cancels due to the presence 
of the normalization factor $N_\xi$.

For an ideal case $\textsf{det}\left(D_{\mn,E}^{-1}(P)\right)=(P_E^2)^4/\xi$ and hence the free energy for $(N_c^2-1)$ massless spin one gluons yields as
\bea
F_g^{\textrm{ideal}} = (N_c^2-1)\sumintb_{P_E}~\ln P_E^2 =(N_c^2-1)\sumintb_{P}~\ln\(- P^2\)= -(N_c^2-1) \frac{\pi^2T^4}{45},
\eea
where $P$ is the four-momentum in Minkowski space and can be written as $P^2=p_0^2-p^2.$

\noindent In presence of thermal background medium~\cite{Lebellac;1996,Kapusta:1989tk,Karmakar:2018aig} one can have
\bea
\textsf{det}\left(D_{\mn,E}^{-1}(P_E)\right) = \frac{P_E^2}{\xi}\left(P_E^2 + \Pi_T\right)^2\left(P_E^2 + \Pi_L\right),
\label{det_tqed}
\eea
with four eigenvalues; respectively $P_E^2$, $(P_E^2 + \Pi_L)$ and two fold degenerate 
$(P_E^2 + \Pi_T)$. Here $\Pi_T$ and $\Pi_L$ are the transverse and longitudinal part of the gluon self-energy in thermal medium. 
Also throughout this manuscript we have considered spacetime dimension $D=4$ with the spatial dimension $d= 3$\footnote{ We will also use $d=3-2\epsilon$ for 
dimensional regularization.}. From now on, we use Minkowski momentum $P$. Eventually the free energy becomes
\bea
F_g^{\textrm{th}} &=& \frac{1}{2}\left[\sumintb_{P}~\ln\(-P^2\) + 2\sumintb_{P}~\ln \left(-P^2 + \Pi_T\right) + \sumintb_{P}~\ln \left(-P^2 + \Pi_L\right)\right] - \sumintb_{P}~\ln \(-P^2\) ,\nn\\
&=& \sumintb_{P}~\ln \left(-P^2+\Pi_T\right) + \frac{1}{2}\sumintb_{P}~\ln \left(1-\frac{\Pi_L}{P^2}\right) \\
&=& (N_c^2-1)\left[(d-1)F_g^T+F_g^L\right], \label{thfq}
\eea
Also, $F_g^T$ and $F_g^L$ are, respectively, the transverse and longitudinal part of the gluon free energy, both of which can be computed with the help of the general structure of gauge boson self-energy evaluated in Refs.~\cite{Lebellac;1996,Kapusta:1989tk,Karmakar:2018aig}. 
 Now, in presence of a hot magnetized medium the general structure of  inverse propagator of a gauge boson is computed in Ref.~\cite{Karmakar:2018aig} and reads as
\bea
\left(\mathcal{D}_{\mn}\right)^{-1} = \frac{P^2}{\xi}\eta_{\mn} + 
\left(P_m^2 - b\right)B_{\mn} + \left(P_m^2 - c\right)R_{\mn} + 
\left(P_m^2 - d\right)Q_{\mn}-a N_{\mn}, 
\label{inverse_prop}
\eea

where 
\bea
P_m^2 = P^2\frac{\xi -1}{\xi}.
\eea
The determinant of inverse of the gauge boson propagator can be evaluated from Eq.~\eqref{inverse_prop} as
\bea
\textsf{det}\left(D_{\mn,E}^{-1}(P)\right) &=& -\frac{P^2}{\xi}\left(-P^2
+c\right)\left\{\left(-P^2+b\right)\left(-P^2+d\right)-a^2\right\}\nn\\
&=& -\frac{P^2}{\xi}\left(-P^2
+c\right)\left(-P^2+\frac{b+d+\sqrt{(b-d)^2+4a^2}}{2}\right)\nn\\
&&\times\left(-P^2+\frac{b+d-\sqrt{(b-d)^2+4a^2}}{2}\right), \label{det_mqed}
\eea
with four eigenvalues: $-P^2/\xi, \, \left(-P^2+c\right), \,\left(-P^2+\frac{b+d+\sqrt{(b-d)^2+4a^2}}{2}\right) \, {\mbox{and}} \, 
\left(-P^2+\frac{b+d-\sqrt{(b-d)^2+4a^2}}{2}\right)$. We note here that instead of a two fold degenerate transverse mode $(-P^2 +\Pi_T)$ in thermal medium in Eq.~\eqref{det_tqed}, now one has two distinct transverse modes, $(-P^2 + c)$ and $(-P^2 + d)$.
Using Eq.~\eqref{det_mqed} in Eq.~\eqref{fe_qed}, the one-loop gluon free energy 
for hot magnetized medium is given by
\bea
F_g = (N_c^2-1)
\left[\mathcal{F}_g^1+\mathcal{F}_g^2+\mathcal{F}_g^3\right], 
\label{free_qed}
\eea
where
\begin{subequations}
\begin{align}
\mathcal{F}_g^1 &= \frac{1}{2}\sumintb_{P}~\ln\left(1-\frac{b+d+\sqrt{(b-d)^2+4a^2}}{2P^2}\right), \label{f1_qed_n}\\
\mathcal{F}_g^2 &= \frac{1}{2}\sumintb_{P} ~\ln\left(-P^2 + c\right),\label{f2_qed_n}\\
\mathcal{F}_g^3 &= \frac{1}{2}\sumintb_{P} ~\ln\left(-P^2 + \frac{b+d-\sqrt{(b-d)^2+4a^2}}{2}\right).\label{f3_qed_n}
\end{align}
\end{subequations}
In this section, we are considering small magnetic field approximation and we are calculating all
the quantities up to $O (eB)^2$ . Within this approximation, Eq.\eqref{f1_qed_n},\eqref{f2_qed_n},\eqref{f3_qed_n} can be approximated as
\begin{subequations}
\begin{align}
\mathcal{F}_g^1 &= \frac{1}{2}\sumintb_{P}~\ln\left(1-\frac{b}{P^2}\right), \label{f1_qed}\\
\mathcal{F}_g^2 &= \frac{1}{2}\sumintb_{P} ~\ln\left(-P^2 + c\right),\label{f2_qed}\\
\mathcal{F}_g^3 &= \frac{1}{2}\sumintb_{P} ~\ln\left(-P^2 + d\right).\label{f3_qed}
\end{align}
\end{subequations}
The various structure functions are obtained in Ref.~\cite{Karmakar:2018aig} for both strong and weak field approximation.
In the following subsections we would discuss the QCD Debye mass and the gluon free energy in weak magnetic field approximation.

%%%%%%%%%%%%%%%%%%%%%%%%%%%%%%%%%%%%%%%%%%%%%%%%%%%%%%%%%%%

\subsection{QCD Debye mass in a magnetized hot and dense medium}
\label{mD_modified}
%%%%%%%%%%%%%%%%%%%%%%%%%%%%%%%%%%%%%%%%%%%%%%%%%%%%%%%%%%%%%%%%%%%%%%%%%%%%%%%%%%%%%%%%%%%%%%%%%%%%%%%%%
The electromagnetic Debye mass in the presence of a magnetic field was computed 
in~\cite{Karmakar:2018aig,Alexandre:2000jc,Bandyopadhyay:2016fyd}.
Below we extend the calculation of Ref.~\cite{Alexandre:2000jc} for QED Debye mass at finite chemical potential. Using Eq.(23) of Ref.~\cite{Alexandre:2000jc} we can straightaway write down 
\bea
\left.\left(m_D^{B}\right)^2\right|_{\rm QED} &=&  \frac{-\alpha T}{\sqrt{\pi}}eB\int_0^\infty du\sqrt{u}\int_{-1}^{1} dv \sum_{l=-\infty}^{\infty}\exp\left\{-u\left(m^2 + W_l^2\right) \right\} \coth\bar{u}\( 2W_l^2-\frac{1}{u}\),
\label{md_qed}
\eea
where $u/v$, $l$ and $\alpha$ represent respectively the proper time, Landau levels and QED coupling constant with $W_l=(2l+1)\pi T -i\mu$ at finite chemical potential.
%%%%
Now, the Poisson resummation~\cite{Alexandre:2000jc} of the Landau levels is represented as 
\bea
\sum_{l=-\infty}^{\infty}\exp^{-a\left(l-z\right)^2} = \(\frac{\pi}{a}\)^{1/2} \sum_{l=-\infty}^{\infty} \exp^{-\pi^2l^2/a-2i\pi z l}.
\label{pr1}
\eea
Taking derivative in both side with respect to  $a$, we obtain
\bea
\sum_{l=-\infty}^{\infty}e^{-a\left(l-z\right)^2}\(l-z\)^2 =\frac{1}{2a} \(\frac{\pi}{a}\)^{1/2} \sum_{l=-\infty}^{\infty} \exp^{-\pi^2l^2/a-2i\pi z l}-\frac{\pi^{5/2}}{a^{5/2}} \sum_{l=-\infty}^{\infty} l^2\exp^{-\pi^2l^2/a-2i\pi z l}.
\label{pr2}
\eea
Using Eqs.~(\ref{pr1}) and~(\ref{pr2}), we can write
\bea
\sum_{l=-\infty}^{\infty}\exp\left(-u  W_l^2\right) \( 2W_l^2-\frac{1}{u}\) &=& \sum_{l=-\infty}^{\infty}e^{-4u\pi^2 T^2 \(l-i\hat{\mu}+1/2\)^2 } \( 8\pi^2 T^2 \(l-i\hat{\mu}+1/2\)^2-\frac{1}{u}\), \nn\\
&=&  \sum_{l=-\infty}^{\infty}e^{-a \(l-z\)^2 }\frac{1}{u} \( 2a \(l-z\)^2-1\),\nn\\
&=& -\frac{2\pi^{5/2}}{a^{3/2}u} \sum_{l=-\infty}^{\infty} l^2\exp^{-\pi^2l^2/a-2i\pi z l},
\label{pr3}
\eea
with $a=4u\pi^2 T^2,\ z=i\hat{\mu}-1/2$, $\hat{\mu}$ being $\mu/2\pi T$. So, Eq.~(\ref{pr3}) becomes
\bea
\sum_{l=-\infty}^{\infty}\exp\left(-u  W_l^2\right) \( 2W_l^2-\frac{1}{u}\)&=& -\frac{1}{4\sqrt{\pi}T^3u^{5/2}} \sum_{l=-\infty}^{\infty} l^2\exp^{-l^2/4uT^2-2\pi \hat{\mu} l} e^{i\pi l},\nn\\
&=& \frac{1}{4\sqrt{\pi}T^3u^{5/2}} \sum_{l=-\infty}^{\infty} (-1)^{l+1}l^2\exp^{-l^2/4uT^2-2\pi \hat{\mu} l},\nn\\
&=& \frac{1}{4\sqrt{\pi}T^3u^{5/2}} \sum_{l=-\infty}^{\infty} (-1)^{l+1}l^2e^{-l^2/4uT^2}\(e^{-2\pi \hat{\mu} l} +e^{2\pi \hat{\mu} l}\),\nn\\
&=& \frac{1}{2\sqrt{\pi}T^3u^{5/2}} \sum_{l=-\infty}^{\infty} (-1)^{l+1}l^2\cosh\(2\pi \hat{\mu} l \)e^{-l^2/4uT^2}.
\label{sum_Wl}
\eea
Using Eq.~(\ref{sum_Wl}) in Eq.~(\ref{md_qed}), we get
\bea
\left.\left(m_D^{B}\right)^2\right|_{\rm QED} &=& 
\frac{\alpha eB}{\pi T^2}\int\limits_0^\infty \frac{du}{u^2} \coth\left(eB u\right)\exp\left(-um^2\right)\nn\\
&&\hspace{1cm}\times\sum\limits_{l=1}^\infty (-1)^{l+1} l^2 \cosh\(2\pi \hat{\mu} l \)\exp\left(-\frac{l^2}{4uT^2}\right).
\label{md_qed3}
\eea 
%%%%%
Changing the variable from $u$ to $x= l^2/(4uT^2)$, we get
\bea
\left.\left(m_D^{B}\right)^2\right|_{\rm QED} &=& 
\frac{e^2\,eB}{\pi^2 T^2}\int\limits_0^\infty e^{-x}dx\sum\limits_{l=1}^\infty 
(-1)^{l+1}\coth\left(\frac{eBl^2}{4xT^2}
\right)\exp\left(-\frac{m^2l^2}{4xT^2}\right).
\label{md_qed2}
\eea
%%%%%%%
Generalizing  this to QCD  we obtain 
the expression for the modified QCD Debye mass at finite chemical potential and 
an arbitrary magnetic field as 
\bea
\left(m_D^{B}\right)^2 &=& \frac{g^2N_cT^2}{3}+\sum\limits_f 
\frac{g^2q_fB}{2\pi^2}\int\limits_0^\infty e^{-x}dx\nn\\
&&\hspace{1cm}\times\sum\limits_{l=1}^\infty 
(-1)^{l+1}\cosh\left(2l\pi\hat{\mu}\right)\coth\left(\frac{q_fBl^2}{4xT^2}
\right)\exp\left(-\frac{m_f^2l^2}{4xT^2}\right),
\label{md_full}
\eea
where the first term in Eq.~(\ref{md_full}) is due to the pure gluonic contributions, which was not present in the case of QED. 
The second term comes from quark loop contribution which is obtained by replacing $m$ with $m_f$ and $e$  with $g$. Also a quark-flavor 
sum with QCD factor as $\frac{1}{2}\sum_f$ is considered.

Now, in the strong magnetic field limit ($m^2_{\textrm{th}} \sim g^2T^2 \le T^2\le q_fB$), i.e., in LLL, neglecting the current quark mass $m_f$, from 
Eq.~\eqref{md_full} we can straightaway reach a simplified expression\footnote{Our fermionic part of Debye mass is different from Ref.~\cite{Singh:2017nfa} by a factor of 2  which was somehow overlooked by the authors of the Ref.~\cite{Singh:2017nfa}	in Matsurbara Sum. We also find the same mismatch with the Ref.~\cite{Hasan:2017fmf}} given as
\bea
\left(m_D^s\right)^2 = \frac{g^2N_cT^2}{3}+\sum\limits_f 
\frac{g^2q_fB}{4\pi^2}.
\label{md_strong}
\eea
We were able to get the same  expression for Debye mass in the strong magnetic field limit as in Eq.~\eqref{md_strong} when 
we calculate the gluon polarization tensor using a  quark propagator in strong field approximation  and take the static limit of 
the zero-zero component of that tensor~\cite{Karmakar:2018aig}.
%%%%%%%%%%%%%%%%%%%%%%%%%%%%%%%%
\begin{center}
\begin{figure}[tbh]
\begin{center}
\subfigure{\includegraphics[scale=0.43]{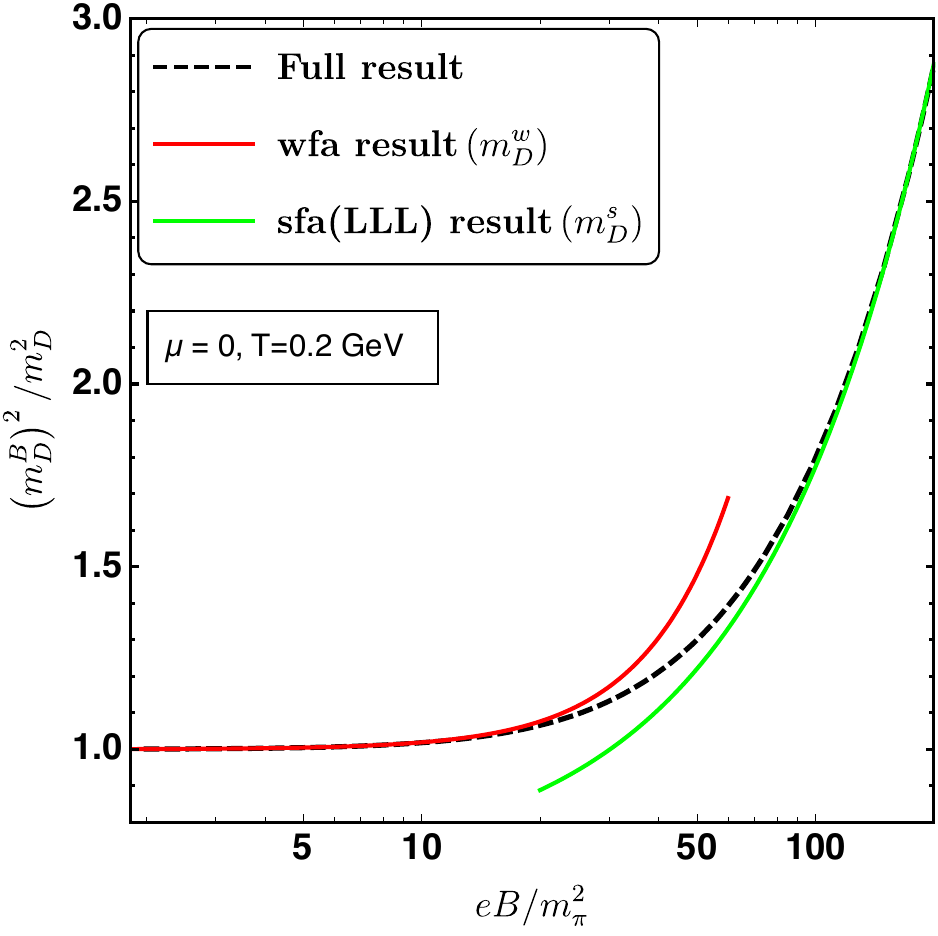} 
{\includegraphics[scale=0.43]{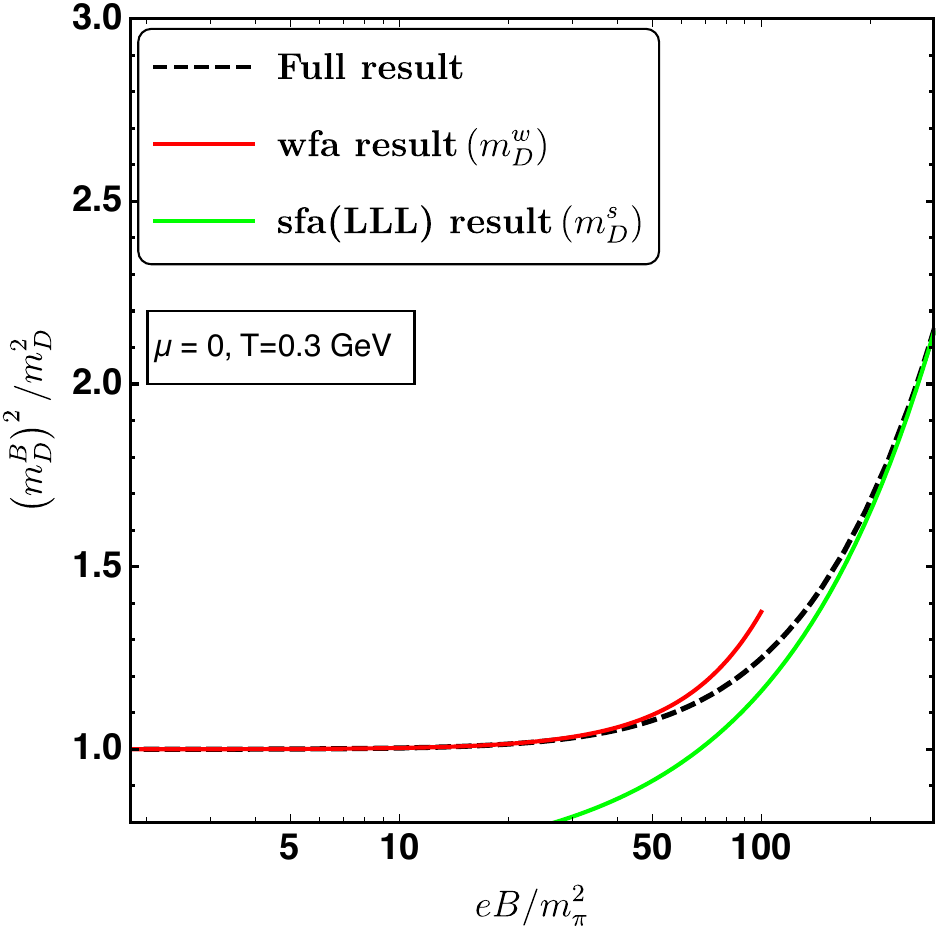} }}
 \caption{Comparison of the scaled one-loop Debye masses in Eqs.(\ref{md_full}), ~\eqref{md_strong}  and 
(\ref{md_wfa}) varying with scaled 
magnetic field for $N_f=3$, $\mu=0$.  Left: $T=200$ MeV, Right: $T=300$ MeV.}
 \label{md_full_vs_expanded}
\end{center}
\end{figure}
\end{center}
%%%%%%%%%%
In the weak field approximation ($T^2 > m^2_{\textrm{th}} > q_fB$), the square of Debye mass can be obtained
from Eq.~(\ref{md_full}) by expanding $\coth\left({q_fBl^2}/{4xT^2}\right) $
as
\bea
\(m_D^w\)^2 &\simeq& 
\frac{g^2 T^2}{3}\left[\left(N_c+\frac{N_f}{2}\right)+6N_f\hat{\mu}
^2\right] \nn\\
&+&\sum\limits_f \frac{g^2(q_fB)^2}{12\pi^2T^2} 
\sum\limits_{l=1}^\infty 
(-1)^{l+1}l^2 \cosh\left(2l\pi\hat{\mu}\right) K_0\left(\frac{m_fl}{T}\right) + 
\mathcal{O}[(q_fB)^4] \nn \\
&=& m_D^2 + \delta m_D^2,
\label{md_wfa}
\eea
where $m_D$ can be identified as the QCD Debye mass in a hot and dense medium in the absence of any external magnetic field and $K_n(z)$ 
represents the modified Bessel function of the second kind. 
%In  Eq.~\eqref{md_wfa} the first term is the Debye mass contribution in the 
%absence of the external magnetic field whereas the second term is 
This second term is the thermomagnetic correction due to the presence of the weak 
external magnetic field. Note that Eq.~(\ref{md_wfa}) is valid only when $\mu\leq m_f$, as the infinite sum over $l$ diverges at $\mu> m_f$.

In Fig.~\ref{md_full_vs_expanded} the full expression in Eq.~(\ref{md_full}), the strong field expression 
in Eq.~\eqref{md_strong} and the weak field expression in Eq.~(\ref{md_wfa})  scaled with $m_D$ are displayed as a function of the magnetic field scaled with squared pion mass. In the strong field limit, (e.g., For $T=200$ MeV, $|eB|/m_\pi^2\ge 10$) 
the weak field result (red colored curve) starts to deviate from the full result (dashed line). However, there is no difference 
between the two in limit $|eB|/m_\pi^2 < 10$, so  it defines the domain of weak field for $T=200$ MeV and  is indeed a good approximation 
to work with Eq.~(\ref{md_wfa}) in the weak field limit at that temperature. On the other hand, the LLL result (green colored line) agrees 
with the full when, e.g., for $T=200$ MeV, $|eB|/m_\pi^2\ge 70$)  MeV. The right panel is for $T=300$ MeV which shows the same behavior.  However, these two plots indicate that the 
domain of applicability for strong ($|eB| > T^2$) and weak ($|eB| < T^2$) fields  changes quantitatively with the change 
in temperature. In between the weak and strong field domain in principle one should work with full expression which is indeed a very involved and difficult task. However, presently we confine ourselves  in the weak  magnetic field limit.

%%%%%%%%%%%%%%%%%%%%%%%%%%%%%%%%%%%%%%%%%%%%%%%%%%%%%%%%%%%%%%%%%%%%%%%%%%%%%%%%%%%%%%%%%%%%%%%%%%%%%%%%

 %%%%%%%%%%%%%%%%%%%%%%%%%%%%%%%%%%%%%%%%%%%%%%%%%%%%%%%%%%
 %%%%%%%%%%%%%%%%%%%%%%%%%%%%%%%%%%%%%%%%%%%%%%%%%%%%%%%%%%
 \subsection{One-loop gluon free energy in weakly magnetized hot medium}
 \label{GFW}

The form factors for the gluonic self-energy in a weakly magnetized medium can be expressed by breaking into thermal and thermomagnetic parts each, as
\bea
b(T,\mu,B)&=& b_0(T,\mu)+b_2(T,\mu,B),\nn\\
c (T,\mu,B)&=& c_0(T,\mu)+c_2(T,\mu,B),\nn\\
d (T,\mu, B)&=& d_0(T,\mu)+d_2(T,\mu,B),\nn
\eea
with
\bea
b_0&=& \frac{m_D^2}{\bar{u}^2}(1-{\cal T}_P), \nn\\
c_0&=& d_0 = \frac{m_D^2}{2p^2}\left[p_0^2-P^2{\cal T}_P\right], \nn\\
b_2 &=& \frac{\delta m_D^2}{\bar{u}^2} + \sum_f\frac{g^2(q_fB)^2}{\bar{u}^2\pi^2}\Biggl[\left(g_k+\frac{\pi m_f-4T}{32m_f^2T}\right)(A_0-A_2)+\left(f_k+\frac{8T-\pi m_f}{128m_f^2T}\right)\(\frac{5}{3}A_0-A_2\)\Biggr],\nn\\
c_2 &=& -\sum_f\frac{4g^2(q_fB)^2}{3\pi^2}g_k + \sum_f\frac{g^2(q_fB)^2}{2\pi^2}\left(g_k+\frac{\pi m_f-4T}{32m_f^2T}\right)\times\Biggl[-\frac{7}{3}\frac{p_0^2}{p_1^2}+\left(2+\frac{3p_0^2}{2p_1^2}\right)A_0\nn\\
&&+\left(\frac{3}{2}+\frac{5p_0^2}{2p_1^2}+\frac{3p_3^2}{2p_1^2}\right)A_2-\frac{3p_0p_3}{p_1^2}A_1-\frac{5}{2}\left(1-\frac{p_3^2}{p_1^2}\right)A_4-\frac{5p_0p_3}{p_1^2}A_3\Biggr],\nn\\
d_2 &=& -\sum_f\frac{g^2(q_fB)^2}{\pi^2}\frac{p^2}{p_1^2}\Bigg[g_k \frac{p_0^2p_3^2}{3p^4} + \left(g_k+\frac{\pi m_f-4T}{32m_f^2T}\right)
\Bigg\{\frac{A_0}{4}-\left(\frac{3}{2}+\frac{p_0^2p_3^2}{p^4}\right)A_2+\frac{5A_4}{4}\Bigg\}\nn\\
&& -\frac{14}{3}f_k \frac{p_0^2p_3^2}{p^4}+ \left(f_k+\frac{8T-\pi m_f}{128m_f^2T}\right)\frac{p_0^2p_3^2}{p^4}(5A_0-A_2)\Bigg]\nn \\ 
&& +\sum_f\frac{g^2(q_fB)^2}{6\pi^2m_fT}~\frac{\left(\frac{3A_1}{2}-A_3\right)}{\left(1+\cosh \frac{m_f}{T}\right)}\frac{p_0p_3}{p_1^2},\label{c2d2_initial}
\eea
where $f_k$, $g_k$ and $A_n= \int\frac{d\Omega}{4\pi}\frac{c^np_0}{p_0-\bm{p\cdot\hat{k}}}$ are calculated in paper-I~\cite{Karmakar:2018aig} with $A_0 \equiv {\cal T}_P$~\footnote{We hereby note that although in paper-I~\cite{Karmakar:2018aig} $f_k$ and $g_k$ are defined with zero chemical potential, it has been checked that within the small $m_f$ approximation (which we are going to use by the virtue of HTLpt), the expressions for $f_k$ and $g_k$ do not change with the inclusion of a finite chemical potential. (see Eq.(\ref{approx_small_mf}))}. 

Therefore, one can write the longitudinal and transverse parts of the gluonic free energy in a weakly magnetized medium respectively as
\bea
\mathcal{F}_g^1&=& \frac{1}{2}\sumintb_P~ \ln \left(1 - \frac{b}{P^2}\right)
          =\underbrace{ -\frac{1}{2}\sumintb_P~ \left(\frac{b_0}{P^2}+\frac{b_0^2}{2P^4}\right) }_{\mbox{Thermal part:\,\, } \mathcal{F}_g^{1T}} 
           \underbrace{- \frac{1}{2}\sumintb_P~ \left(\frac{b_2}{P^2}+\frac{b_0b_2}{P^4}\right) }_{\mbox{\,\,\, Thermomagnetic part:\ \,} \mathcal{F}_g^{1B}}-\cdots \nn\\
           &=& \mathcal{F}_g^{1T} + \mathcal{F}_g^{1B}
           \label{fg1_initial}
\eea
and
\bea
\mathcal{F}_g^2+\mathcal{F}_g^3 &=& \left[\sumintb_P~ \ln \(-P^2\) + \frac{1}{2}\sumintb_P~\ln \left(1-\frac{c}{P^2}\right)+ 
\frac{1}{2}\sumintb_P~\ln \left(1-\frac{d}{P^2}\right)\right]\nn\\
&=&  \underbrace{-\frac{\pi^2 T^4}{45} - \sumintb_P~\left(\frac{c_0}{P^2}+\frac{c_0^2}{2P^4}\right)}_{\mbox{Thermal part:}\ \mathcal{F}_g^{2T}+\mathcal{F}_g^{3T}} 
\underbrace{-\frac{1}{2}\sumintb_P~ (c_2+d_2)\left(\frac{1}{P^2}+\frac{c_0}{P^4}\right)}_{\mbox{Thermomagnetic part:}\ \mathcal{F}_g^{2B}+\mathcal{F}_g^{3B}}  \nn \\
&=&\left( \mathcal{F}_g^{2T}+\mathcal{F}_g^{3T} \right )+\left(\mathcal{F}_g^{2B}+\mathcal{F}_g^{3B}\right), \label{2t+2b}
\eea 
where we have kept terms upto $\mathcal{O}(q_fB)^2$ in high $T$ expansion in the same spirit as done for quark case.

Using Eqs.~\eqref{fg1_initial} and \eqref{2t+2b} and sum-integrals from Appendix~\ref{BS} the total thermal part ({\it{i.e.}}, magnetic field independent terms) can  straightaway be written as
\bea
&&\mathcal{F}_g^{1T}+\mathcal{F}_g^{2T}+\mathcal{F}_g^{3T} = -\frac{\pi^2 T^4}{45}-\frac{1}{2}\sumintb_P~ \left(\frac{b_0}{P^2}+\frac{b_0^2}{2P^4}\right)- \sumintb_P~\left(\frac{c_0}{P^2}+\frac{c_0^2}{2P^4}\right)\nn\\
&&=  -\frac{\pi^2 T^4}{45} -\frac{m_D^2}{2}\sumintb_P~ \frac{1}{P^2}-  \frac{m_D^4}{8}\sumintb_P~\Bigg[\frac{1}{P^4}+\frac{2}{p^2P^2}-\frac{6{\cal T}_p}{p^4} -\frac{2{\cal T}_p}{p^2P^2}+\frac{3{\cal T}_p^2}{p^4}\Bigg]\nn\\
&&=-\frac{\pi^2 T^4}{45}+\frac{m_D^2T^2}{24}-\frac{m_D^4}{128\pi^2}\left(\frac{\Lambda}{4\pi T}\right)^{2\eps}\left[\frac{1}{\eps}+2\gamma_E+\frac{2\pi^2}{3}-7\right]
\label{fwg_htl}
\eea
In the following two subsubsections we evaluate the thermomagnetic parts of the gluon free energy. 

\subsubsection{Longitudinal part - $\mathcal{F}_g^{1B}$}
\label{LF}
The  thermomagnetic  contribution  of the longitudinal part can subsequently be expressed from Eqs.~\eqref{fg1_initial} and \eqref{2t+2b}  as
\bea
\mathcal{F}_g^{1B} &=&-\frac{1}{2}\sumintb_P~ \left(\frac{b_2}{P^2}+\frac{b_0b_2}{P^4}\right) = \frac{m_D^2\delta m_D^2}{2}\sumintb_P~ \frac{{\cal T}_P}{p^4}+\sum_f\frac{g^2(q_fB)^2}{2\pi^2} \times \nn\\
&&\Biggl[\left(g_k+\frac{\pi m_f-4T}{32m_f^2T}\right)\left(\sumintb_P~\frac{{\cal T}_P}{p^2}-\sumintb_P~\frac{A_2}{p^2}-m_{D}^2\sumintb_P~\frac{(1-{\cal T}_P)({\cal T}_P-A_2)}{p^4}\right) +\nn\\
&&\left(f_k+\frac{8T-\pi m_f}{128m_f^2T}\right)\left(\frac{5}{3}\sumintb_P\frac{{\cal T}_P}{p^2}-\sumintb_P\frac{A_2}{p^2}\!-\!m_{D}^2\sumintb_P\frac{(1-{\cal T}_P)(\frac{5}{3}{\cal T}_P-A_2)}{p^4}\right)\Biggr]\nn\\
&=& -\frac{m_D^2\delta m_D^2}{(4\pi)^2}\left(\frac{\Lambda e^{\gamma_E}}{4\pi T}\right)^{2\eps}\left(\frac{1}{2\eps}+\ln 2\right)-\sum_f\frac{g^2(q_fB)^2}{2\pi^2}\left(\frac{\Lambda }{4\pi T}\right)^{2\eps} \Biggl[\left(g_k+\frac{\pi m_f-4T}{32m_f^2T}\right) \nn\\
&&\Bigg(\frac{T^2}{72}\left[\frac{2}{\eps}+11.046\right]-\frac{2m_{D}^2e^{2\gamma_E \eps}}{9(4\pi)^2}\left[\frac{1-4\ln 2}{\eps}-5.326\right]\Bigg)+\left(f_k+\frac{8T-\pi m_f}{128m_f^2T}\right) \nn\\
&& \Bigg(\frac{T^2}{72}\left[\frac{4}{\eps}+21.759\right]-\frac{2m_{D}^2e^{2\gamma_E \eps}}{9(4\pi)^2}\left[\frac{2(1-4\ln 2)}{\eps}-10.589\right]\Bigg)\Biggr].
\label{fwg_b_longi}
\eea

By virtue of HTLpt, we now make a small $m_f$ approximation within which
\bea
g_k &\approx& -2f_k \approx \frac{1}{8m_f^2}.
\label{approx_small_mf}
\eea
and subsequently  Eq.(\ref{fwg_b_longi}) becomes
\bea 
\mathcal{F}_g^{1B}&=& -\frac{m_D^2\delta m_D^2}{(4\pi)^2}\left(\frac{\Lambda}{4\pi T}\right)^{2\eps}\left(\frac{1}{2\eps}+\ln 2+\gamma_E\right)-\sum_f\frac{g^2(q_fB)^2}{(12\pi)^2}\frac{\pi T}{32m_f}\left(\frac{\Lambda }{4\pi T}\right)^{2\eps} \nn\\
&&\Bigg\{\left(\frac{1}{\eps}+5.606\right)+\frac{3\hat{m}_{D}^2}{4}\left(\frac{8(4\ln 2-1)}{3\eps}+19.7467\right)\Bigg\},
\label{fwg_b_longi_smf}
\eea
with divergence of ${\cal O}(1/\epsilon)$.
%%%%%%%%%%%%%%%%%%%%%%%%%%%%%%%%%%%%%%%%%%%%%%%%%%%%%%

\subsubsection{Transverse part - $\mathcal{F}_g^{2B}$ and $\mathcal{F}_g^{3B}$}
\label{TF}

From Eq.\eqref{2t+2b} it is evident that to evaluate a thermomagnetic contribution for the transverse part of the gluonic free energy in a weakly magnetized medium one needs to compute the following sum-integrals:
\bea
\sumintb_P~\frac{c_2+d_2}{P^2} &=&  \sum_f\frac{g^2(q_fB)^2}{\pi^2}\Biggl[(14f_k-g_k)\sumintb_P\frac{p_0^2p_3^2}{3p_1^2p^2P^2}-\frac{4}{3}g_k
\sumintb_P\frac{1}{P^2}+\left(g_k+\frac{\pi m_f-4T}{32m_f^2T}\right)\sumintb_P\nn\\
\Biggl\{-\frac{7}{6}\frac{p_0^2}{p_1^2P^2}\!\!&+&\!\!\! \frac{A_0}{P^2} \left(1+\frac{3p_0^2-p^2}{4p_1^2}\right)\!+\!\frac{A_2}{P^2}\left(\frac{5p_0^2+9p^2}{4p_1^2}+\frac{p_0^2p_3^2}{p_1^2p^2}\right)-\frac{5A_4}{2P^2}-\frac{p_0p_3}{2p_1^2P^2}(3A_1+5A_3)\Bigg\}\nn\\
-\Bigl(f_k\!\! &+& \!\!\frac{8T-\pi m_f}{128m_f^2T}\Bigr)\!\sumintb_P\frac{p_0^2p_3^2(5A_0-A_2)}{p_1^2p^2P^2}\!+\! \!\frac{\left(1+\cosh \frac{m_f}{T}\right)^{-1}}{6m_fT} \sumintb_P\frac{\left(\frac{3A_1}{2}-A_3\right)p_0p_3}{p_1^2P^2}\Biggr],
\eea
%%%%%%%%%%%%%%%%%%%%%%%%%%%%%%%%%%%%%%%%%%%%%%%%%%%%%%%%%%%%%%%%%%
and 
\bea
&&\sumintb_P~\frac{c_0(c_2+d_2)}{P^4} = \sum_f\frac{g^2(q_fB)^2m_{D}^2}{2\pi^2}\Biggl[(14f_k-g_k)\sumintb_P\frac{p_0^2p_3^2(p_0^2-A_0P^2)}{3p_1^2p^4P^4}-\frac{4}{3}g_k\sumintb_P\frac{p_0^2-A_0P^2}{p^2P^4}\nn\\
&&+\left(g_k+\frac{\pi m_f-4T}{32m_f^2T}\right)\sumintb_P \left(\frac{p_0^2-A_0P^2}{p^2P^2}\right)\Biggl\{-\frac{7}{6}\frac{p_0^2}{p_1^2P^2}+\frac{A_0}{P^2}
\Bigg(1+\frac{3p_0^2-p^2}{4p_1^2}\Bigg)\nn\\
&&+\frac{A_2}{P^2}\Bigg(\frac{5p_0^2+9p^2}{4p_1^2}+\frac{p_0^2p_3^2}{p_1^2p^2}\Bigg)-\frac{5A_4}{2P^2}-\frac{p_0p_3}{2p_1^2P^2}
\Big(3A_1+5A_3\Big)\Bigg\}-\Bigl(f_k+\frac{8T-\pi m_f}{128m_f^2T}\Bigr)\nn\\
&&\sumintb_P\ \frac{p_0^2p_3^2(5A_0-A_2)(p_0^2-A_0P^2)}{p_1^2p^4P^4}+\frac{\left(1+\cosh \frac{m_f}{T}\right)^{-1}}{6m_fT}\sumintb_P\frac{\left(\frac{3A_1}{2}-A_3\right)p_0p_3(p_0^2-A_0P^2)}{p_1^2p^2P^4}\Biggr].
\eea
Hence, using various specific HTL sum-integrals listed in Appendix~\ref{BS}, the master sum-integrals listed in Appendix~\ref{master_trans} along with angular integrations  listed in Appendix~\ref{ang_av}, one obtains
\bea
&&\mathcal{F}_g^{2B}+\mathcal{F}_g^{3B} = - \frac{1}{2}\sumintb_P~ (c_2+d_2)\left(\frac{1}{P^2}+\frac{c_0}{P^4}\right)\nn\\
&&=-\sum_f\frac{g^2(q_fB)^2T^2}{144\pi^2}\left(\frac{\Lambda}{4\pi T}\right)^{2\eps}\Biggl[(14f_k-g_k)\Bigg\{\left(\frac{1}{\epsilon }+24\ln G\right)+3\hat{m}_D^2\Bigg[\frac{1-\ln 2}{ \epsilon^2}+\frac{1}{\epsilon }\Big(4-\frac{\pi ^2}{6}\nn\\
   &&-\ln ^2(2)-2 \gamma_E  (\ln 2-1)-\ln 2\Big)+4.38\Bigg]\Bigg\}+g_k\Bigg\{8+3\hat{m}_D^2\Bigg[\frac{1}{ \epsilon }\left(4-4\ln 4\right) -2.79\Bigg]\Bigg\}\nn\\
&&+\left(g_k+\frac{\pi m_f-4T}{32m_f^2T}\right)\Biggl\{\frac{3}{8 \epsilon ^2}+\frac{1}{4\epsilon }\Big(36 \ln G+2+15 \ln 2\Big)+20.83+3\hat{m}_D^2\Bigg[\frac{1}{\eps^2}\Big(-\frac{319}{20}+\pi ^2\nn\\
&&+\frac{89 \ln 2}{10}\Big) +\frac{1}{600 \epsilon }\Big(3600 \zeta (3)-37658+2900 \pi ^2+5340 \ln ^2(2)+4736 \ln 2+30 \gamma_E  \big(-638\nn\\
&&+40   \pi ^2+356 \ln 2\big)\Big)+7.18\Bigg]\Bigg\}-\Bigl(f_k\!\!+\!\!\frac{8T-\pi m_f}{128m_f^2T}\Bigr)\Bigg\{\frac{3}{2}\(\frac{1+8\ln 2}{\epsilon }+45.68\)+3\hat{m}_D^2\nn\\
&&\Bigg[\frac{1}{10
   \epsilon ^2}\Big(29+10 \pi ^2-128 \ln 2\Big)+\frac{1}{25 \epsilon }\Big(150 \zeta (3)+564+\frac{125 \pi ^2}{6}+5 \gamma_E  \big(29+10 \pi ^2-128 \ln 2\big)\nn\\
   &&-4\ln 2 (147+80 \ln 2)\Big)+58.01\Bigg]\Bigg\}+ \frac{\left(1+\cosh \frac{m_f}{T}\right)^{-1}}{6m_fT}\Bigg\{\frac{3\ln 2-4}{2\eps}-3.92+3\hat{m}_D^2\nn\\
   &&\Bigg[\frac{1}{40 \epsilon ^2}\Big(11+5\pi ^2-92 \ln 2\Big)+
\frac{1}{600 \epsilon }\Big(450\zeta (3)+4671-200 \pi ^2-1380\ln ^2(2)-4032 \ln 2\nn\\
&&+30\gamma_E(11+5\pi^2-92\ln 2)\Big)-1.86\Bigg]\Bigg\}\Biggr].
\label{fwg_b_trans}
\eea

%%%%%%%%%%%%%%%%%%%%%%%%%%%%%%%%%%%%%%%%%%%%%%%%%%%%%%%%
Similar to the longitudinal part, using small $m_f$ approximations in Eq.(\ref{fwg_b_trans}), we get 
\bea
&&\hspace{-0.5cm}\mathcal{F}_g^{2B}+\mathcal{F}_g^{3B}
= \sum_f\frac{g^2(q_fB)^2}{(12\pi)^2}\frac{T^2}{m_f^2}\left(\frac{\Lambda}{4\pi T}\right)^{2\eps}\nn\\
&\times& \Bigg[\frac{1}{\epsilon }+4.97+3\hat{m}_D^2\Bigg\{\frac{1-\ln 2}{ \epsilon^2}+\frac{1}{\epsilon }\bigg(\frac{7}{2}-\frac{\pi ^2}{6}-\ln ^2(2)-2 \gamma_E  (\ln 2-1)\bigg)+4.73\Bigg\}\Bigg]\nn\\
&-& \sum_f\frac{g^2(q_fB)^2}{(12\pi)^2}\frac{\pi T}{32m_f}\left(\frac{\Lambda}{4\pi T}\right)^{2\eps}\Biggl\{\frac{3}{8\epsilon ^2}+\frac{1}{\epsilon }\(\frac{13}{8}+\frac{3}{4}\frac{\zeta'(-1)}{\zeta(-1)}+\frac{27}{4}\ln 2\)+37.96\nn\\
&+&\frac{3}{4}\hat{m}_D^2\Bigg[\frac{1}{\eps^2}\(5\pi ^2-\frac{609}{10}+\frac{114 \ln 2}{5}\) +\frac{1}{ \epsilon }\Big(30 \zeta (3)-\frac{17137}{75}+\frac{121}{6} \pi ^2+\frac{114}{5}\ln ^2(2)\nn\\
&+&\frac{604}{75} \ln 2+ \gamma_E  \big(10\pi ^2-\frac{609}{5}+\frac{228}{5} \ln 2\big)\Big)+86.73\Bigg]\Bigg\}- \sum_f\frac{g^2(q_fB)^2}{(12\pi)^2}\frac{T}{12m_f}\Bigg\{\frac{3\ln 2-4}{2\eps}\nn\\
&-&3.92+3\hat{m}_D^2\Bigg[\frac{1}{40 \epsilon ^2}\Big(11+5\pi ^2-92 \ln 2\Big)+
\frac{1}{600 \epsilon }\Big(450\zeta (3)+4671-200 \pi ^2-1380\ln ^2(2) \nn\\
&-&4032\ln 2+30\gamma_E(11+5\pi^2-92\ln 2)\Big)-1.86\Bigg]\Bigg\}.
\label{fwg_b_trans_smf}
\eea

We note that the gluon free energy also contains thermomagnetic correction starting from $\mathcal O[|q_fB|^2]$ as the form factors start with $\mathcal O[|q_fB|^2]$.

Due to high temperature expansion within the HTL approximation, there also appears a soft contribution~\cite{Andersen:2002ey} as 
\bea
\mathcal{F}_{g}^{soft} = -\frac{1}{12\pi}(m_D^w)^3T.
\label{fwg_longi_soft}
\eea
where $m_D^w$ is given in Eq.(\ref{md_wfa}). This contains contributions from both the thermal and thermomagnetic parts.
%%%%%%%%%%%%%%%%%%%%%%%%%%%%%%%%%%%%%%%%%%%%%%%%%%%%%%%%%%%%%%%%%%%%%%%%%%%%%%%%%%%%%%%%%%%%%%%%%%%%%%%%%%%%%%%%%%%%%%%%%%%%%%%%%%%%%%%%%%%%%%%%%%%%%%%%%%%%%%%%%%%%%%%%%%%%%%%%%%%%%%%%%%%%%

\section{Total free energy and pressure in weak Field approximation}
\label{RFPW}
\subsection{Free energy}
\label{RF}
Finally, the total one-loop free energy of a weakly magnetized  hot medium as written in Eq.~\eqref{total_fe} reads as
\bea
F &=& F_q+F_g+F_0 + \Delta {\mathcal E}_0,
\label{total_fef}
\eea
where the quark part of the free energy $F_q$ has both HTL ({\it{viz.}}., magnetic field independent) part as well as the thermomagnetic correction as obtained in Eq.~(\ref{free_quark}). Similarly, the gluonic part has also HTL part ($F_g^{\textrm{HTL}}$) plus the thermomagnetic correction ($F^B_g$).
Thus the total gluonic free energy $F_g$ can be written from Eqs.~(\ref{fwg_htl},~\ref{fwg_b_longi},~\ref{fwg_b_trans},~\ref{fwg_longi_soft}) as 
\bea
&&F_g =d_A\left[\mathcal{F}_g^{1T}+\mathcal{F}_g^{2T}+\mathcal{F}_g^{3T}+\mathcal{F}_{g}^{soft}+\mathcal{F}_g^{1B}+\mathcal{F}_g^{2B}+\mathcal{F}_g^{3B}\right]+\Delta {\mathcal E}_0\nn\\
&&= -\frac{d_A\pi^2 T^4}{45}\left[1-\frac{15}{2}\hat{m}_D^2+30(\hat{m}_D^w)^3+\frac{45}{8}\hat{m}_D^4\(2\ln\frac{\hat{\Lambda}}{2}-7+2\gamma_E+\frac{2\pi^2}{3}\)\right]\nn\\
&&+d_A\Bigg[-\frac{m_D^2\delta m_D^2}{(4\pi)^2}\left(\frac{\Lambda}{4\pi T}\right)^{2\eps}\left(\frac{1}{2\eps}+\ln 2+\gamma_E\right)+\sum_f\frac{g^2(q_fB)^2}{(12\pi)^2}\frac{T^2}{m_f^2}\left(\frac{\Lambda}{4\pi T}\right)^{2\eps}\Bigg[\frac{1}{\epsilon }+4.97\nn\\
&&+3\hat{m}_D^2\Bigg\{\frac{1-\ln 2}{ \epsilon^2}+\frac{1}{\epsilon }\bigg(\frac{7}{2}-\frac{\pi ^2}{6}-\ln ^2(2)-2 \gamma_E  (\ln 2-1)\bigg)+4.73\Bigg\}\Bigg]\nn\\
&&- \sum_f\frac{g^2(q_fB)^2}{(12\pi)^2}\frac{\pi T}{32m_f}\left(\frac{\Lambda}{4\pi T}\right)^{2\eps}\Biggl[\Biggl\{\frac{3}{8\epsilon ^2}+\frac{1}{\epsilon }\(\frac{21}{8}+\frac{3}{4}\frac{\zeta'(-1)}{\zeta(-1)}+\frac{27}{4}\ln 2\)+43.566\nn\\
&&+\frac{3}{4}\hat{m}_D^2\Bigg[\frac{1}{\eps^2}\(5\pi ^2-\frac{609}{10}+\frac{114 \ln 2}{5}\) +\frac{1}{ \epsilon }\bigg( 30 \zeta (3)-\frac{5779}{75}+\frac{121}{6} \pi ^2+\frac{114}{5}\ln ^2(2)\nn\\
&&+\frac{468}{25} \ln 2+ \gamma_E  \(10\pi ^2-\frac{609}{5}+\frac{228}{5} \ln 2\)\bigg)+106.477\Bigg]\Bigg\}+\frac{8}{3\pi}\Bigg\{\frac{3\ln 2-4}{2\eps}-3.92\nn\\
&&+3\hat{m}_D^2\Bigg[\frac{1}{40 \epsilon ^2}\Big(11+5\pi ^2-92 \ln 2\Big)+
\frac{1}{\epsilon }\(\frac{3}{4}\zeta (3)+\frac{1557}{200}- \frac{\pi ^2}{3}-\frac{23}{10}\ln ^2(2)\right.\nn\\
&&-\left.\frac{168}{25} \ln 2+\gamma_E\(\frac{11}{20}+\frac{\pi^2}{4}-\frac{23}{5}\ln 2\)\)-1.86\Bigg]\Bigg\}\Biggr]\Bigg],\label{free_gluon_b}
\eea
where we have also used the HTL counterterm~\cite{Haque:2012my}  as given in Eq.~\eqref{htl_count}, $d_A=N_c^2-1$, $\hat{m}_D = m_D/2\pi T$ and $\hat{m}_D^w = m_D^w/2\pi T$.

%%%%%%%%%%%%%%%%%%%%%%%%%%%%%%%%%%%%%%%%%%%%%%%%%%%%%%%%
The magnetic field dependent gluonic free energy has both ${\cal O}(1/\epsilon)$ (UV) and ${\cal O}(1/\epsilon^2)$ (both colinear and UV) divergences. Now, the external magnetic field $B$ dependent divergences present in Eqs.(\ref{free_quark}) and (\ref{free_gluon_b}) can be removed~\cite{Andersen:2014xxa} by redefining the magnetic field contribution in the tree-level free energy as
\bea
F_0&=&\frac{B^2}{2}\rightarrow \frac{B^2}{2}\left[1 -\frac{4N_cg^4C_F^2}{9\epsilon}\frac{M_B^4}
{B^2} +\frac{m_D^2\delta m_D^2}{\eps(4\pi)^2B^2} -\sum_f\frac{g^2q_f^2}{(12\pi)^2}\frac{2T^2}{m_f^2} \Bigg[\frac{1}{\eps}\right.\nn\\
&+&\left.3\hat{m}_D^2\Bigg\{\frac{1-\ln 2}{ \epsilon^2}+\frac{1}{\epsilon}\bigg(\frac{7}{2}-\frac{\pi ^2}{6}-\ln ^2(2)-2 \(\gamma_E +\ln\frac{\hat\Lambda}{2}\) (\ln 2-1)\bigg)\Bigg\}\right]\nn\\
&+& \sum_f\frac{g^2q_f^2}{(12\pi)^2}\frac{\pi T}{16m_f}\Biggl[\Biggl\{\frac{3}{8\epsilon ^2}+\frac{1}{\epsilon }\(\frac{21}{8}+\frac{3}{4}\frac{\zeta'(-1)}{\zeta(-1)}+\frac{27}{4}\ln 2+\frac{3}{4}\ln\frac{\hat\Lambda}{2}\)\nn\\
&+&\frac{3}{4}\hat{m}_D^2\Bigg[\frac{1}{\eps^2}\(5\pi ^2-\frac{609}{10}+\frac{114 \ln 2}{5}\) +\frac{1}{ \epsilon }\bigg( 30 \zeta (3)-\frac{5779}{75}+\frac{121}{6} \pi ^2+\frac{114}{5}\ln ^2(2)\nn\\
&+&\frac{468}{25} \ln 2+ \(\gamma_E +\ln\frac{\hat\Lambda}{2}\) \(10\pi ^2-\frac{609}{5}+\frac{228}{5} \ln 2\)\bigg)\Bigg]\Bigg\}+\frac{8}{3\pi}\Bigg\{\frac{3\ln 2-4}{2\eps}\nn\\
&+&3\hat{m}_D^2\Bigg[\frac{1}{40 \epsilon ^2}\Big(11+5\pi ^2-92 \ln 2\Big)+
\frac{1}{\epsilon }\(\frac{3}{4}\zeta (3)+\frac{1557}{200}- \frac{\pi ^2}{3}-\frac{23}{10}\ln ^2(2)\right.\nn\\
&-&\left.\frac{168}{25} \ln 2+ \(\gamma_E +\ln\frac{\hat\Lambda}{2}\)\(\frac{11}{20}+\frac{\pi^2}{4}-\frac{23}{5}\ln 2\)\)\Bigg]\Bigg\}\Biggr] . \label{free_tree}
\eea
So, the renormalized total free energy becomes
\bea
F=F_q^r + F_g^r
\eea
where,
\bea
F_q^r &=& N_c N_f\Bigg[ 
-\frac{7\pi^2T^4}{180}\left(1+\frac{120\hat\mu^2}{7}+\frac{240\hat\mu^4}{7}
\right)+\frac{g^2C_FT^4}{48}\left(1+4\hat{\mu}^2\right)\left(1+12\hat{\mu}^2\right)\nn\\
%%%%%%%%%%%%%%
&&+\, \frac{g^4C_F^2T^4}{
768\pi^2}\left(1+4\hat{\mu}^2\right)^2\left(\pi^2-6\right)+\frac{g^4C_F^2}{27N_f}
M_B^4 \bigg(12 \ln \frac{ \hat\Lambda}{2}-6\aleph(z)+\frac{36 \zeta (3)}{\pi^2}\nn\\
&&-2 -\frac{72}{\pi^2}\bigg)\Bigg],
\eea
and
\bea
\hspace{-1cm}\frac{F_g^r}{d_A}&=&-\frac{\pi^2 T^4}{45}\left[1-\frac{15}{2}\hat{m}_D^2+30(\hat{m}_D^w)^3+\frac{45}{8}\hat{m}_D^4\(2\ln\frac{\hat{\Lambda}}{2}-7+2\gamma_E+\frac{2\pi^2}{3}\)\right]\nn\\
&-&\pi^2T^4\hat{m}_D^2\delta \hat{m}_D^2\left(\gamma_E+\ln\hat\Lambda\right)+\sum_f\frac{g^2(q_fB)^2}{(12\pi)^2}\frac{T^2}{m_f^2}\Bigg[4.97+2\ln\frac{\hat\Lambda}{2}\nn\\
&+&3\hat{m}_D^2\Bigg\{2\(1-\ln 2\)\ln^2\frac{\hat\Lambda}{2}+2\bigg(\frac{7}{2}-\frac{\pi ^2}{6}-\ln ^2(2)-2 \gamma_E  (\ln 2-1)\bigg)\ln\frac{\hat\Lambda}{2}+4.73\Bigg\}\Bigg]\nn\\
&-& \sum_f\frac{g^2(q_fB)^2}{(12\pi)^2}\frac{\pi T}{32m_f}\Biggl[\Biggl\{\frac{3}{4}\ln^2\frac{\hat\Lambda}{2}+2\ln\frac{\hat\Lambda}{2}\(\frac{21}{8}+\frac{3}{4}\frac{\zeta'(-1)}{\zeta(-1)}+\frac{27}{4}\ln 2\)+43.566\nn\\
&+&\frac{3}{4}\hat{m}_D^2\Bigg[2\ln^2\frac{\hat\Lambda}{2}\(5\pi ^2-\frac{609}{10}+\frac{114 \ln 2}{5}\) + 2\ln\frac{\hat\Lambda}{2}\bigg( 30 \zeta (3)-\frac{5779}{75}+\frac{121}{6} \pi ^2+\frac{114}{5}\ln ^2(2)\nn\\
&+&\frac{468}{25} \ln 2+ \gamma_E  \(10\pi ^2-\frac{609}{5}+\frac{228}{5} \ln 2\)\bigg)+106.477\Bigg]\Bigg\}+\frac{8}{3\pi}\Bigg\{\(3\ln 2-4\)\ln\frac{\hat\Lambda}{2}-3.92\nn\\
&+&3\hat{m}_D^2\Bigg[\frac{1}{20}\ln^2\frac{\hat\Lambda}{2}\Big(11+5\pi ^2-92 \ln 2\Big)+
2\ln\frac{\hat\Lambda}{2}\(\frac{3}{4}\zeta (3)+\frac{1557}{200}- \frac{\pi ^2}{3}-\frac{23}{10}\ln ^2(2)\right.\nn\\
&-&\left.\frac{168}{25} \ln 2+\gamma_E\(\frac{11}{20}+\frac{\pi^2}{4}-\frac{23}{5}\ln 2\)\)-1.86\Bigg]\Bigg\}\Biggr].
\eea

\subsection{Pressure}
\label{RP}
The expression for the pressure of hot and dense QCD matter in one-loop HTLpt  
in the presence of a weak magnetic 
field can now be written directly from the one-loop free energy as 
\bea
P(T,\mu,B,\Lambda) = -F(T,\mu,B,\Lambda),
\eea
whereas the ideal gas limit of the pressure reads as
\bea
P_{\text{Ideal}}(T,\mu) = \frac{B^2}{2}+
N_cN_f\frac{7\pi^2T^4}{180}\left(1+\frac{120}{7}\hat{\mu}^2+\frac{240}{7}\hat{
\mu}^4\right) + (N_c^2-1)\frac{\pi^2T^4}{45}.
\eea

\section{Strong coupling and scales}
\label{alpha_ayala}

\begin{center}
\begin{figure}[t]
 \begin{center}
\includegraphics[scale=0.57]{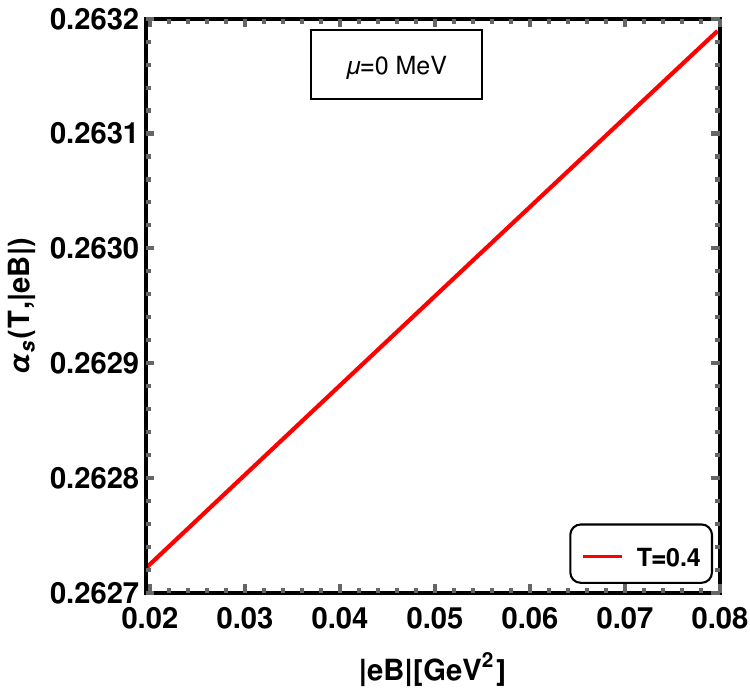} 
\includegraphics[scale=0.55]{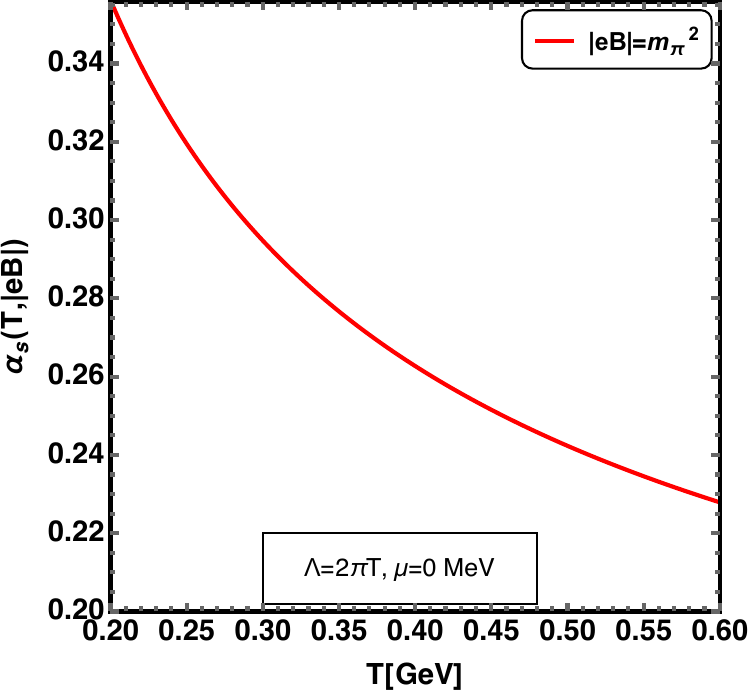}
\caption{ Left: variation of the  one-loop QCD coupling with weak magnetic field, $|eB|$ for $T=0.4$GeV. Right: variation with temperature, $T$ for 
$|eB|=m^2_\pi$ .}
  \label{1loop_coupling}
 \end{center}
\end{figure}
\end{center}

The one-loop running coupling which evolves with both the momentum transfer and the magnetic field is recently obtained in Ref.~\cite{Ayala:2018wux} as
\bea
\alpha_s(\Lambda^2, |eB|)&=&\frac{\alpha_s(\Lambda^2)}{1+b_1\, \alpha_s(\Lambda^2)
\ln\left(\frac{{\Lambda^2}}{{\Lambda^2\ +\ |eB|}}\right)},
\eea
in the domain $|eB| < \Lambda^2$ where the one-loop running coupling at renormalization scale reads as
\bea
\alpha_s(\Lambda^2)&=&\frac{1}
{b_1\, \ln\left({\Lambda^2}/{\Lambda_{\overline{{\rm MS}}}^2}\right)},
\eea
with $b_1= \frac{11N_c-2N_f}{12\pi}$,  $\Lambda_{\overline{{\rm MS}}}=176~{\rm MeV}$ \cite{Beringer:1900zz} at 
$\alpha_s(1.5 {\mbox{GeV}})=0.326 $ for $N_f=3$. 
We note here that we choose separate renormalization scales  for gluon 
$\Lambda =\Lambda_g$, for quark $\Lambda =\Lambda_q$, which are chosen at their central values, respectively as $\Lambda_g=2\pi T$ and $\Lambda_q =2\pi \sqrt{T^2+\mu^2/\pi^2}$. The renormalization scales can be varied by a factor of 2  with respect to its central value. On the other hand  the magnetic field strength can also be varied as long as  $|eB| > \Lambda^2$ for strong field and $|eB| < \Lambda^2$ for weak field approximation for  a given temperature \textit{vis-\`a-vis} the renormalization scale, as discussed in Sec.~\ref{scale}. The left panel of Fig.~\ref{1loop_coupling} displays running of $\alpha_s$ with $|eB|$ as the central value of the renormalization scale $ \Lambda_g=\Lambda_q =2 \pi T$ GeV for $T=0.4$ GeV. This indicates a slow increase of $\alpha_s$ with the increase of $|eB|$ in the domain $|eB| < \Lambda^2$.  On the other hand the right panel of Fig.~\ref{1loop_coupling} exhibits  running of $\alpha_s$ with $T$  for 
$|eB|=m_{\pi}^2$. This behavior also consolidates that $\alpha_s$ runs very slowly with $|eB|$. 
%%%%%%%%%%%%%%%%
 %%%%%%%%%%%%%%%%%%%%%%%%%%%%%%%%%%%%%%%

%%%%%%%%%%%%%%%%
\section{Comparison with Full Results }
\label{full}
\begin{figure}[t]
	\includegraphics[scale=.6]{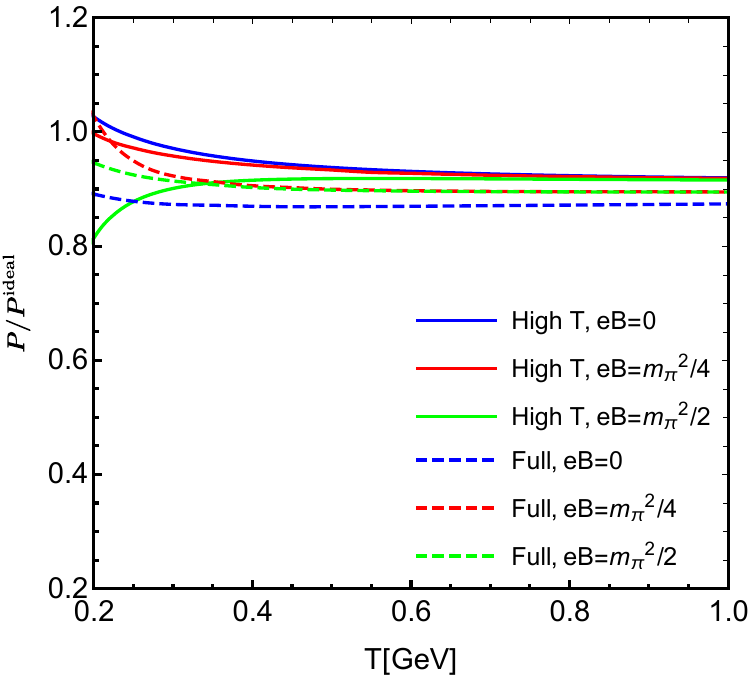}
	\caption{Comparison of pressure with and without high $T$ expansion for $eB= 0,\, m_\pi^2/4$ and $m_\pi^2/2$ with $N_f=2$ . ``Full" represents without high $T$ expansion.}
	\label{compare_plot}
	\end{figure}
In this section we try to justify the use of high $T$ expansion in our calculation by comparing it with the full results (without high $T$ expansion) in the spirit of HTL perturbation theory. The full result is computed numerically and the detailed calculations are given in Appendix~\ref{wohighT}.  Figure~\ref{compare_plot} displays a comparison between the scaled pressure of QCD matter computed with and without high temperature expansion for various field strengths. The solid lines represent results from high $T$ expansion whereas dashed lines correspond to the results  of without high $T$ expansion.  As can be seen from the figure, the two plots (i.e. with and without high $T$ expansion) for a given magnetic field strength do not differ much except at low $T$. Thus we can use the high $T$ expansion to get the analytical expression of the pressure.  Although  this high $T$ expansion is not in the spirit 
of HTL perturbation theory, it nevertheless is very effective (especially in higher order loops), because in this way one can bypass the tiresome work of quasiparticle pole as well as Landau damping calculations. For simplicity, this high $T$ expansion in the absence of a magnetic field  has  widely been used in the literature for leading order, next-to-leading-order, and next-to-next-to-leading-order in HTL. In the same spirit we have used such high $T$ expansion here in the presence of a magnetic field and hereafter we show the results with high $T$ expansion.

%%%%%%%%%%%%%%%%%
\section{Results}
\label{RES}
%%%%%%%%%%%%%%%%%%%%%%%%%%%%%%%%%%%%%%%%%%
% \subsection{Weak Field}
\begin{center}
\begin{figure}[tbh]
 \begin{center}
 \includegraphics[scale=0.55]{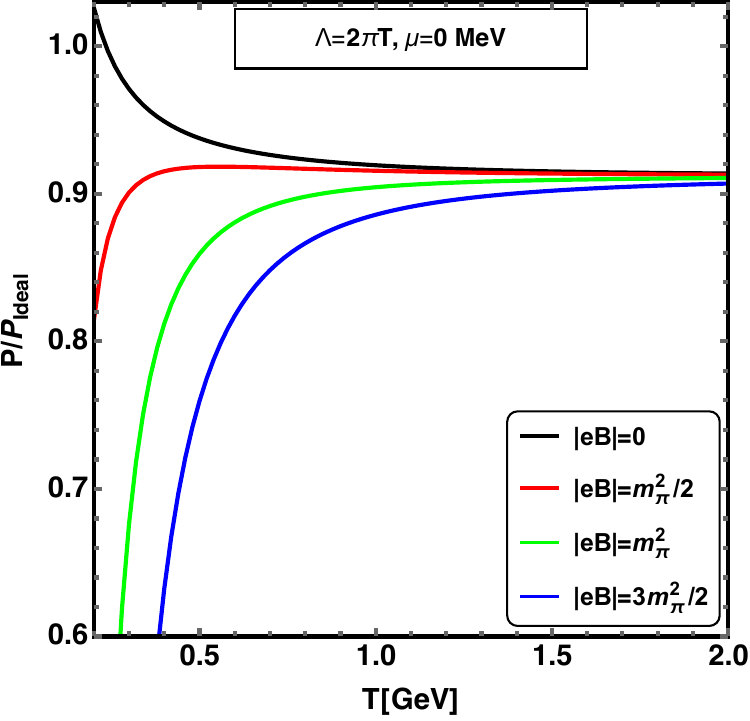}
\includegraphics[scale=0.55]{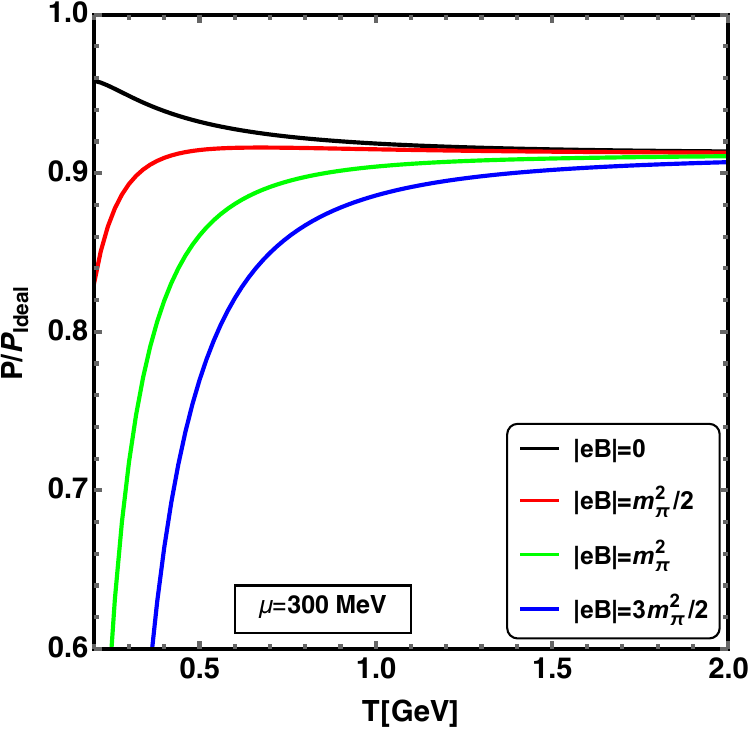} 
 \caption{Variation of the scaled one-loop pressure with temperature for 
$N_f=2$ with $\mu=0$ (left) for $\mu=300$ MeV (right) in the presence of weak 
magnetic fields of various strengths, $|eB|=0$, $m^2_\pi/2, \  m^2_\pi$ and $3m^2_\pi/2$. In the right panel  for $\mu\ne 0$, the renormalization scales are defined  in the text in Sec.~\ref{alpha_ayala}.}
  \label{1loop_pressure}
 \end{center}
\end{figure}
\end{center}
In this section we discuss the main results of this paper. In Fig.~\ref{1loop_pressure} we display the temperature variation of scaled pressure with that of ideal gas value for hot and dense  magnetized  QCD matter in one-loop HTLpt within weak field approximation for different values of field strengths, $|eB|= 0, \, \ m_{\pi}^2/2, \, m_{\pi}^2 \, \ {\mbox{and}} \ \, 3m_{\pi}^2/2$. The left panel shows vanishing chemical potential, $\mu=0$  whereas the right panel shows $\mu=0.3$ GeV. We note that for $|eB|=0$ one gets back usual one-loop HTLpt pressure~\cite{najmul13,najmul12,najmul11,sylvain2,sylvain1,andersen3,andersen2,andersen1,Haque:2018eph}.  From both plots one can observe that the low $T$ ($< 0.8$ GeV) behavior of the pressure is strongly affected by the presence of magnetic field whereas at high $T$ ($\ge 0.8$ GeV) it almost remains unaffected as the temperature becomes the dominant scale because of weak field approximation  $|eB| < m^2_{\textrm{th}}<T^2$. Nevertheless, we also note a specific difficulty one encounters with HTLpt ($|eB|=0$). This has to do with the fact that the  
one-loop HTLpt introduces an overcounting of some contributions~\cite{najmul13,najmul12,najmul11,sylvain2,sylvain1,andersen3,andersen2,andersen1} in strong coupling ($g$). This is because the loop expansion and the coupling expansion are not symmetrical in HTLpt. So, at each loop order in HTLpt the result is  an infinite series in $g$. At leading order in HTLpt one only gets the correct perturbative coefficients for $g^0$ and $g^3$ when one expands in power of $g$. Thus, for a given order in $g$ higher loop orders contribute and this is only corrected by extending the calculation up to higher loop-orders~\cite{Andersen:2002ey,Andersen:2003zk,Haque:2012my,najmul2qns, 
3loopglue1,3loopglue2,3loopqcd1,3loopqcd2,3loopqcd3,najmul3,Haque:2014rua}. We also note that the pressure is slightly reduced  in the presence of a $\mu$ (right panel) than that of $\mu=0$, for a given $|eB|$. 
%%%%%%%%%%%%%%%%%%%%%
\begin{center}
\begin{figure}[t]
 \begin{center}
 \includegraphics[scale=0.55]{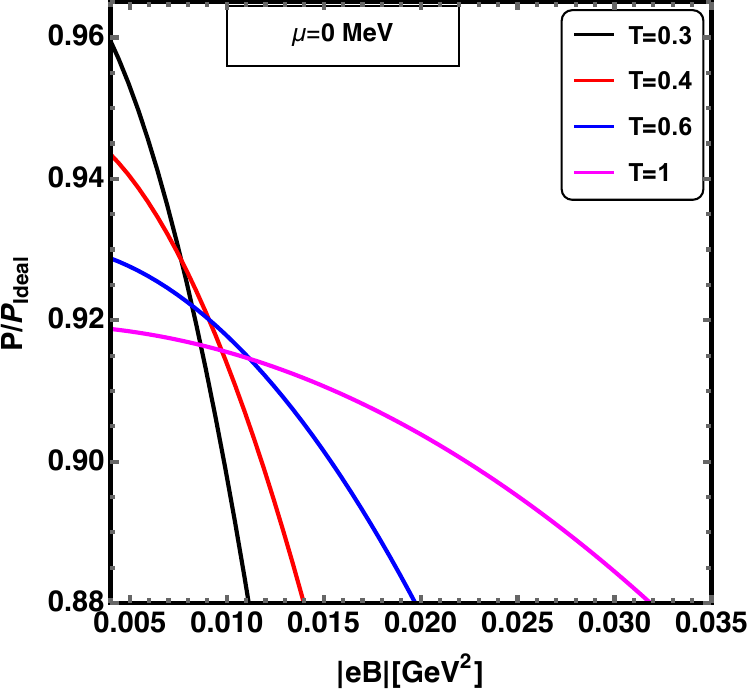} 
\includegraphics[scale=0.55]{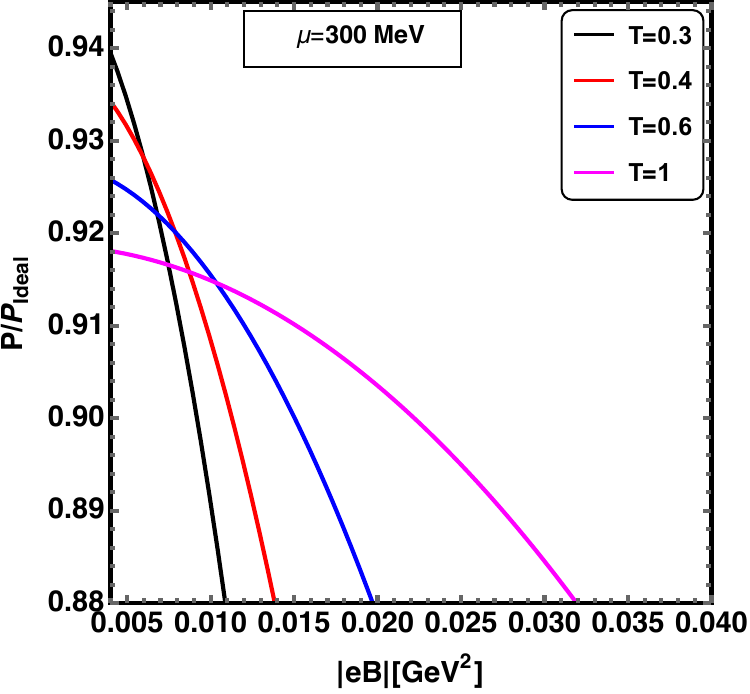} 
 \caption{Variation of the scaled one-loop  pressure with magnetic field for 
$N_f=2$ with $\mu=0$ (left) and $\mu=300$ MeV (right) for $T=$ (0.3, 0.4, 0.6 and 1) GeV. The renormalization scales are defined  in the text in Sec.~\ref{alpha_ayala}.}
  \label{1loop_pressure_eB}
 \end{center}
\end{figure}
\end{center}
%%%%%%%%%%%%%%%%
It is seen from Fig.~\ref{1loop_pressure_eB} that the slope of the curve decreases with the increase of $T$. So, Fig.~\ref{1loop_pressure_eB} also consolidates the fact that in weak field approximation the effect of the magnetic field diminishes with increase of $T$. This indicates that the magnetic field is the dominant scale at low $T$ and becomes negligible at high $T$.   
%%%%%%%%%
\begin{center}
\begin{figure}[tbh]
 \begin{center}
 \includegraphics[scale=0.55]{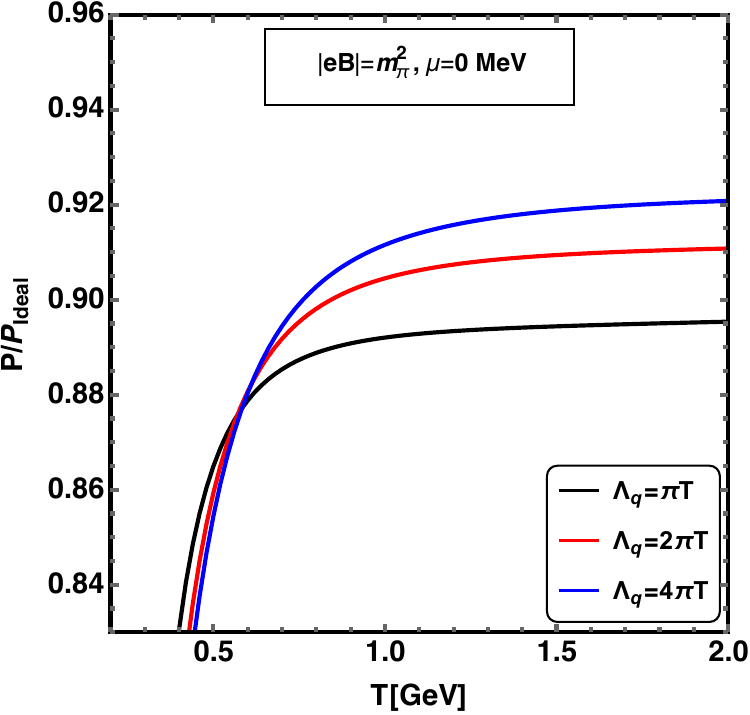} 
\includegraphics[scale=0.55]{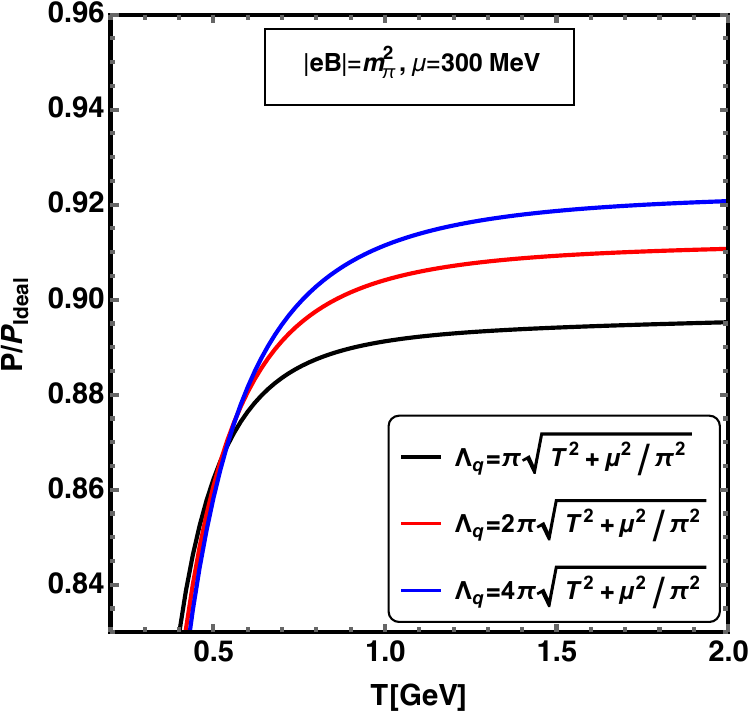} 
 \caption{Variation of the scaled one-loop pressure with temperature for 
$N_f=2$ with $\mu=0$ (left) and $\mu=300$ MeV (right) in the presence of a weak 
magnetic field of strength $eB = m_\pi^2$ for different values of renormalization 
scale  of gluons, $\Lambda_g=\pi T, \, 2\pi T, \ \textrm{and}\, 4\pi T$ and scale of quarks is given in the inset.}
  \label{1loop_pressure_vl}
 \end{center}
\end{figure}
\end{center}
%%%%%%%%%%%%%%%%%%%%%%%%%%%%%%%%%%%%%%%%%%
To check the sensitivity with the renormalization scale $\Lambda_{q,g}$ 
in Fig.~\ref{1loop_pressure_vl} we have displayed the temperature variation of 
the scaled one-loop pressure in the presence of a constant weak magnetic field by 
varying $\Lambda_{q,g}$ by a factor of two around its central value for both zero and finite chemical potential. It is found to depend moderately on the renormalization scale $\Lambda_{q,g}$. One may need higher loops calculation and log resummation to further reduce the renormalization scale dependent band.

\section{Conclusion}
\label{conclusion}
In this paper we presented a systematic framework based on the general structure of two-point functions of a fermion and a gauge boson to evaluate the QCD pressure in nontrivial backgrounds, \textit{viz.}, when both heat bath and magnetic field are considered together. This framework has been applied to the case when the heat bath is weakly magnetized. The total pressure of a magnetized hot and dense deconfined QCD matter is the sum of three contributions coming from (a) the quark part, (b) the gluonic part and (c) and the tree-level free energy due to the constant magnetic field. We note that the presence of an external magnetic field affects both the fermion and gluon effective two-point (self-energy and propagator) functions. We have also used strong coupling that runs through both renormalization scale and magnetic field strength. Although gluons are electrically charge neutral, they are mostly affected through the quark loop because quark propagators get modified in the presence of a background magnetic field. Based on the most general structure of the effective two-point functions, the quark propagator in our earlier calculation~\cite{Das:2017vfh} and gluon propagator in paper-I~\cite{Karmakar:2018aig}, we obtain the QCD Debye screening mass, gluon and quark free energy in one-loop HTLpt in the presence of a weak field approximation. The divergences appeared therein are taken care of by redefining the magnetic field  in the tree-level free energy term and through an HTL counterterm. Using high temperature expansion we found finite results which are also completely analytic and  gauge independent but depend on renomalization scale and magnetic field strength. We have again compared the results with the numerically evaluated full results thereby justifying the use of the high temperature expansion. We have also discussed in details the modification of QCD Debye mass which depends on three scales, {\it viz.}, the thermal quark mass, temperature and the magnetic field. The weak field pressure is strongly affected  at low $T$($<0.8$ GeV) beyond which the HTL result takes over. We have checked the sensitivity of the pressure on the various scales, {\it{viz.}}, the renormalization and magnetic field strength. The result is sensitive to renormalization scale as it produces band while varying its value by a factor of 2. The sensitivity of pressure on the magnetic field is strong at low $T$ and negligible at high $T$. We have also outlined a general drawback with one-loop HTLpt that introduces an overcounting of some contributions, as a remedy of which one needs to push the calculation to higher loop orders.

\section{Acknowledgment}
The authors gratefully acknowledge a valuable discussion with J.~Kapusta on a technical issue in the calculation. The authors would also like to acknowledge useful discussions with Arghya Mukherjee, Jens O.~Andersen, and Michael Strickland. A.B., B.K., and M.G.M. were funded by the Department of Atomic Energy, India via the project TPAES. AB was also partially supported by the National Post Doctoral Program CAPES, Government of Brazil. N.H. gratefully acknowledges financial support from the Alexander von Humboldt  Foundation, Germany, and also from the Department of Atomic Energy, India.

%%%%%%%%%%%%%%%%%%%%%%%%%%%%%%%%%%%%%%%%%%%%%%%%%%%%%%%%%%%%%%%%%%%%%%%%%%%%%%%%
%%%%%%%%%%%%%%%%%%%%%%%%%

\appendix 

\section{Fermionic sum-integrals}
\label{a_sumint}
%%%%%%%%%%
The dimensionally regularized sum-integrals are defined as,
\bea
\sumintf_{\{ P\} } &=& \left(\frac{e^{\gamma_E}\Lambda^2}{4\pi}\right)^\epsilon 
T\sum\limits_{\substack{p_0=i\omega_n\\ \omega_n=(2n+1)\pi 
T-i\mu}}\int\frac{d^{d-2\epsilon}p}{(2\pi)^{d-2\epsilon}},
\eea
where $\Lambda$ can be identified as the $\overline{MS}$ renormalization scale which also 
introduces the factor $\left(\frac{e^{\gamma_E}}{4\pi}\right)^\epsilon$ 
along with it, with $\gamma_E$ being the Euler-Mascheroni constant. Before listing all of  
the 
sum-integrals used in our paper, we note that they are interrelated among themselves 
via  
\bea
\sumintf_{\{ P\} }\frac{1}{P^4}= -\frac{d-2}{2}\sumintf_{\{ P\} 
}\frac{1}{p^2P^2} = 
\frac{d-5}{d-4}\sumintf_{\{ P\} }\frac{\mathcal{T}_P}{P^4}.
\eea
\subsection{Simple one-loop sum-integrals}
The list of fermionic sun-integrals needed are
%%%%%%%%
 %%
\bea
\sumintf_{\{P\}}\frac{1}{P^2}&=&\frac{T^2}{24}\left(\frac{\Lambda}{4\pi 
T}\right)^{2\epsilon}\left[1+12\hat\mu^2+2\epsilon
\left(1+12\hat\mu^2
       +12\aleph(1,z)\right)\right],\\
%%%%%%%%%%%%%%%%
\sumintf_{\{ P\} 
}\frac{1}{P^4}&=&\frac{1}{\left(4\pi\right)^2}\left(\frac{\Lambda}{4\pi 
T}\right)^{2\epsilon}\Bigg[\frac{1}
{\epsilon}-\aleph(z)\Bigg],\\
%%%%%%%%%%%%%%%%%%%%
\sumintf_{\{P\}}\frac{p^2}{P^6}&=&-\frac{3}{4}\frac{1}{\left(4\pi\right)^2}
\left(\frac{\Lambda}{4\pi T}\right)^{2\epsilon}
\left[\frac{1}{\epsilon}-\frac{2}{3}
            -\aleph(z)\right], \\
%%%%%%%%%%%%%%%%%%%%%
\sumintf_{\{P\}}\frac{1}{p^2P^2}&=&-\frac{2}{(4\pi)^2}\left(\frac{\Lambda}
{4\pi T}\right)^{2\epsilon}\left[\frac{1}{\epsilon} \!+\!2\!-\!\aleph(z)\right], \\
%%%%%%%%%%%%%%%%%%%%%
\sumintf_{\{P\}}\frac{p_3^2}{p^2P^4}&=&\frac{1}{3(4\pi)^2}\left(\frac{
\Lambda}{4\pi T}\right)^{2\epsilon}\Bigg[\!\frac{1}{\epsilon} \!+\! \frac{2}{3}\!
-\!\aleph(z)\Bigg], \\
%%%%%%%%%%%%%%%%%%%%%
\sumintf_{\{P\}}\frac{p_3^2}{p^4P^2}&=&-\frac{2}{3(4\pi)^2}\!\left(\frac{\Lambda}
{4\pi T}\right)^{2\epsilon}\left[\!\frac{1}{\epsilon} \!+\!\frac{8}{3}\!-\!\aleph(z)\right], 
%%%%%%%%%%%%%%%%%%%%%
\eea
For some frequently occurring combinations of special functions we applied the 
following abbreviations 
\bea
\zeta'(x,y) &\equiv& \partial_x \zeta(x,y), \label{}\\
\aleph(n,z) &\equiv& \zeta'(-n,z)+\(-1\)^{n+1}\zeta'(-n,z^{*}),\label{alephnz} \\
\aleph(z) &\equiv& \Psi(z)+\Psi(z^*),
\eea
where $n$, in Eq.~\eqref{alephnz}, is assumed to be an integer and $z$ a general complex number, here 
$z = 1/2-i\hat{\mu}$.
Here $\zeta$ and $\Psi$ denote the Riemann zeta function and the digamma function respectively and $\Psi$ can be expressed as
\bea
\Psi(z)&\equiv&\frac{\Gamma'(z)}{\Gamma(z)}.
\eea
Below we enlist the values of the function $\aleph$ as required for our calculation. Although the following list  are given at small values of $\mu/T$, in the actual plot we calculate $\aleph$ for any value of $\mu$ using {\it Mathematica}.
%\subsection{At small \texorpdfstring{$\mu/T$}{mu/T}}
\bea
 \aleph(z)&=&-2\gamma_E-4\ln 
2+14\zeta(3)\hat{\mu}^2-62\zeta(5)\hat{\mu}^4+254\zeta(7)\hat{\mu}^6+{\cal 
O}(\hat{\mu}^8),\label{aleph}\\
 \aleph(1,z)&=&-\frac{1}{12}\(\ln2-\frac{\zeta'(-1)}{\zeta(-1)}\) - 
\(1-2\ln2-\gamma_E\)\hat{\mu}^2-\frac{7}{6}\zeta(3)\hat{\mu}^4\nn\\
&&+\ \frac{31}{15}\zeta(5)\hat{
\mu}^6
+{\cal O}(\hat{\mu}^8)
\eea

\subsection{HTL one-loop sum-integrals for weak field case}
%%%%
We also need some more difficult one-loop sum-integrals that involve the angular average defined earlier in Eq.~\eqref{ang_avg}. For brevity, henceforth we will use the notation $c=\cos\theta$ for single angular average and $c_i=\cos\theta_i$ for multiple angular averages. We list the sum-integrals below. The expressions for the respective angular averages appearing in the process and denoted by $\Delta_i$'s are given in Appendix \ref{ang_av}.
%%%%%%%%%

\bea
\sumintf_{\{P\}}\frac{1}{P^4}{\cal T}_P &=& \frac{d-4}{d-5}\sumintf_{\{ P\} 
}\frac{1}{P^4}, \\
\sumintf_{\{P\}}\frac{1}{p^2P^2}{\cal T}_P 
&=&-\frac{2\Delta_3}{d-2} \sumintf_{\{P\}}\frac{1}{P^4},\\
\sumintf_{\{P\}}\frac{1}{p^2P^2}{\cal T}_P^2 
&=&-\frac{2\Delta_4''}{d-2} \sumintf_{\{P\}}\frac{1}{P^4},\\
\sumintf_{\{P\}}\frac{1}{p_0^2P^2}{\cal T}_P^2 
&=&-\frac{2\Delta_3''}{d-2} \sumintf_{\{P\}}\frac{1}{P^4},\\
\sumintf_{\{P\}}\frac{1}{P^4}{\cal T}_P^2 &=&\left(\frac{d-4}{d-2} \Delta_0'' - \frac{2}{d-2} \Delta_{10}\right) 
\sumintf_{\{P\}}\frac{1}{P^4},\\
\sumintf_{\{P\}}\frac{p_3^2}{p_0^2P^4}{\cal T}_P^2 
&=&\left(\frac{\Delta_0''}{d} - \frac{2}{d(d-2)}\Delta_{11}\right)\sumintf_{\{P\}}\frac{1}{P^4},\\
\sumintf_{\{P\}}\frac{ p_3^2 }{p^4P^2} {\cal T}_P^2&=& \Delta_4''\Delta_0'\sumintf_{\{P\}}\frac{1}{p^2P^2}=-\frac{2}{d-2}\Delta_4''\Delta_0'\sumintf_{\{P\}}\frac{1}{P^4},\\
\sumintf_{\{P\}}\frac{ p_3^2}{p^4P^2}{\cal T}_P&=&\Delta_3\Delta_0'\sumintf_{\{P\}}\frac{1}{p^2P^2}=-\frac{2}{d-2}\Delta_3\Delta_0'\sumintf_{\{P\}}\frac{1}{P^4},\\
\sumintf_{\{P\}}\frac{p_3^2}{p^2P^4}{\cal T}_P^2&=&\Delta_0'' \sumintf_{\{P\}}\bigg(\frac{p_3^2}{p^2P^4}+\frac{p_3^2}{p^4P^2}\bigg)+\Delta_{10}\sumintf_{\{P\}} \frac{p_3^2}{p^4P^2}\\
&&=\Bigg(\frac{d-4}{d-2}\Delta_0'' \Delta_0'-\frac{2}{d-2}\Delta_{10}\Delta_0'\Bigg)\sumintf_{\{P\}}\frac{1}{P^4},\\
\sumintf_{\{P\}}\frac{p_3^2}{p^2P^4}{\cal T}_P&=&(1+\Delta_0)\Delta_0'\bigg(\sumintf_{\{P\}}\frac{1}{P^4}+\sumintf_{\{P\}}\frac{1}{p^2P^2}\bigg)-\Delta_3'\Delta_0'\sumintf_{\{P\}}\frac{1}{p^2P^2}\\
&&=\Bigg(\frac{d-4}{d-2}(1+\Delta_0)\Delta_0'+\frac{2}{d-2}\Delta_3'\Delta_0'\Bigg)\sumintf_{\{P\}}\frac{1}{P^4}.
\eea

%%%%%%%%%%%%%%%%%%%%%%%%%%%%%%%%%%%%
%%%%%%%%%%%%%%%%%%%%%%%%%%%%%%

%%%%%%%%%%%%%%%%%%%%%%%%%%%%%%%%%%%%%%%%%%%%%%%%%%%%%%%%%%%%%%%%%%%%%%%%%%%%%%%%%%%%%%%%%%%%%%%%%%%%%%%%%%
\section{Bosonic sum-integrals}
\label{BS}

\subsection{Simple one-loop sum-integrals}
To evaluate the sum-integrals over the external bosonic momenta, we use the following master formula
\bea
\sumintb_P ~\frac{p_\perp^ip_3^j}{p^mP^n} = \left\langle c^j 
(1-c^2)^{\frac{i}{2}}\right\ranglec\sumintb_P~\frac{1}{p^{m-i-j}P^n},
\eea
thus eventually requiring the following basis integrals
\bea
\sumintb_P \frac{1}{P^2} &=& -\frac{T^2}{12}\left(\frac{\Lambda}{4\pi T}\right)^{2\eps}\Bigg[1+2\eps\left(1+\frac{\zeta'(-1)}{\zeta(-1)}\right)+
\eps^2 \Bigg(4 \gamma_E  \left(1+\frac{\zeta'(-1)}{\zeta(-1)} -\frac{\gamma_E}{2}- \ln 2\pi\right)\nn\\
&&+4 \ln 2\pi\left(1+\frac{\zeta'(-1)}{\zeta(-1)}\right)+\frac{12 \zeta
   ''(2)}{\pi ^2}+\frac{\pi^2}{12}-2 \ln ^2(2 \pi )\Big)\Bigg]+\mathcal{O}[\eps]^3,\\
%%%%%%%
\sumintb_P \frac{1}{p^2P^2} &=& -\frac{2}{(4\pi)^2}\left(\frac{\Lambda}{4\pi T}\right)^{2\eps}\Bigg[\frac{1}{\eps}+2\gamma_E+2
+\eps\left(4+4\gamma_E+\frac{\pi^2}{4}-4\gamma_1\right)\nn\\
&&+\frac{\epsilon ^2}{6} \Big(3 \left(8 \left(-2 \gamma _1+\gamma _2+2\right)+\pi ^2+\gamma_E
\left(16+\pi ^2\right)\right)-14 \zeta
   (3)\Big)\Bigg]+\mathcal{O}[\eps]^3,\\
 %%%%%%%%%%
\sumintb_P \frac{1}{P^4} &=&  \frac{1}{(4\pi)^2}\left(\frac{\Lambda}{4\pi T}\right)^{2\eps}\left[\frac{1}{\eps}
+2\gamma_E+\eps\left(\frac{\pi^2}{4}-4\gamma_1\right)+\frac{\epsilon ^2 }{6}\left(24 \gamma _2-14 \zeta (3)+3 \gamma_E 
\pi ^2\right)\right]\nn\\
&&+\mathcal{O}[\eps]^3.
\eea

\subsection{HTL one-loop sum-integrals}
Similarly as in the fermionic part, here also we list the one-loop bosonic HTL integrals required for our computation. We define the following master HTL integrals which is largely required to compute the HTL sum-integrals appearing in the expression for free energies. 
%%%%%%%%%%%%%%%
\begin{enumerate}
 \item 
\bea
\sumintb_P~\frac{p_\perp^i~p_3^j~p_0^k}{p^m}~{\cal T}_P &=& \langle c^{m-i-j-d} \rangle_c \sumintb_P~\frac{p_\perp^ip_3^j~p_0^{k+2}}{p^mP^2} \nn\\ 
&=& \langle c^{m-i-j-d} \rangle_c \sumintb_P~\frac{p_\perp^ip_3^j}{p^{m-k-2}P^2}\nn\\
&=& \langle c_1^{m-i-j-d} \rangle_{c_1}\langle c_2^j(1-c^2)^{\frac{i}{2}}\rangle_{c_2} \sumintb_P~\frac{1}{p^{m-k-i-j-2}P^2}
\eea
 \item 
\bea
\sumintb_P~\frac{p_\perp^i~p_3^j~p_0^k}{p^mP^2}~{\cal T}_P &=& \left\langle \frac{1-c^{m+2-i-j-d}}{1-c^2} 
\right\rangle_c \sumintb_P~\frac{p_\perp^ip_3^j~p_0^{k+2}}{p^{m+2}P^2}\nn\\
&=&\left\langle \frac{1-c^{m+2-i-j-d}}{1-c^2} \right\rangle_c \sumintb_P~\frac{p_\perp^ip_3^j}{p^{m-k}P^2}\nn\\
&=& \left\langle \frac{1-c_1^{m+2-i-j-d}}{1-c_1^2} \right\rangle_{c_1} \langle c_2^j(1-c_2^2)^{\frac{i}{2}}\rangle_{c_2}
\sumintb_P~\frac{1}{p^{m-k-i-j}P^2}
\eea
 \item 
\bea
\sumintb_P~\frac{p_\perp^i~p_3^j~p_0^k}{p^mP^4}~{\cal T}_P &=& \left\langle \frac{1}{1-c^2} \right\rangle_c
\sumintb_P~\frac{p_\perp^ip_3^j~p_0^{k+2}}{p^{m+2}P^4}-\left\langle \frac{1-c^{m+4-i-j-d}}{(1-c^2)^2} \right\rangle_c 
\sumintb_P~\frac{p_\perp^ip_3^j~p_0^{k+2}}{p^{m+4}P^2}\nn\\
&=& \left\langle \frac{1}{1-c^2} \right\rangle_c
\sumintb_P~\frac{p_\perp^ip_3^j}{p^{m-k}P^4}+\left\langle \frac{1}{1-c^2} \right\rangle_c
\sumintb_P~\left(\frac{k}{2}+1\right)\frac{p_\perp^ip_3^j}{p^{m+2-k}P^2} \nn\\
&&- \left\langle \frac{1-c^{m+4-i-j-d}}{(1-c^2)^2} \right\rangle_c 
\sumintb_P~\frac{p_\perp^ip_3^j}{p^{m-k+2}P^2}\nn\\
&=& \left\langle \frac{1}{1-c_1^2} \right\rangle_{c_1} \langle c_2^j(1-c_2^2)^{\frac{i}{2}}\rangle_{c_2}
\sumintb_P~\frac{1}{p^{m-k-i-j}P^4}\nn\\
&&+\left\langle \frac{1}{1-c_1^2} \right\rangle_{c_1}
\langle c_2^j(1-c_2^2)^{\frac{i}{2}}\rangle_{c_2} \sumintb_P~\left(\frac{k}{2}+1\right)\frac{1}{p^{m+2-k-i-j}P^2} \nn\\
&&- \left\langle \frac{1-c_1^{m+4-i-j-d}}{(1-c_1^2)^2} \right\rangle_{c_1} \langle c_2^j(1-c_2^2)^{\frac{i}{2}}\rangle_{c_2} 
\sumintb_P~\frac{1}{p^{m-k-i-j+2}P^2}
\eea
 \item 
\bea
\sumintb_P~\frac{p_\perp^i~p_3^j~p_0^k}{p^m}~{\cal T}_P^2 &=& \left\langle \frac{c_1^{m+2-i-j-d}-c_2^{m+2-i-j-d}}{c_1^2-c_2^2}
\right\rangle_{c_1,c_2} \sumintb_P~\frac{p_\perp^ip_3^j~p_0^{k+4}}{p^{m+2}P^2}\nn\\
&=&  \left\langle \frac{c_1^{m+2-i-j-d}-c_2^{m+2-i-j-d}}{c_1^2-c_2^2}
\right\rangle_{c_1,c_2} \sumintb_P~\frac{p_\perp^ip_3^j}{p^{m-k-2}P^2}\nn\\
&=& \left\langle \frac{c_1^{m+2-i-j-d}-c_2^{m+2-i-j-d}}{c_1^2-c_2^2}
\right\rangle_{c_1,c_2} \!\!\! \!\!\!\langle c_3^j(1-c_3^2)^{\frac{i}{2}}\rangle_{c_3}  \!\!\!\sumintb_P~\frac{1}{p^{m-k-i-j-2}P^2}
\eea
 \item 
\bea
\sumintb_P~\frac{p_\perp^i~p_3^j~p_0^k}{p^mP^2}~{\cal T}_P^2 &=& \left\langle \frac{1-c_1^{m+4-i-j-d}}{(1-c_1^2)(c_1^2-c_2^2)}+c_1 
\leftrightarrow c_2 \right\rangle_{c_1,c_2} \sumintb_P~\frac{p_\perp^ip_3^j~p_0^{k+4}}{p^{m+4}P^2}\nn\\
&=& \left\langle \frac{1-c_1^{m+4-i-j-d}}{(1-c_1^2)(c_1^2-c_2^2)}+c_1 
\leftrightarrow c_2 \right\rangle_{c_1,c_2} \sumintb_P~\frac{p_\perp^ip_3^j}{p^{m-k}P^2}\nn\\
&=& \left\langle \frac{1-c_1^{m+4-i-j-d}}{(1-c_1^2)(c_1^2-c_2^2)}+c_1 
\leftrightarrow c_2 \right\rangle_{c_1,c_2}\langle c_3^j(1-c_3^2)^{\frac{i}{2}}\rangle_{c_3} \sumintb_P~\frac{1}{p^{m-k-i-j}P^2}
\eea
\end{enumerate}
%%%%%%%%%%%%%%%
\subsubsection{HTL sum-integrals required for longitudinal part}
\begin{enumerate}
%1.
\item 
\bea
\sumintb_P~\frac{A_2}{p^2} &=& \sumintb_P~\frac{1}{2}\left[{\cal T}_P\left(\frac{1}{p^2}-\frac{p_3^2}{p^4}\right) +(1-{\cal T}_P)\left(\frac{p_0^2}{p^4}-\frac{3p_0^2p_3^2}{p^6}\right)\right]\nn\\
&=& \Bigg[\frac{1}{2}\langle c_1^{2-d} - c_1^{4-d} \rangle_{c_1} \langle 1 - c_2^2 \rangle_{c_2}+\langle c_1^{4-d} \rangle_{c_1}\langle c_2^2 \rangle_{c_2} \Bigg]\sumintb_P~\frac{1}{P^2}\nn\\
&=& -\(\frac{\Lambda}{4\pi T}\)^{2\eps}\frac{T^2}{72} \left[\frac{1}{\eps}-\frac{1}{3}+2 \frac{\zeta'(-1)}{\zeta(-1)} + 2\ln 2 \right].
\eea
%2.
\item 
\bea
\sumintb_P~\frac{A_2}{p^4} &=& \sumintb_P~\frac{1}{2}\left[{\cal T}_P\left(\frac{1}{p^4}-\frac{p_3^2}{p^6}\right) +(1-{\cal T}_P)\left(\frac{p_0^2}{p^6}-\frac{3p_0^2p_3^2}{p^8}\right)\right]\nn\\
&=& \Bigg[\frac{1}{2}\langle c_1^{4-d} - c_1^{6-d} \rangle_{c_1} \langle 1 - c_2^2 \rangle_{c_2}+\langle c_1^{6-d} \rangle_{c_1}\langle c_2^2 \rangle_{c_2} \Bigg]\sumintb_P~\frac{1}{p^2P^2}\nn\\
&=& -\(\frac{\Lambda e^{\gamma_E}}{4\pi T}\)^{2\eps}\frac{1}{3(4\pi)^2} \left[\frac{1}{\eps}+\frac{1}{6}\(1+ 12\ln 2\) \right].
\eea
%3.
\item
\bea
\sumintb_P~\frac{{\cal T}_p A_2}{p^4}&=& \sumintb_P~\frac{1}{2}\left[{\cal T}_P^2\left(\frac{1}{p^4}-\frac{p_3^2}{p^6}\right) +({\cal T}_P-{\cal T}_P^2)\left(\frac{p_0^2}{p^6}-\frac{3p_0^2p_3^2}{p^8}\right)\right]\nn\\
&=&\Bigg[\frac{1}{2}\langle c_1^{6-d} \rangle_{c_1}\langle 1 - 3c_2^2 \rangle_{c_2}+\frac{1}{2}\(\Delta_4'''-\Delta_5'''\)\langle 1 - c_3^2 \rangle_{c_3} +\Delta_5'''\langle c_3^2 \rangle_{c_3} \Bigg]\sumintb_P~\frac{1}{p^2P^2}\nn\\
&=& -\(\frac{\Lambda e^{\gamma_E}}{4\pi T}\)^{2\eps}\frac{2}{9(4\pi)^2} \Bigg[\frac{1+2\ln 2}{\eps}+\frac{1}{60}\Big(-91+ 8\ln 2\(59+15\ln 2\) \Big)\Bigg].
\eea
\end{enumerate}
Here different $\Delta$'s are the nontrivial angular averages given in Appendix \ref{ang_av}.
\subsubsection{HTL sum-integrals required for transverse part}
\label{master_trans}
\begin{enumerate}
  \item 
 \bea
 &&\sumintb_P~\frac{A_1p_0p_3}{p_1^2P^2} = -\sumintb_P~\frac{p_0^2p_3^2}{p_1^2p^2P^2} \left(1-{\cal T}_P\right)= \left[\Delta_3\Delta_0-\Delta_0 \right]\sumintb_P~\frac{1}{P^2}\nn\\
 =&&\!\!  \(\frac{\Lambda}{4\pi T}\)^{2\eps}\frac{T^2}{24} \left[\frac{\ln 2 -1}{\eps}+\frac{\pi^2}{6}-2+(\ln 2)^2+2(\ln 2-1)\frac{\zeta'(-1)}{\zeta(-1)}\right].
 \eea
%  2.
  \item 
 \bea
 &&\sumintb_P~\frac{A_3p_0p_3}{p_1^2P^2} = \sumintb_P~\frac{p_0^2p_3^2}{2p^2p_1^2P^2}\(1-\frac{5}{3}\frac{p_3^2}{p^2}\)-\sumintb_P~\frac{3p_0^2p_3^2\(1-{\cal T}_P\)}{2p^2p_1^2P^2}\(\frac{p_1^2}{p^2}-\frac{p_0^2}{p^2}+\frac{5}{3}\frac{p_0^2}{p^2}\frac{p_3^2}{p^2}\)\nn\\
 &&=\Bigg[\frac{1}{2}\Delta_0-\frac{11}{6}\Delta_1+\frac{3}{2}\Delta_3\Delta_0-\frac{3}{2}\Delta_4\Delta_0-\frac{3}{2}\Delta_3\Delta_1+\frac{5}{2}\Delta_4\Delta_1\Bigg]\sumintb_P~\frac{1}{P^2}\nn\\
 &&=\!\(\frac{\Lambda}{4\pi T}\)^{2\eps}\frac{T^2}{144} \Bigg[\frac{6\ln 2-5}{\eps}+\pi^2-\frac{55}{3}+2\ln 2 (3\ln 2+5)-2(5-6\ln 2)\frac{\zeta'(-1)}{\zeta(-1)}\Bigg].\nn\\
 \eea
%  3.
 \item 
 \bea
&&\sumintb_P~\frac{A_2p_0^2p_3^2}{p_1^2p^2P^2} = \sumintb_P~\frac{1}{2}\left[\frac{{\cal T}_Pp_0^2p_3^2}{p_1^2p^2P^2}\left(1-\frac{p_3^2}{p^2}\right) +\frac{(1-{\cal T}_P)p_0^2p_3^2}{p_1^2p^2P^2}\left(\frac{p_0^2}{p^2}-\frac{3p_0^2p_3^2}{p^4}\right)\right]\nn\\
&&=\Bigg[\Delta_0-3\Delta_1+\Delta_3\Delta_0-\Delta_4\Delta_0-\Delta_3\Delta_1+3\Delta_4\Delta_1\Bigg]\sumintb_P~\frac{1}{2P^2}\nn\\
 &&=\Bigg[2\Delta_3\Delta_1+\left(1- \Delta_7\right)\left(\Delta_0-3\Delta_1\right)\Bigg]\sumintb_P~\frac{1}{2P^2}\nn\\
&=& \(\frac{\Lambda}{4\pi T}\)^{2\eps}\frac{T^2}{48}\Bigg[\frac{2\ln2-1}{\eps}+\frac{\pi^2}{3}-5+2\ln2\(\ln2+\frac{5}{3}\)\nn\\
&&+\ 2\(2\ln2-1\)\frac{\zeta'(-1)}{\zeta(-1)}\Bigg] +\mathcal{O}(\eps)
\eea
% 4.
\item 
\bea
&&\hspace{-.5cm}\sumintb_P~\frac{A_2(5p_0^2+9p^2)}{4p_1^2P^2} \nn\\
&=& \sumintb_P~\frac{1}{2}\left[\frac{{\cal T}_P(5p_0^2+9p^2)}{4p_1^2P^2}\left(1-\frac{p_3^2}{p^2}\right) +\frac{(1-{\cal T}_P)(5p_0^2+9p^2)}{4p_1^2P^2}\left(\frac{p_0^2}{p^2}-\frac{3p_0^2p_3^2}{p^4}\right)\right]\nn\\
&=&\Bigg[\frac{7}{2}-7\Delta_0-\frac{5}{4}\Delta_7-\frac{9}{4}\Delta_9+\frac{5}{2}\Delta_4\Delta_0+\frac{9}{2}\Delta_3 \Delta_0\Bigg]\sumintb_P~\frac{1}{2P^2}\nn\\
%%%%%%
&=&\(\frac{\Lambda}{4\pi T}\)^{2\eps}\frac{7T^2}{96}\Bigg[\frac{2\ln2-1}{\eps} +\frac{\pi^2}{3}-\frac{79}{14}+2\ln2\(\ln2+1\)\nn\\
&&+\ 2\(2\ln2-1\)\frac{\zeta'(-1)}{\zeta(-1)}\Bigg] +\mathcal{O}(\eps)
 \eea
 %%%%%%%%%
% 5.
\item 
 \bea
 &&\hspace{-.5cm}\sumintb_P~\frac{A_4}{P^2} \nn\\
 &=& \sumintb_P~\frac{3}{8P^2}\left(1-\frac{p_3^2}{p^2}\right)^2-\sumintb_P~\frac{p_0^2}{8p^2P^2}\left(1-\frac{5p_3^2}{p^2}\right)^2+\sumintb_P~\frac{5}{3}\frac{p_0^2}{p^2P^2}\frac{p_3^4}{p^4}\nn\\
&-&\frac{3}{8}\sumintb_P~\left\{\!\left(1-\frac{p_0^2}{p^2}\right)^2
-\frac{2p_3^2}{p^2} \left(1-\frac{3p_0^2}{p^2}\right)^2+\frac{p_3^4}{p^4} \left(1-\frac{5p_0^2}{p^2}\right)^2+\frac{8p_0^4}{p^4}\frac{p_3^2}{p^2} \left(1-\frac{5p_3^2}{3p^2}\right)\!\right\}\! \nn\\
&&\times\left(\frac{1-{\cal T}_P}{P^2}\right)\nn\\
&=&\frac{3}{8}\Bigg[\frac{2}{3}+\frac{4}{3}\Delta_0'-\frac{50}{9}\Delta_1'+\Delta_2-2\Delta_3+\Delta_4-2\Delta_2\Delta_0'-10\Delta_4\Delta_0'
+12\Delta_3\Delta_0'\nn\\
&&+\Delta_2\Delta_1'+\frac{35}{3}\Delta_4\Delta_1'-10\Delta_3\Delta_1'\Bigg]\sumintb_P~\frac{1}{P^2}\nn\\
&=&\(\frac{\Lambda}{4\pi T}\)^{2\eps}\frac{T^2}{120}\Bigg[\frac{1}{\eps}-\frac{13}{30}+2\frac{\zeta'(-1)}{\zeta(-1)}\Bigg]
\eea
 %6.
 \item 
 \bea
 &&\sumintb_P~\frac{A_1p_0p_3(p_0^2-{\cal T}_PP^2)}{p_1^2p^2P^4} = -\sumintb_P~\frac{p_0^2p_3^2(p_0^2-{\cal T}_PP^2)}{p_1^2p^4P^4} \left(1-{\cal T}_P\right)\nn\\
 &&= -\sumintb_P~\frac{p_0^4p_3^2}{p_1^2p^4P^4}+\sumintb_P~\frac{p_0^2p_3^2}{p_1^2p^4P^2}{\cal T}_P+\sumintb_P~\frac{p_0^4p_3^2}{p_1^2p^4P^4}{\cal T}_P-\sumintb_P~\frac{p_0^2p_3^2}{p_1^2p^4P^2}{\cal T}_P^2\nn\\
 &&=\Bigg[(1+\Delta_0)\Delta_0-\Delta_0\Bigg]\sumintb_P~\frac{1}{P^4}+\Bigg[-2\Delta_0+\Delta_4\Delta_0+3(1+\Delta_0)\Delta_0
 -\Delta_5'\Delta_0\nn\\
 &&-\Delta_5''\Delta_0\Bigg]\sumintb_P~\frac{1}{p^2P^2}\nn\\
   %%%%%%%%%%
   &=&\(\frac{\Lambda e^{\gamma_E}}{4\pi T}\)^{2\eps}\frac{1}{12\(4\pi\)^2}\Bigg[\frac{-13+\pi^2+4\ln 2}{\eps^2}+\frac{1}{\eps}\bigg\{-\frac{137}{3}+4\pi^2+\frac{4}{3}\ln2\(3\ln2-4\)\nn\\
   &&+6\zeta(3)\bigg\} - 0.81947\Bigg].
 \eea
 %7.
 \item 
 \bea
 &&\sumintb_P~\frac{A_3p_0p_3(p_0^2-{\cal T}_PP^2)}{p_1^2p^2P^4} = \sumintb_P~\frac{p_0^2p_3^2(p_0^2-{\cal T}_PP^2)}{2p^4p_1^2P^4}\(1-\frac{5}{3}\frac{p_3^2}{p^2}\)\nn\\
 &&-\sumintb_P~\frac{3p_0^2p_3^2~\(p_0^2-{\cal T}_PP^2\)\(1-{\cal T}_P\)}{2p^4p_1^2P^4}\(1-\frac{p_0^2}{p^2}-\frac{p_3^2}{p^2}+\frac{5}{3}\frac{p_0^2}{p^2}\frac{p_3^2}{p^2}\)\nn\\
 &&=\Bigg[(1+\Delta_0)\Delta_1+\frac{1}{2}\Delta_0-\frac{11}{6}\Delta_1\Bigg]\sumintb_P~\frac{1}{P^4}
 +\Bigg[\frac{5}{2}\Delta_0-\frac{37}{6}\Delta_1-\frac{1}{2}\Delta_4\left(\Delta_0-\frac{5}{3}\Delta_1\right)\nn\\
 &&-\frac{3}{2}(1+\Delta_0)\Delta_0+\frac{11}{2}(1+\Delta_0)\Delta_1-\Delta_6'\Delta_1
 -\frac{3}{2}\left(\Delta_5'-\Delta_6'\right)\left(\Delta_0-\Delta_1\right) \nn\\
&& +\frac{3}{2}\left(\Delta_4-\Delta_5\right)\left(\Delta_0-\Delta_1\right) -\frac{3}{2}\left(\Delta_5''-\Delta_6''\right)
 \left(\Delta_0-\Delta_1\right)\nn\\
 &&+\Delta_5\Delta_1-\Delta_6''\Delta_1\Bigg]\sumintb_P~\frac{1}{p^2P^2}\nn\\
 &&= \(\frac{\Lambda e^{\gamma_E}}{4\pi T}\)^{2\eps}\frac{1}{60(4\pi)^2}\Bigg[\frac{-103+5\pi^2+76\ln 2}{\eps^2}+\frac{1}{15\eps}\Big\{-7473+550\pi^2\nn\\
   && +12\ln 2(118+95\ln 2)  +450\zeta(3)\Big\}+21.3892\Bigg] + \mathcal{O}[\eps].
 \eea
 %8.
 \item 
\bea
&&\sumintb_P~\frac{A_2p_0^2p_3^2(p_0^2-{\cal T}_PP^2)}{p_1^2p^4P^4} = \sumintb_P~\frac{1}{2}\Bigg[\frac{{\cal T}_Pp_0^2p_3^2(p_0^2-{\cal T}_PP^2)}{p_1^2p^4P^4}\left(1-\frac{p_3^2}{p^2}-\frac{p_0^2}{p^2}+\frac{3p_0^2p_3^2}{p^4}\right)\nn\\
&& +\frac{p_0^2p_3^2(p_0^2-{\cal T}_PP^2)}{p_1^2p^4P^4}\left(\frac{p_0^2}{p^2}-\frac{3p_0^2p_3^2}{p^4}\right)\Bigg]\nn\\
&&=\frac{1}{2}\Bigg[(1+\Delta_0)\left(3\Delta_1+\Delta_0'-\Delta_0\right)+\Delta_0-3\Delta_1\Bigg]\sumintb_P~\frac{1}{P^4}
+\frac{1}{2}\Bigg[3\Delta_0-9\Delta_1-\Delta_5\left(\Delta_0-3\Delta_1\right)\nn\\
&&+(1+\Delta_0)\left(12\Delta_1+3\Delta_0'-4\Delta_0\right)-\Delta_5'\Delta_0'+\Delta_6'\left(\Delta_0-3\Delta_1\right)
-\Delta_5''\Delta_0'\nn\\
&&+\Delta_6''\left(\Delta_0-3\Delta_1\right)\Bigg]\sumintb_P~\frac{1}{p^2P^2}\nn\\
&&= \(\frac{\Lambda e^{\gamma_E}}{4\pi T}\)^{2\eps}\frac{1}{60(4\pi)^2}\Bigg[\frac{-83+5\pi^2+56\ln 2}{\eps^2}+\frac{1}{15\eps}\Big\{-5893+500\pi^2 \nn\\
   &&+\ln 2(556+840\ln 2)+450\zeta(3)\Big\}+ 73.7496\Bigg] + \mathcal{O}[\eps].
 \eea
% 9.
\item 
\bea
&&\sumintb_P~\frac{A_2(5p_0^2+9p^2)(p_0^2-{\cal T}_PP^2)}{4p_1^2p^2P^4} = \sumintb_P~\frac{1}{2}\Bigg[\frac{{\cal T}_P(p_0^2-{\cal T}_PP^2)(5p_0^2+9p^2)}{4p_1^2p^2P^4}\times \nn\\
&&\left(1-\frac{p_3^2}{p^2}-\frac{p_0^2}{p^2}+\frac{3p_0^2p_3^2}{p^4}\right)+\frac{(p_0^2-{\cal T}_PP^2)(5p_0^2+9p^2)}{4p_1^2p^2P^4}\left(\frac{p_0^2}{p^2}-\frac{3p_0^2p_3^2}{p^4}\right)\Bigg]\nn\\
&&=\frac{7}{4}\Bigg[1-2\Delta_0+(1+\Delta_0)\left(1-(1+\Delta_0)+3\Delta_0\right)\Bigg]\sumintb_P~\frac{1}{P^4}+\frac{1}{2}\Bigg[
\frac{33}{4}\left(1-2\Delta_0\right)\nn\\
&&+\frac{1}{4}(1+\Delta_0)\left(33-47(1+\Delta_0)+141\Delta_0\right)-\frac{5}{4}\Delta_5\left((1+\Delta_0)-3\Delta_0\right)-\frac{9}{4}\Delta_4\nn\\
 &&\left((1+\Delta_0)-3\Delta_0\right)-\frac{1}{4}\Delta_5'\left(5-9(1+\Delta_0)+27\Delta_0\right)+\frac{5}{4}\Delta_6'
 \left((1+\Delta_0)-3\Delta_0\right)\nn\\
 &&-\frac{9}{4}\Delta_4'-\frac{1}{4}\Delta_5''\left(5-9(1+\Delta_0)+27\Delta_0\right)+\frac{5}{4}\Delta_6''\left((1+\Delta_0)-3\Delta_0\right)
 \nn\\
 &&-\frac{9}{4}\Delta_4''\Bigg]\sumintb_P~\frac{1}{p^2P^2}\nn\\
&&= \(\frac{\Lambda e^{\gamma_E}}{4\pi T}\)^{2\eps}\frac{1}{24(4\pi)^2}\Bigg[\frac{-100+7\pi^2+46\ln 2}{\eps^2}+\frac{1}{15\eps}\Big\{-6679+495\pi^2\nn\\
   && +2\ln 2 (599+345\ln 2)+630\zeta(3)\Big\}+14.5448\Bigg] + \mathcal{O}[\eps].
\eea
% 10.
\item 
 \bea
 &&\sumintb_P~\frac{A_4(p_0^2-{\cal T}_PP^2)}{p^2P^4} = \sumintb_P~\frac{3(p_0^2-{\cal T}_PP^2)}{8p^2P^4}\left(1-\frac{p_3^2}{p^2}\right)^2-\sumintb_P~\frac{p_0^2(p_0^2-{\cal T}_PP^2)}{8p^4P^4}\left(1-\frac{5p_3^2}{p^2}\right)^2\nn\\
&&+\sumintb_P~\frac{5}{3}\frac{p_0^2(p_0^2-{\cal T}_PP^2)}{p^4P^4}\frac{p_3^4}{p^4}-\frac{3}{8}\sumintb_P~\left(\frac{(1-{\cal T}_P)(p_0^2-{\cal T}_PP^2)}{p^2P^4}\right) \nn\\
&&\times\left\{\!\left(1-\frac{p_0^2}{p^2}\right)^2
-\frac{2p_3^2}{p^2} \left(1-\frac{3p_0^2}{p^2}\right)^2+\frac{p_3^4}{p^4} \left(1-\frac{5p_0^2}{p^2}\right)^2+\frac{8p_0^4}{p^4}\frac{p_3^2}{p^2} \left(1-\frac{5p_3^2}{3p^2}\right)\!\right\}\!\nn\\
&&=\Bigg[\frac{1}{4}+\frac{\Delta_0'}{2}-\frac{25\Delta_1'}{12}+(1+\Delta_0)\Delta_1'\Bigg]\sumintb_P~\frac{1}{P^4}
+\Bigg[\frac{1}{8}+\frac{19\Delta_0'}{4}-\frac{205\Delta_1'}{24}+(1+\Delta_0)\nn\\
&&\left(7\Delta_1'-3\Delta_0'\right)
+\Delta_5\left(\frac{3}{8}-\frac{15}{4}\Delta_0'+\frac{35}{8}\Delta_1'\right)-\Delta_4\left(\frac{5}{8}-\frac{13}{4}\Delta_0'+\frac{55}{24}
 \Delta_1'\right)\nn\\
 &&-\frac{3}{8}\Delta_4'\left(1-2\Delta_0'+\Delta_1'\right)+\frac{3}{8}\Delta_5'\left(2-12\Delta_0'+10\Delta_1'\right)
 -\frac{3}{8}\Delta_6'\left(1-10\Delta_0'+\frac{35}{3}\Delta_1'\right)\nn\\
 &&-\frac{3}{8}\Delta_4''\left(1-2\Delta_0'+\Delta_1'\right)+\frac{3}{8}\Delta_5''\left(2-12\Delta_0'+10\Delta_1'\right)\nn\\
&& -\frac{3}{8}\Delta_6''\left(1-10\Delta_0'+\frac{35}{3}\Delta_1'\right)\Bigg]\sumintb_P~\frac{1}{p^2P^2}\nn\\
   &=&\(\frac{\Lambda e^{\gamma_E}}{4\pi T}\)^{2\eps}\frac{1}{30(4\pi)^2}\Bigg[\frac{3-\pi^2+12\ln 2}{\eps}+25-\frac{16\pi^2}{15}+\frac{12\ln2}{5}\(5\ln2-7\)-6\zeta(3)\Bigg] .
 \eea
 \end{enumerate}

\section{c integrations}
\label{ang_av}

In this section we note down the following angular averages, which appeared throughout this paper due to the angular integrals $A_n$ ($A_0\equiv \mathcal{T}_P$). The symbol $\langle\rangle_c$ depicts the standard definition given in Ref~\cite{Andersen:2002ey}.
\bea
\Delta_0 &=& \left\langle \frac{c^2}{1-c^2}\right\rangle_c = 
-\frac{1}{2\eps}+\mathcal{O}[\epsilon]^3 \\
\Delta_0'&=& \left\langle c^2\right\rangle_c=\frac{1}{3}+\frac{2\epsilon}{9}+\frac{4\eps^2}{27}+\mathcal{O}[\epsilon]^3 \\
\Delta_0''&=& =\left\langle\frac{1}{(1-c_1^2)(1-c_2^2)}\right\rangle_{c_1,c_2} = 
\frac{(d-2)^2}{(d-3)^2} = \frac{1}{4\eps^2} 
- \frac{1}{\eps}  + 1 \\
\Delta_1 &=& \left\langle \frac{c^4}{1-c^2}\right\rangle_c = -\frac{1}{2\eps}-\frac{1}{3}-\frac{2\eps}{9}-\frac{4 \eps^2}{27}+\mathcal{O}[\epsilon]^3\\
\Delta_1'&=& \left\langle c^4\right\rangle_c=\frac{1}{5}+\frac{16 \epsilon}{75}+\mathcal{O}[\epsilon]^2 \\
\Delta_2 &=& \left\langle \frac{1-c^{2-d}}{1-c^2}\right\ranglec = 1-\frac{1}{2\epsilon}\\
\Delta_3 &=& \left\langle \frac{1-c^{4-d}}{1-c^2}\right\ranglec = 
\ln 2+\(\frac{\pi^2}{6}-(2-\ln 2)\ln 2\)\eps\nn\\
&&~~~~+\left\{\frac{2}{3}(\ln 2)^2(\ln 2 -3)+\frac{\pi^2}{3}(\ln 2-1)+\zeta(3)\right\}\eps^2+\mathcal{O}[\eps]^3. \\
\Delta_3' &=& \left\langle \frac{1-c^{4-d}}{(1-c^2)^2}\right\ranglec = -\frac{1}{4\eps}+\frac{1}{4}+\frac{3\eps}{4}-\frac{3\eps^2}{4}+\mathcal{O}[\eps]^3\,\\
\Delta_3'' &=& \left\langle \frac{1-c_1^{4-d}}{(1-c_1^2)(c_1^2-c_2^2)}+c_1\lrarrow c_2\right\rangleci \!\!\!\!= 
-\frac{\pi^2}{12} + \left(\frac{\pi^2}{3} - \frac{\zeta(3)}{2}\right)\eps + 
\mathcal{O}(\eps^2). \\
\Delta_4 &=& \left\langle \frac{1-c^{6-d}}{1-c^2}\right\ranglec = 
\(\frac{1}{2}+\ln 2\) + \(\frac{\pi^2}{6}-1-(1-\ln 2)\ln 2\)\eps +\epsilon ^2 \Big(\zeta (3)\nn\\
&&+\frac{1}{3} \ln 2 \big\{\ln 2 (2\ln 2-3)-6\big\}+\frac{1}{6}
\pi ^2 (2\ln 2 -1)\Big)+ \mathcal{O}[\epsilon]^3
\\
\Delta_4' &=& \left\langle \frac{1-c^{6-d}}{(1-c^2)^2}\right\ranglec 
= 
-\frac{3}{4 \epsilon }+\frac{5}{4}-\ln 2+\epsilon  \left(\frac{3}{4}-\frac{\pi ^2}{6}-\ln ^2(2)+\ln 4\right)
+\frac{1}{12} \epsilon ^2 \Big(-12 \zeta (3)\nn\\
&&-9-\ln ^3(4)+6 \ln ^2(4)-2 \pi ^2 (\ln 4-2)\Big)+ \mathcal{O}[\epsilon]^3\\
\Delta_4'' &=& \left\langle \frac{1-c_1^{6-d}}{(1-c_1^2)(c_1^2-c_2^2)}+c_1\lrarrow c_2\right\rangleci \!\!\!\!= 
-\frac{\pi^2}{12} + \ln 4 + \left(\frac{\pi^2}{3} - \ln4 (2 - \ln2) -  
\frac{\zeta(3)}{2}\right) \eps + \mathcal{O}(\eps^2)\\
%%%%%%%%
\Delta_4''' &=& \left\langle \frac{c_1^{6-d}-c_2^{6-d}}{c_1^2-c_2^2}\right\rangleci \nn\\
 &=& \frac{1}{3}(1+2\ln2) + \frac{2}{9}\(-5+\ln 2\(5+3\ln 2\)\)\eps+\mathcal{O}[\eps]^2\\
 %%%%%%%
\Delta_5 &=& \left\langle \frac{1-c^{8-d}}{1-c^2}\right\ranglec = \frac{3}{4}+\ln2+\frac{ \epsilon }{12} \left(2 \pi ^2+3 ((\ln (4)-1) \ln (4)-5)\right)\nn\\
&+&
\(\zeta (3)-\frac{1}{2}-\frac{\pi ^2}{12}+\frac{\ln2}{6}  \left(2 \pi ^2-15+4 \ln^2 2-3 \ln2\right)\)\epsilon ^2 \\
%%%%%%%%%
\Delta_5' &=& \left\langle \frac{1-c^{8-d}}{(1-c^2)^2}\right\ranglec = 
-\frac{5}{4
   \epsilon }  +\frac{7}{4}-\ln 4 +\epsilon  \left(\frac{7}{4}-\frac{\pi ^2}{3}-\frac{\ln ^2(4)}{2}+\ln 8\right)
   +\frac{\epsilon ^2 }{12}\Big(-24 \zeta (3)\nn\\
   &&-9+\pi ^2 (6-4 \ln 4)+\ln 4 (12+(9-2 \ln 4) \ln 4)\Big)+\mathcal{O}[\eps]^3
 \\
\Delta_5'' &=& \left\langle \frac{1-c_1^{8-d}}{(1-c_1^2)(c_1^2-c_2^2)}+c_1\lrarrow c_2\right\rangleci = 
\frac{1}{12}\left(4-\pi^2+32 \ln 2\right) \nn\\ && ~~~~+\frac{1}{18}\(-20+6\pi^2-52\ln 2+48(\ln 2)^2-9\zeta(3)\)\eps +0.469927\eps^2+\mathcal{O}[\epsilon]^3\\
\Delta_5''' &=& \left\langle \frac{c_1^{8-d}-c_2^{8-d}}{c_1^2-c_2^2}\right\rangleci\nn\\
&=& \frac{1}{10}(3+\ln 16)+ \frac{1}{150}\(-107+2\ln 2\(97+30\ln 2\)\)\eps + \mathcal{O}[\eps]^2\\
\Delta_6' &=& \left\langle \frac{1-c^{10-d}}{(1-c^2)^2}\right\ranglec = 
2-\frac{7}{4\eps}+\frac{1}{4} \epsilon  \left(12-2 \pi ^2+(7-3 \ln 4) \ln 4\right)-\ln 8+\frac{1}{24} \epsilon ^2 \Big(-72 \zeta (3)\nn\\
&&-6+2 \pi ^2 (7-6 \ln 4)+3 \ln 4 (18+(7-2 \ln 4) \ln 4)\Big)+\mathcal{O}[\eps]^3\\
\Delta_6'' &=& \left\langle \frac{1-c_1^{10-d}}{(1-c_1^2)(c_1^2-c_2^2)}+c_1\lrarrow c_2\right\rangleci \nn\\
&=& 
\frac{1}{60}\left(38-5\pi^2+184 \ln 2\right) + \frac{1}{450}\(-821+150\pi^2+(-359+690\ln 2)2\ln 2-225\zeta(3)\)\eps\nn\\
&&~~~~~~+1.04576\eps^2+ \mathcal{O}[\epsilon]^3\\
\Delta_7 &=& \left\langle c^{2\eps +1} \right\ranglec = \frac{1}{2} + \mathcal{O}[\epsilon]\\
\Delta_8 &=& \left\langle c^{2\eps +3} \right\ranglec = \frac{1}{4} + \mathcal{O}[\epsilon]\\
\Delta_9 &=& \left\langle c^{2\eps -1} \right\ranglec = \frac{1}{2\eps}-1+\ln 2 + \mathcal{O}[\epsilon]
\eea

 \bea
 \Delta_{10}&=&\left\langle\frac{c_1^{3+2\epsilon}-c_1^2}{(c_1^2-c_2^2)(1-c_1^2)^2}
   -\frac{c_2^{3+2\epsilon}-c_2^2}{(c_1^2-c_2^2)(1-c_2^2)^2}
\right\rangle_{c_1,c_2} . 
\label{delta5_2}
\eea

Unlike other $\Delta_i$ functions, computation of $\Delta_{10}$ is not straightforward
and cannot be done directly analytically in {\it Mathematica} in $\eps\rightarrow 0$.
After calculating the angular average in Eq.~(\ref{delta5_2}), we end up with
the following equation.
\bea
\Delta_{10}&=& -\frac{1}{\pi \Gamma(1-\epsilon)^2} \Gamma(-1 - \epsilon)^2 \Gamma(3/2 - 
\epsilon)^2\nn\\
&\times&\Bigg[\frac{1}{3} 4^{-1 - \epsilon}
   e^{-i \pi \epsilon} (1 + 3 \epsilon + 2 \epsilon^2)\ \Gamma(1 + 2 \epsilon) 
\Bigg\{3 e^{2 i \pi \epsilon} F\left(\mycom{\epsilon, 
   \frac{3}{2} + \epsilon}{\frac{1}{2}}\Bigg| 1\right)\nn\\
   &+& \left(2 + e^{2 i \pi \epsilon}\right) \Bigg(3 F\left(\mycom{\frac{3}{2} + 
\epsilon, 2 + \epsilon}{\frac{1}{2}}\Bigg| 1\right) -6 (3 + 2 \epsilon)
    F\left(\mycom{2 + \epsilon, \frac{5}{2} + \epsilon}{\frac{3}{2}}\Bigg| 
1\right) \nn\\
&+& (5 + 2 \epsilon)(3 + 2 \epsilon) F\left(\mycom{2 + \epsilon, 
 \frac{7}{2} + \epsilon}{\frac{5}{2}}\Bigg| 1\right]\Bigg)\Bigg\} 
 -\frac{\pi}{\Gamma[-1/2 - \epsilon]^2}\Bigg\{-F\left(\mycom{1, 
        \frac{3}{2} + \epsilon}{-\frac{1}{2}  - \epsilon}, 
        \Bigg| 1\right) \nn\\
&-& \frac{3 + 2 \epsilon}{1 + 2 \epsilon} F\left(\mycom{1, \frac{5}{2} + 
\epsilon}{\frac{1}{2} - \epsilon}\Bigg| 1\right) +
    \frac{1}{(1+2 \epsilon)^2} F\left(\mycom{1,\frac{3}{2},\frac{1}{2}+\epsilon }{ 
-\frac{1}{2},\frac{1}{2}-\epsilon}\Bigg|1\right) 
    + \frac{3}{1-4 \epsilon^2} F\left(\mycom{1,\frac{5}{2},\frac{3}{2}+\epsilon}{ 
\frac{1}{2},\frac{3}{2}-\epsilon}\Bigg|1\right)\Bigg\}\Bigg]\!.\hspace{1.2cm}
\label{delta5_3}
\eea

Eq.~(\ref{delta5_3}) cannot be expanded directly at small $\eps$ in {\it Mathematica}. 
So, we use the following technique to expand Eq.~(\ref{delta5_3}) at small $\eps$.
In Eq.~(\ref{delta5_3}), $F$ represents the generalized hypergeometric function.
The generalized hypergeometric function of type $_pF_q$ is
an analytic function of one variable with $p+q$ parameters. Here,
 the parameters are functions of $\epsilon$, so the list of
parameters sometimes gets so lengthy and the standard notation
for these functions becomes cumbersome. We therefore introduce
a more compact notation as
\bea
F\left(\mycom{\alpha_1,\alpha_2,\cdots \alpha_n}{ \beta_1,\cdots 
\beta_{n-1}}\Bigg|1\right)= 
\ _pF_q\left(\alpha_1,\alpha_2,\cdots \alpha_p; \beta_1,\cdots \beta_q;z\right)
\eea
It is not possible to directly expand $_pF_q$ at small $\epsilon$. So, we 
will use the following procedure to expand $_pF_q$ in the series of $\epsilon$.

In Eq.~(\ref{delta5_3}), there are two types of hypergeometric function {\it 
{\it{viz.}}.} $_2F_1$ and $_3F_2$. 
$_2F_1$ can be expanded in small $\epsilon$ if one uses the following relation.
\bea
F\left(  \alpha_1,\alpha_2; \beta_1; 1 \right)= \frac{ \Gamma(\beta_1) \Gamma(\beta_1 
- \alpha_1 - \alpha_2)}{ \Gamma(\beta_1 - \alpha_1) \Gamma(\beta_1 - \alpha_2) } \,
\eea

To expand $_3F_2$ type of hypergeometric function, we can try the following power series representation for the generalized hypergeometric function as
\begin{equation}
F\left( \mycom{ \alpha_1,\alpha_2,\ldots,\alpha_p
	}{ \beta_1,\ldots,\beta_q } \Bigg| z \right)
\;=\; \sum_{n=0}^\infty\frac{ (\alpha_1)_n (\alpha_2)_n \cdots (\alpha_p)_n
	}{ (\beta_1)_n \cdots (\beta_q)_n n! } z^n
	\, ,
\label{ps-pFq}
\end{equation}
%
%{\bf [ejpm] }
where $(a)_b$ is Pochhammer's symbol:
\begin{equation}
(a)_b = \frac{\Gamma(a+b)}{ \Gamma(a)} \,.
\label{Poch}
\end{equation}
%
%{\bf [ejp]}
The power series converges for $|z|<1$.
For $z=1$, it converges if ${\rm Re}\ s > 0$, where
\begin{equation}
s\;=\; \sum_{i=1}^{p-1} \beta_i -  \sum_{i=1}^p \alpha_i\, .
\label{s-def}
\end{equation}

In Eq.~(\ref{delta5_3}), both $_3F_2$ has negative $s$ value for $\epsilon 
\rightarrow 0$. So, we will use the following
relation to change the parameters to make $s$ value positive.
\bea
F\left( \mycom{ \alpha_1,\alpha_2,\alpha_3
	}{ \beta_1,\beta_2 } \Bigg| 1 \right) = \frac{ \Gamma(\beta_1) 
\Gamma(\beta_2) \Gamma(s)
	}{ \Gamma(\alpha_1+s) \Gamma(\alpha_2+s) \Gamma(\alpha_3)} \,
F\left( \mycom{ \beta_1-\alpha_3, \beta_2-\alpha_3, s
	}{ \alpha_1+s, \alpha_2+s } \Bigg| 1 \right) \,
	\label{3F2-1}
\eea

\begin{itemize}
\item 
{\large{\bf Expansion of $\prescript{}{3}F_2(1,\frac{3}{2},\frac{1}{2}+\epsilon; 
 -\frac{1}{2},\frac{1}{2}-\epsilon; 1)$}}
 \noindent
 
 Here, $s=\beta_1+\beta_2-\alpha_1-\alpha_2-\alpha_3= -3-2\epsilon < 0$ at 
$\epsilon\rightarrow 0$.
 So, we use Eq.~(\ref{3F2-1}) to change the parameter.
 %%%%
 \bea
  F\left(\mycom{1,\frac{3}{2},\frac{1}{2}+\epsilon}{ 
-\frac{1}{2},\frac{1}{2}-\epsilon}\Bigg|1\right) 
  = \frac{\Gamma\left(-\frac{1}{2}\right)\Gamma\left(\frac{1}{2}-\epsilon\right) 
\Gamma\left(3-2\epsilon\right)}
  {\Gamma\left(-2 - 
2\epsilon\right)\Gamma\left(-\frac{3}{2}-2\epsilon\right)\Gamma\left(\frac{1}{2}
+\epsilon\right)}
  F\left(\mycom{-1-\epsilon,-2\epsilon,-3-2\epsilon}{ -2 - 2\epsilon,-\frac{3}{2} 
- 2\epsilon}\Bigg|1\right)\nn\\
 \label{3F2_ini}
 \eea
For the hypergeometric function that appears on the right hans side of the above equation, 
$s=1/2+9\epsilon>0$ at $\epsilon\rightarrow 0$ , 
so the power series expansion will converge.

So,
 \bea
 &&F\left(\mycom{-1-\epsilon,-2\epsilon,-3-2\epsilon}{ 
-2+2\epsilon,-\frac{3}{2}+2\epsilon}\Bigg|1\right)\nn\\
 &=& \sum_{n=0}^\infty \frac{\Gamma\left(-1-\eps + 
n\right)\Gamma\left(-2\epsilon+n\right)
 \Gamma\left(-3-2\epsilon+n\right)\Gamma\left(-2 - 
2\epsilon\right)\Gamma\left(-\frac{3}{2} - 2\epsilon\right)}
  {\Gamma\left(-2 - 2\eps +n\right)\Gamma\left(-\frac{3}{2} - 
2\epsilon+n\right)\Gamma\left(-1-\eps\right)
  \Gamma\left(-2\epsilon\right) \Gamma\left(-3-2\epsilon\right)}\frac{1}{n!}.\hspace{1.2cm}
  \eea
  %%%%%%%%%%%%%%%%%%%%%%%%%%%%%%%%%%%%%%%%%%%%%%%%%%%%%%
  
In the above sum, if one expand the term within summation at small $\epsilon$, 
one ends up with an expression that diverges
for $n=0$ to $n=3$. To avoid that we will do the summation before the expansion 
from $n=0$ to $n=3$. From $n=4$, we will perform
the summation after expansion in small $\epsilon$.

So,
\bea &&{}_3F_2\left(-1-\epsilon,-2\epsilon,-3-2\epsilon; 
-2+2\epsilon,-\frac{3}{2}+2\epsilon;1\right)\nn\\
  &=& \left(\sum_{n=0}^3+\sum_{n=4}^\infty\right)\frac{\Gamma\left(-1-\eps + 
n\right)\Gamma\left(-2\epsilon+n\right)
  \Gamma\left(-3-2\epsilon+n\right)\Gamma\left(-2 - 
2\epsilon\right)\Gamma\left(-\frac{3}{2} - 2\epsilon\right)}
  {\Gamma\left(-2 - 2\eps +n\right)\Gamma\left(-\frac{3}{2} - 2\epsilon+n\right)
  \Gamma\left(-1-\eps\right)\Gamma\left(-2\epsilon\right) 
\Gamma\left(-3-2\epsilon\right)}\frac{1}{n!}\nn\\
  &=&\Bigg[ 1 - \frac{14}{3}\eps +\frac{64}{9}\eps^2 -\frac{652}{27}\eps^3 + 
\mathcal{O}\left(\epsilon^4\right) \Bigg]  
  + \sum_{n=4}^\infty\Bigg[ \frac{8 \sqrt{\pi }\  \Gamma (n-1)}{n(n-3) \Gamma 
\left(n-\frac{3}{2}\right)}\ \epsilon^2\nn\\
 &-& \frac{8  \sqrt{\pi }\ \Gamma (n) \left(3 \gamma_E + 11 + 
\frac{9}{n-3}+\frac{9}{n-2}+\frac{6}{n-1}+15 \psi ^{(0)}(n-3)
   -12 \psi ^{(0)}(2 n-4)\right)\epsilon ^3}{3 n(n-3) (n-1) \Gamma 
\left(n-\frac{3}{2}\right)} + \mathcal{O}
   \left(\epsilon^4\right)\Bigg]\nn\\
 &=& 1 - \frac{14}{3}\eps +\frac{64}{9}\eps^2 -\frac{652}{27}\eps^3  +
  \frac{16}{3}{}_4F_3\left(1, 1, 3, 4; 2, 5 , \frac{5}{2}; 1\right)\eps^2 \nn\\
  &+& \frac{1}{18} \Bigg[32 (3 \gamma_E -11 + 6\ln2) \, 
_3F_2\left(1,1,4;\frac{5}{2},5;1\right)
  +32 (6 \gamma_E -11 + 12\ln 2) \, _3F_2\left(1,1,3;\frac{3}{2},4;1\right)\nn\\
  &+& 252 \zeta (3)-36 \gamma_E  \pi ^2-51 \pi ^2-640 \gamma_E + 824-72 \pi ^2 \ln 
2-1280 \ln 2\Bigg]\eps^3 + \mathcal{O}
   \left(\epsilon^4\right)
  \label{3F2_interm}
  \eea
The hypergeometric function of type $_{p+1}F_{q+1}$
has an integral representation in terms of the hypergeometric function
of type $_pF_q$:
\bea
&&\hspace{-12mm}
\int_0^1 dt \, t^{\nu-1} (1-t)^{\mu-1} \,
F\left( \mycom{ \alpha_1,\alpha_2,\ldots,\alpha_p
        }{ \beta_1,\ldots,\beta_q  } \Bigg| tz \right)
 =\, \frac{ \Gamma(\mu) \Gamma(\nu)}{ \Gamma(\mu+\nu)} \,
F\left( \mycom{ \alpha_1,\alpha_2,\ldots,\alpha_p,\nu
        }{ \beta_1,\ldots,\beta_q,\mu+\nu} \Bigg| z \right)\!. \hspace{1cm} 
\label{int-pFq}
\eea
Using Eq.~(\ref{int-pFq}), Eq.~(\ref{3F2_interm}) can be simplified as
  \bea &&{}_3F_2\left(-1-\epsilon,-2\epsilon,-3-2\epsilon; 
-2+2\epsilon,-\frac{3}{2}+2\epsilon;1\right)\nn\\
  &=& 1 - \frac{14}{3}\eps +\frac{64}{9}\eps^2 -\frac{652}{27}\eps^3 
  + \frac{16}{3}\int\limits_0^1 dt F\left(\mycom{ 1, 3, 4 }{ 5 
,\frac{5}{2}}\Bigg|t\right)\eps^2 \nn\\
   && +
  \frac{1}{18} \Bigg[32 (3 \gamma_E -11 + 6\ln2) \times 3\int\limits_0^1 dt\ t^2\ 
{}_2F_1\left( 1, 1;\frac{3}{2}; t\right)
  \nn\\
  &&+32 (6 \gamma_E -11 + 12\ln 2) \times 4\int\limits_0^1 dt\ t^3\ {}_2F_1\left( 
1, 1;\frac{5}{2}; t\right)\nn\\
  &&+252 \zeta (3)+824-51 \pi ^2-4(160+9\pi ^2)( \gamma_E +2 \ln 2)\Bigg]\eps^3 + 
\mathcal{O}\left(\eps^4\right)\nn\\
    &=& 1 - \frac{14}{3}\eps +\frac{64}{9} \eps^2 -\frac{652}{27}\eps^3  + 
\left(\pi^2 +\frac{112}{9}\right) \eps^2 
    +\left(14 \zeta (3)-\frac{2}{27} \left(746+63 \pi ^2\right)\right)\eps^3 + 
\mathcal{O}\left(\eps^4\right)\nn\\
  &=& 1 - \frac{14}{3}\eps + \left(\pi^2 +\frac{176}{9}\right)\eps^2 
+\left(-\frac{2144}{27}
         - \frac{14 \pi^2}{3} + 14 \zeta(3)\right) \eps^3 +\mathcal{O}(\eps^4)
 \eea
 
 So, Eq.~(\ref{3F2_ini}) can be rewritten as
 \bea
  F\left(1,\frac{3}{2},\frac{1}{2}+\epsilon; 
-\frac{1}{2},\frac{1}{2}-\epsilon;1\right)
  &=& 
\frac{\Gamma\left(-\frac{1}{2}\right)\Gamma\left(\frac{1}{2}-\epsilon\right) 
\Gamma\left(3-2\epsilon\right)}
  {\Gamma\left(-2 - 
2\epsilon\right)\Gamma\left(-\frac{3}{2}-2\epsilon\right)\Gamma\left(\frac{1}{2}
+\epsilon\right)}
 \nn\\
 &&\hspace{-4cm}\times
 \left[1 - \frac{14}{3}\eps +  \left(\pi^2 +\frac{176}{9}\right)\eps^2 
+\left(-\frac{2144}{27}
         - \frac{14 \pi^2}{3} + 14 \zeta(3)\right) \eps^3  
+\mathcal{O}(\eps^4)\right]\hspace{1cm} 
 \eea
 
 \item 
 {\large{\bf Expansion of ${}_3F_2\left(1,\frac{5}{2},\frac{3}{2}+\epsilon; 
\frac{1}{2},\frac{3}{2}-\epsilon; 1\right)$ }}
\noindent
 
 Following a similar procedure, we can write
 \bea
  {}_3F_2\left(1,\frac{5}{2},\frac{3}{2}+\epsilon; 
\frac{1}{2},\frac{3}{2}-\epsilon; 1\right) 
  &=& 
\frac{\Gamma\left(\frac{1}{2}\right)\Gamma\left(\frac{3}{2}-\epsilon\right) 
\Gamma\left(3-2\epsilon\right)}
  {\Gamma\left(-2 - 
2\epsilon\right)\Gamma\left(-\frac{1}{2}-2\epsilon\right)\Gamma\left(\frac{3}{2}
+\epsilon\right)}
  \nn\\
  &&\hspace{-4cm}\times\left[1 - \frac{10}{3}\eps  + \left(\pi^2 + 
\frac{40}{3}\right)\eps^2 
  +\left(-\frac{160}{3} - \frac{10 \pi ^2}{3} +14 \zeta (3)\right) \epsilon ^3 + 
\mathcal{O}(\eps^4)\right]\hspace{.7cm}
 \eea
 
 Adding all the contributions, we can write Eq.~(\ref{delta5_3}) as
\bea
 \Delta_{10}&=&\left\langle\frac{c_1^{3+2\epsilon}-c_1^2}{(c_1^2-c_2^2)(1-c_1^2)^2}
   -\frac{c_2^{3+2\epsilon}-c_2^2}{(c_1^2-c_2^2)(1-c_2^2)^2}
\right\rangle_{c_1,c_2}  \nn\\
&=& -\frac{1}{8\eps^2}+\frac{1}{4\eps} + \frac{1}{24}\left(18-\pi^2\right) 
+\frac{1}{12} (2 \pi^2 - 3 (9 + \zeta(3))) \eps
+\mathcal{O}(\eps^2)
\eea
\end{itemize}

The remaining $c$ integration $\Delta_{11}$ can be written as 
\bea
\Delta_{11}&=&\left\langle\frac{c_1^{1+2\epsilon}-c_1^2}{(c_1^2-c_2^2)(1-c_1^2)^2}
   -\frac{c_2^{1+2\epsilon}-c_2^2}{(c_1^2-c_2^2)(1-c_2^2)^2}
\right\rangle_{c_1,c_2} \nn\\
&=& \Delta_0''+\Delta_{10}-\Delta_3''\nn\\
&=& \frac{1}{8\eps^2}-\frac{3}{4\eps} + 
\frac{1}{24}\left(\pi^2+42\right)+\frac{1}{12} ( 3 (\zeta(3)-9)-2 \pi^2) \eps 
+\mathcal{O}(\eps^2)
\eea
%%%%%%%%%%%%%%%%%%%%%%%%%%
%%%%%%%%%%%%%%%%%%%%%%%%%%
\section{Free energy without high $T$ expansion}
\label{wohighT}
\subsection{Quark part}
Free energy of quark is given by
\bea
F_q&=& 
-N_c\sum_f\int\frac{d^4P}{(2\pi)^4}~\ln\left({\mbox{det}}\left[S_{\textrm{eff}}^{-1}(P)\right]
\right). \label{quark_F}
\eea

Argument of the logarithm can also be factorized as 
\bea
{\mbox{det}}\left[S_{\textrm{eff}}^{-1}\right] &=& (\mathcal{C}^2p_0^2-\mathcal{D}^2p^2+\mathcal{B}'^2-\mathcal{C}'^2)^2-4(p_0 
\mathcal{B}'\mathcal{C}+p_3 \mathcal{C}'\mathcal{D})^2\nn\\
&=& \left[\left(\mathcal{C}p_0-\mathcal{B}'\right)^2-\left(\left(\mathcal{D}p_3+\mathcal{C}'\right)^2+\mathcal{D}^2p_\perp^2\right) \right] \times \nn\\
&& ~~~\left[\left(\mathcal{C}p_0+\mathcal{B}'\right)^2-\left(\left(\mathcal{D}p_3-\mathcal{C}'\right)^2+\mathcal{D}^2p_\perp^2\right) \right] \nn\\
&=& \left[\left(\left(1+\mathcal{A}\right)p_0+\mathcal{B}-\mathcal{B}'\right)^2-\left(\left((1+\mathcal{A})p_3+\mathcal{C}'\right)^2+(1+\mathcal{A})^2p_\perp^2\right) \right] \times \nn\\
&& ~~~\left[\left(\left(1+\mathcal{A}\right)p_0+\mathcal{B}+\mathcal{B}'\right)^2-\left(\left((1+\mathcal{A})p_3-\mathcal{C}'\right)^2+(1+\mathcal{A})^2p_\perp^2\right) \right] \nn\\
&=& L_+L_-R_+R_- ,\nn
\label{LR}
\eea
with
\bea
L_\pm&=& \left(\left(1+\mathcal{A}\right)p_0+\mathcal{B}-\mathcal{B}'\right)\mp\sqrt{\left((1+\mathcal{A})p_3+\mathcal{C}'\right)^2+(1+\mathcal{A})^2p_\perp^2}, \nn\\
R_\pm&=& \left(\left(1+\mathcal{A}\right)p_0+\mathcal{B}+\mathcal{B}'\right)\mp\sqrt{\left((1+\mathcal{A})p_3-\mathcal{C}'\right)^2+(1+\mathcal{A})^2p_\perp^2} ,
\eea

and
\bea
\mathcal{A}(p_0,p) &=& -\frac{m_{\rm th}^2}{p^2}Q_1\(\frac{p_0}{p}\),\nn\\
\mathcal{B}(p_0,p) &=& \frac{m_{\rm th}}{p}\left[\frac{p_0}{p}Q_1\(\frac{p_0}{p}\)-Q_0\(\frac{p_0}{p}\)\right],\nn\\
\mathcal{B}'(p_0,p,\theta) &=& -\frac{m_{\rm eff}^2\  p_z}{p^2}Q_1\(\frac{p_0}{p}\),\nn\\
\mathcal{C}'(p_0,p) &=& \frac{m_{\rm eff}^2\  }{p}Q_0\(\frac{p_0}{p}\).
\eea
Using these in Eq.\eqref{quark_F} one gets
\bea
F_{\text{q}}&=& -N_c\sum_f\int\frac{d^4P}{(2\pi)^4}~\ln L_+L_-R_+R_-\nn\\
&=& -N_cN_f\int\frac{d^4P}{(2\pi)^4}~\ln P^4-N_c\sum_f\int\frac{d^4P}{(2\pi)^4}~\ln\frac{L_+L_-R_+R_-}{P^4}\nn\\
&=&F^q_{\rm{Free}} + F^q_{\rm{QP}} + F^q_{\rm{LD}},
\eea
where the free part is obtained as
\be
F^q_{\text{Free}}=  -N_cN_f \frac{7\pi^2T^4}{180},
\ee

The quasiparticle (QP) part $ F_{\rm{QP}}$ is given as
\bea
F^q_{\text{QP}}&=&-N_c\sum_f\int\frac{d^3p}{(2\pi)^3}\oint \frac{dp_0}{4\pi i}\ln\frac{ D_+D_-}{P^4} \tanh{\frac{\beta p_0}{2}}\nn\\
&=& - N_c\sum_f\int\frac{d^3p}{(2\pi)^3}\oint \frac{dp_0}{4\pi i}\bigg[\ln  L_++\ln  L_-+\ln  R_++\ln  R_--\ln P^4\bigg] \tanh{\frac{\beta p_0}{2}}\nn\\
&=& -N_c\sum_f\int\frac{d^3p}{(2\pi)^3}\bigg[I_1+I_2+I_3+I_4- I_0\bigg].\label{qp_int}
\eea
One has the following dispersion  solutions for a quark in a thermomagnetic medium:

\noindent $L_+$ has solutions at $p_0=\omega_L^+,\ -\omega_R^-$\,,\\
$L_-$ has solutions at  $p_0=\omega_L^-,\ -\omega_R^+$\,,\\
$R_+$ has solutions at $p_0=\omega_R^+,\ -\omega_L^-$\,,\\
$R_-$ has solutions at  $p_0=\omega_R^-,\ -\omega_L^+$\,.\\

Now one can calculate the various integrations in Eq.\eqref{qp_int} as
\bea
I_1&=&\oint \frac{dp_0}{4\pi i}\ln  L_+ \tanh{\frac{\beta p_0}{2}}\nn\\
&=& -\frac{2}{\beta}\oint \frac{dp_0}{4\pi i}\frac{ L_+'}{L_+} \ln \cosh{\frac{\beta p_0}{2}}\nn\\
&=& \frac{T}{-2\pi i}\oint dp_0\frac{ L_+'}{L_+} \ln \cosh{\frac{\beta p_0}{2}}.
\eea

The integral has poles at $p_0=\omega_L^+,\ -\omega_R^-$. We now calculate the residues as
\bea
R\rvert_{p_0=\pm\omega_i}&=& \frac{ L_+'(p_0,p)}{L_+(\pm\omega_i,p)+(p_0\mp\omega_i)L_+'(\pm\omega_i,p)} \ln \cosh{\frac{\beta p_0}{2}}\rvert_{p_0=\pm\omega_i}\nn\\
&=& \ln \cosh{\frac{\pm \beta \omega_i }{2}}= \frac{\beta \omega_i}{2}+\ln{\big(1+e^{-\beta \omega_i}\big)}.
\eea
Now one can write
\bea
I_1&=&\frac{T}{-2\pi i}(-2\pi i)\left[\frac{\beta }{2}\big(\omega_L^+\ +\omega_R^-\big)+\ln{\big(1+e^{-\beta \omega_L^+}\big)}+\ln{\big(1+e^{-\beta \omega_R^-}\big)}\right]\nn\\
&=& \frac{1}{2}\big(\omega_L^+\ +\omega_R^-\big)+T\ln{\big(1+e^{-\beta \omega_L^+}\big)}+T\ln{\big(1+e^{-\beta \omega_R^-}\big)},
\eea
\bea
I_2&=&\frac{1 }{2}\big(\omega_L^-\ +\omega_R^+\big)+T\ln{\big(1+e^{-\beta \omega_L^-}\big)}+T\ln{\big(1+e^{-\beta \omega_R^+}\big)},
\eea
\bea
I_3&=&\frac{1}{2}\big(\omega_R^+\ +\omega_L^-\big)+T\ln{\big(1+e^{-\beta \omega_R^+}\big)}+T\ln{\big(1+e^{-\beta \omega_L^-}\big)},
\eea
\bea
I_4&=&\frac{1}{2}\big(\omega_L^+\ +\omega_R^-\big)+T\ln{\big(1+e^{-\beta \omega_L^+}\big)}+T\ln{\big(1+e^{-\beta \omega_R^-}\big)},
\eea
\bea
I_0&=&2p+4T \ln{\big(1+e^{-\beta p}\big)}.
\eea
As found $I_1=I_3$ and $I_2=I_4$.

Now one can write the quasiparticle part of the quark free energy as
\bea
F^q_{\text{QP}}&=& -N_c \sum_f\int \frac{d^3p}{(2\pi)^3}\bigg[\bigg(\omega_L^++\omega_R^++\omega_L^-+\omega_R^--2p\bigg)+2T \ln (1+e^{-\beta \omega_L^+})\nn\\
&&+2T \ln (1+e^{-\beta \omega_R^+})+2T \ln \frac{1+e^{-\beta \omega_L^-}}{1+e^{-\beta p}}+2T \ln \frac{1+e^{-\beta \omega_R^-}}{1+e^{-\beta p}}\bigg].\label{qp}
\eea

At large momentum limit, the QP modes are obtained as
\bea
\omega_L^+ &\approx& p+ \frac{1}{p}\(m_{\rm th}^2+m_{\rm eff}^2\,\frac{p_z}{p}\)\nn\\ &&\hspace{1cm}-\ \frac{1}{2p^3}\left[\(m_{\rm th}^2+m_{\rm eff}^2\frac{p_z}{p}\)^2-\frac{m_{\rm eff}^4p_\perp^2}{p^2}\log\frac{2p^2}{m_{\rm th}^2+m_{\rm eff}^2\,\frac{p_z}{p}}\right]\log\frac{2p^2}{m_{\rm th}^2+m_{\rm eff}^2\,\frac{p_z}{p}},\nn\\
%%%%
\omega_R^+ &\approx& p+ \frac{1}{p}\(m_{\rm th}^2-m_{\rm eff}^2\,\frac{p_z}{p}\)\nn\\
&&\hspace{1cm} -\frac{1}{2p^3}\left[\(m_{\rm th}^2-m_{\rm eff}^2\frac{p_z}{p}\)^2-\frac{m_{\rm eff}^4p_\perp^2}{p^2}\log\frac{2p^2}{m_{\rm th}^2-m_{\rm eff}^2\,\frac{p_z}{p}}\right]\log\frac{2p^2}{m_{\rm th}^2-m_{\rm eff}^2\,\frac{p_z}{p}}. \nn
\eea
As can be seen, the terms in the parentheses in Eq.\eqref{qp} have divergences in the high momentum limit and one needs counterterms to regulate them. We regulate as below
through the subtraction method following Ref.~\cite{Andersen:1999va}. 

Now we construct the $\mathcal{F}_{q,\,\rm{QP}}^{sub}$ for $\omega_L^+$ poles as
\bea
\mathcal{F}_{q,\,QP}^{sub,L^+}&=& -N_c\sum_f\!\int\limits_{\bm p}\Bigg[\sqrt{p^2+2m_{\rm th}^2+\frac{2m_{\rm eff}^2p_z}{p}} \nn\\
&-& \frac{\(m_{\rm th}^2+\frac{m_{\rm eff}^2p_z}{p}\)^2}{2\(p^2+2m_{\rm th}^2+\frac{2m_{\rm eff}^2p_z}{p}\)^{3/2}}\left\{\log\frac{2\(p^2+2m_{\rm th}^2+\frac{2m_{\rm eff}^2p_z}{p}\)}{m_{\rm th}^2+\frac{m_{\rm eff}^2p_z}{p}}\right.\nn\\
&\times&\left.\left.\(1+\frac{m_{\rm eff}^4p_\perp^2}{4p^4\left\{m_{\rm th}^2+\frac{m_{\rm eff}^2p_z}{p}\right\}}
-\frac{m_{\rm eff}^4p_\perp^2}{4p^2\left\{m_{\rm th}^2+\frac{m_{\rm eff}^2p_z}{p}\right\}^2}
\log\frac{2\(p^2+2m_{\rm th}^2+\frac{2m_{\rm eff}^2p_z}{p}\)}{m_{\rm th}^2+\frac{m_{\rm eff}^2p_z}{p}}\) -1   \right\}\right],\qquad
\label{qp_subL+}
\eea
and for the $\omega_R^+$ poles as
\bea
\mathcal{F}_{q,\,QP}^{sub,R^+}&=& -N_c\sum_f\!\int\limits_{\bm p}\left[\sqrt{p^2+2m_{\rm th}^2-\frac{2m_{\rm eff}^2p_z}{p}}\right.\nn\\
&-& \frac{\(m_{\rm th}^2-\frac{m_{\rm eff}^2p_z}{p}\)^2}{2\(p^2+2m_{\rm th}^2-\frac{2m_{\rm eff}^2p_z}{p}\)^{3/2}}\left\{\log\frac{2\(p^2+2m_{\rm th}^2-\frac{2m_{\rm eff}^2p_z}{p}\)}{m_{\rm th}^2-\frac{m_{\rm eff}^2p_z}{p}}\right.\nn\\
&\times&\left.\left.\(1+\frac{m_{\rm eff}^4p_\perp^2}{4p^4\left\{m_{\rm th}^2-\frac{m_{\rm eff}^2p_z}{p}\right\}}
-\frac{m_{\rm eff}^4p_\perp^2}{4p^2\left\{m_{\rm th}^2-\frac{m_{\rm eff}^2p_z}{p}\right\}^2}
\log\frac{2\(p^2+2m_{\rm th}^2-\frac{2m_{\rm eff}^2p_z}{p}\)}{m_{\rm th}^2-\frac{m_{\rm eff}^2p_z}{p}}\) -1   \right\}\right].\qquad
\label{qp_subR+}
\eea

A regularized quasiparticle part is obtained by subtracting Eqs.~\eqref{qp_subL+} and~\eqref{qp_subR+} from~\eqref{qp} as
\bea
F_{\text{q,\,QP}}^{\rm ren}&=& -N_c \sum_f\int \frac{d^3p}{(2\pi)^3}\bigg[2T \ln (1+e^{-\beta \omega_L^+}) +2T \ln (1+e^{-\beta \omega_R^+})+2T \ln \frac{1+e^{-\beta \omega_L^-}}{1+e^{-\beta p}}\nn\\
&&+2T \ln \frac{1+e^{-\beta \omega_R^-}}{1+e^{-\beta p}}\bigg] - N_c\sum_f\int \frac{d^3p}{(2\pi)^3} \bigg(\omega_L^+- \mathcal{F}_{q,\,qp}^{sub,L^+} +\omega_R^+-\mathcal{F}_{q,\,qp}^{sub,R^+}\bigg)\nn\\
&-&N_c\sum_f\int \frac{d^3p}{(2\pi)^3} \bigg(\omega_L^-+\omega_R^--2p\bigg).\label{FqQPren}
\eea

%\begin{figure}
%	\includegraphics[scale=.7]{qp_compare}
%	\caption{Total QP including renormalized zero point energy compared with vanishing magnetic field case}
%\end{figure}
In the high momentum limit, $\omega_{L,R}^-$ approaches to the light cone and $(\omega_L^-+\omega_R^--2p)$ vanishes. 
Also, the logarithmic terms that appear as a ratio in the above equation also vanish. The Stefan-Boltzmann limit of the above equation becomes
\bea
F_{\text{QP}}^{\rm SB}&=& -2N_c N_f\int \frac{d^3p}{(2\pi)^3}\bigg[p+2T \ln \left(1+e^{-\beta p}\right)\bigg].
\eea
Dropping the vacuum contribution, we get
\bea
F_{\text{QP}}^{\rm SB}&=&-N_c N_f\frac{7\pi^2T^4}{180}.
\eea

Now, the Landau damping (LD) part $F^q_{\rm{LD}}$ of the free energy can be obtained as
\bea
F^q_{LD} &=&- \frac{2N_c}{\pi}\sum_f \int\frac{d^3p}{(2\pi)^3} \int\limits_0^p dp_0 \left(n_F(p_0)-\frac{1}{2}\right)\bigg[{\rm Arg} \( L_+\)+{\rm Arg} \( R_+\)+{\rm Arg} \(L_-\)+{\rm Arg} \(R_-\)\bigg]_{p_0<p} \nn\\
&=& \frac{2N_c}{\pi}\sum_f \int\frac{d^3p}{(2\pi)^3} \int\limits_0^p dp_0 \left(n_F(p_0)-\frac{1}{2}\right)\left[\Theta_{LD}^L+\Theta_{LD}^R\right],
\label{F_qLD}
\eea
where
\bea
\Theta_{LD}^L=\tan^{-1}\frac{-{\rm Im}\(L_+L_-\)}{{\rm Re}\(L_+L_-\)}=\tan^{-1}\frac{-{\rm Im}\(L^2\)}{{\rm Re}\(L^2\)},\\
\Theta_{LD}^R=\tan^{-1}\frac{{-\rm Im}\(R_+R_-\)}{{\rm Re}\(R_+R_-\)}=\tan^{-1}\frac{{-\rm Im}\(R^2\)}{{\rm Re}\(R^2\)}.
\eea
Now we obtain the relevant contributions as
\bea
{\rm Im}\(L^2\)&=& \frac{\pi m_{\rm th}^4}{p^2}\left[\frac{p_0}{p} \left(1-\mu ^2 \xi^2\right)-\frac{\mu\,\xi\(p^2-p_0^2\)}{m_{\rm th}^2}+\frac{1}{2} \left\{\left(1-\frac{p_0^2}{p^2}\right)-\mu^2 \left(1-\frac{\xi^2 p_0^2}{p^2}\right)\right\}L_g\right] ,\\
%%%%%%%
{\rm Re}\(L^2\)&=& -\left[p^2-p_0^2+2m_{\rm th}^2+\frac{2 m_{\rm eff}^2\,  \xi\, p_0}{p} +\frac{m_{\rm th}^4}{p^2}\left\{\left(1-\mu ^2 \xi ^2\right)- \left(\frac{p_0 \left(1-\mu ^2 \xi ^2\right)}{p}-   \frac{\mu\, \xi\(p^2-p_0^2\)}{m_{\rm th}^2}\right)L_g\right.\right.\nn\\
&&\left.\left.
-\frac{ \left(1-\mu ^2\right) p^2-p_0^2 \left(1-\mu ^2 \xi ^2\right)}{4 p^2}\left(L_g^2-\pi ^2\right)\right\}\right],
\eea

\bea
{\rm Im}\(R^2\)&=& \frac{\pi m_{\rm th}^4}{p^2}\left[\frac{p_0}{p} \left(1-\mu ^2 \xi^2\right)+ \frac{\mu\,\xi\(p^2-p_0^2\)}{m_{\rm th}^2}+\frac{1}{2} \left\{\left(1-\frac{p_0^2}{p^2}\right)-\mu^2 \left(1-\frac{\xi^2 p_0^2}{p^2}\right)\right\}L_g\right] ,\\
%%%%%%%
{\rm Re}\(R^2\)&=& -\left[p^2-p_0^2+2m_{\rm th}^2-\frac{2 m_{\rm eff}^2\, \xi  p_0}{p} +\frac{m_{\rm th}^4}{p^2}\left\{\left(1-\mu ^2 \xi ^2\right)- \left(\frac{p_0 \left(1-\mu ^2 \xi ^2\right)}{p}+\frac{\mu\,\xi\(p^2-p_0^2\)}{m_{\rm th}^2}\right)L_g\right.\right.\nn\\
&&\left.\left.
-\frac{ \left(1-\mu ^2\right) p^2-p_0^2 \left(1-\mu ^2 \xi ^2\right)}{4 p^2}\left(L_g^2-\pi ^2\right)\right\}\right],
\eea
with $\xi=\cos\theta;\, \mu=\frac{m_{\rm eff}^2}{m_{\rm th}^2},\ L_g=\log\frac{p+p_0}{p-p_0}$.

Now, because of the term $-\frac{1}{2}$ with the distribution function in Eq.~\eqref{F_qLD}, the LD part will be UV diverging and we need suitable subtraction term to regularize it. We choose the following subtraction terms~\cite{Andersen:1999va} for $L$ and $R$ branches as
\bea
F_{\rm LD}^{\rm sub, L}&=& -2m_{\rm th}^4\int\limits_{\bm p}\int dp_0 \left[\frac{p_0\(1-\mu^2\xi^2-\frac{p}{p_0}\,\frac{\mu\,\xi(p^2-p_0^2)}{m_{\rm th}^2}\)}{p^3\left(p^2-p_0^2+2m_{\rm th}^2+\frac{2m_{\rm eff}^2p_0\xi}{p}\right)}\right.\nn\\
&&\left.\hspace{4cm} +\ \frac{\left(1+\mu ^2\xi ^2\right) \left(p^2-p_0^2\right)-\mu^2 \left(1-\xi ^2\right) p^2}{2p^4\left(p^2-p_0^2+2m_{\rm th}^2+\frac{2m_{\rm eff}^2p_0\xi}{p}\right)}\log \frac{p+p_0}{p-p_0} \right],\\
%%%%%%%%
F_{\rm LD}^{\rm sub, R}&=& -2m_{\rm th}^4\int\limits_{\bm p}\int\! dp_0 \left[\frac{p_0\(1-\mu^2\xi^2+\frac{p}{p_0}\,\frac{\mu\,\xi(p^2-p_0^2)}{m_{\rm th}^2}\)}{p^3\left(p^2-p_0^2+2m_{\rm th}^2-\frac{2m_{\rm eff}^2p_0\xi}{p}\right)} \right.\nn\\
&&\left.\hspace{4cm} +\  \frac{\left(1+\mu ^2\xi ^2\right) \left(p^2-p_0^2\right)-\mu^2 \left(1-\xi ^2\right) p^2}{2p^4\left(p^2-p_0^2+2m_{\rm th}^2-\frac{2m_{\rm eff}^2p_0\xi}{p}\right)}\log \frac{p+p_0}{p-p_0} \right].
\eea
Total LD part of the quark free energy becomes
\bea
F_{q,\,\rm LD}^{\rm ren}
&=& \frac{N_c}{2\pi^3}\sum_f \int\limits_0^{\infty} p^2dp \int\limits_0^p dp_0 \left[\frac{1}{e^{\beta p_0}+1}-\frac{1}{2}\right]\left[\theta_{LD}^L+\theta_{LD}^R\right] - F_{\rm LD}^{\rm sub, L} - F_{\rm LD}^{\rm sub, R}.\label{FqLDren}
\eea
%Total free energy of quark without high $T$ expansion 
%\bea
%F_q^{\text{full}}=F_{\text{q,\,QP}}^{\text{ren}}+F_{\text{q,\,LD}}^{\text{ren}},
%\eea
%where pressure is defines as $P=-F$.

\subsection{Gluonic part}
Free energy of a gluon within the weak magnetic field approximation up to $\mathcal O(eB)^2$ is given by

\bea
F_g&=&d_A\(\mathcal F^1_g+\mathcal F^2_g+\mathcal F^3_g\)
\eea
where $\mathcal F^1_g$, $\mathcal F^2_g$ and $\mathcal F^3_g$ are defined in Eqs.~\eqref{f1_qed},\eqref{f2_qed} and \eqref{f3_qed}.
Similar to the quark case, we get the quasiparticle contribution of the $b$ mode of gluon as
\bea
\mathcal F_{g,QP}^1&=&\int_{\bf p} \bigg[T\ln\big(1-e^{-\beta \omega_b}\big)+\frac{\omega_b}{2}\bigg].\label{b_qp}
\eea
We can also write the quasiparticle contribution of $c$ and $d$ mode as
\bea
\mathcal F_{g,QP}^2&=&\int_{\bf p} \bigg[T\ln\big(1-e^{-\beta \omega_c}\big)+\frac{\omega_c}{2}\bigg],\label{c_qp}\\
\mathcal F_{g,QP}^3&=&\int_{\bf p} \bigg[T\ln\big(1-e^{-\beta \omega_d}\big)+\frac{\omega_d}{2}\bigg].\label{d_qp}
\eea
%We have omitted the $\eta$ term as this goes to zero in dimensional regularization.

So the total quasiparticle contribution of the gluon free energy is given by
\bea
\mathcal F_{g,QP}&=& d_A\int\limits_{\bf p} \bigg[T\ln\frac{1-e^{-\beta \omega_b}}{1-e^{-\beta p}}+\frac{\omega_b-p}{2}+T\ln\big(1-e^{-\beta \omega_c}\big)+T\ln\big(1-e^{-\beta \omega_d}\big)+\frac{\omega_c+\omega_d}{2}\bigg].\label{F_gqp}\ \ \ \ \ \ 
\eea
In the above equation, all the logarithmic terms are converging integral. $\frac{\(\omega_b-p\)}{2}$ is also UV converging as $\omega_b$ approaches to light-cone at large momentum. The last term is UV diverging and we need to regulate that with appropriate counterterms. Below, we write down the asymptotic form of $\omega_c$ at large momentum as
\bea
&&\hspace{-0.5cm}\omega_c^{\rm asy}\nn\\
&\approx& p+\frac{m_D^2}{4p}+\frac{m_D^4 }{32 p^3}\left(3-2 \log\frac{8 p^2}{m_D^2}\right)\nn\\
&+&\sum_f\frac{q_fB^2 g^2 \left[\(\pi m_f - 4T+32 m_f^2 T g_k\) \(190 \cos2\theta_p + 125 \cos4\theta_p -27\) -8192m_f^2 T g_k \right]}{12288 \pi ^2 m_f^2 p T} \nn\\
&-&\sum_f\frac{g^2q_fB^2 \(10 \cos2\theta_p+5 \cos4\theta_p-23\) \left(32m_f^2 T g_k+\pi m_f - 4T\right) \log \frac{8p^2}{m_D^2}}{ 2048 m_f^2~p\, T\, \pi^2}\nn\\
&+&\sum_f\frac{m_D^2g^2q_fB^2  \left[\left(32m_f^2 T g_k+\pi m_f - 4T\right) (970 \cos2\theta_p + 1055\cos4\theta_p-201)-16384m_f^2 T g_k \right]}{49152 \pi ^2 m_f^2 p^3 T}\nn\\
%%%
&-&\sum_f\frac{m_D^2g^2q_fB^2  \left[\left(32m_f^2 T g_k+\pi m_f - 4T\right) (670 \cos2\theta_p + 545\cos4\theta_p-591)-8192m_f^2 T g_k \right]\log \frac{8 p^2}{m_D^2}}{49152 \pi ^2 m_f^2 p^3 T}\nn\\
%%%
&+&\sum_f\frac{m_D^2g^2q_fB^2  \(10\cos2\theta_p+5 \cos4\theta_p-23\) \left(32m_f^2 T g_k+\pi m_f-4 T\right) \log ^2\frac{8 p^2}{m_D^2}}{8192 \pi ^2 m_f^2 p^3 T}
\eea

Similarly, we can write down the asymptotic form of $\omega_d$ at large momentum as
\bea
&&\hspace{-0.5cm}\omega_d^{\rm asy} \nn\\
&\approx& p+\frac{m_D^2}{4p}+\frac{m_D^4}{32p^3}\left(3-2\ln{\frac{8p^2}{m_D^2}}\right)
+\sum_f\frac{g^2(q_fB)^2}{6144 \pi^2 m_f^2 p T}\Bigg[\frac{128 m_f(13-11 \cos2\theta_p) \cot ^2\theta_ p}{3 \left(\cosh(m_f/T)+1\right)}+27 \pi m_f \nn\\
&+&  60T + (356 T-71 \pi m_f)\cos2\theta_p + 8 (16 T-7 \pi m_f)\csc ^2\theta_p+ 8m_f^2 T \Big\{32  (53-9 \cos2\theta_p)f_k \cot^2\theta_p\nn\\
&-& (193 + 308 \cos2\theta_p - 53\cos4\theta_p)g_k \csc^2\theta_p\Big\}+12\log\frac{8 p^2}{m_D^2}\Bigg\{\frac{32 m_f (\cos 2 \theta_p-2) \cot^2\theta_p}{3 \left(\cosh \left(\frac{m_f}{T}\right)+1\right)} + 46 T-8 \pi  m_f\nn\\
%%%%%%%%%%
&+& \left(\pi  m_f-6 T\right) \left(\cos 2 \theta_p+8 \csc^2\theta_p\right) -16 m_f^2 T \Big(4 (9-\cos2 \theta_p) f_k \cot^2\theta_p + g_k \left(9-\cos2\theta_p - 8\csc ^2\theta_p\right)\Big)\Bigg\} \Bigg]\nn\\
%%%%%%
&-&\sum_f\frac{m_D^2g^2(q_fB)^2\csc^2\theta_p}{98305 \pi^2 m_f^2 p^3 T}\Bigg[\frac{512 m_f (64-77\cos 2 \theta_p) \cos^2\theta_p}{3 \left(\cosh \left(m_f/T\right)+1\right)} -4 \cos 2\theta_p (247 \pi m_f -796 T)-621 \pi m_f \nn\\
&+& 2172T +  (713 \pi m_f-3308 T)\cos4\theta_p - 32m_f^2 T \Big\{32f_k (57 \cos 2\theta_p-233) \cos ^2\theta_p+g_k (1436 \cos 2\theta_p\nn\\
&-&599 \cos 4\theta_p+955)\Big\}+ \bigg\{32 m_f^2 T \left(128f_k (9 \cos2 \theta_p - 53) \cos ^2\theta_p + g_k (1124 \cos 2 \theta_p-257 \cos4 \theta_p+733)\right)\nn\\
&-&\frac{2048 m_f (13-11 \cos2\theta_p)\cos ^2\theta_p}{3 \left(\cosh \left(m_f/T\right)+1\right)}+20 (77 \pi  m_f-404 T) \cos 2 \theta_p+ (1604T - 329 \pi  m_f)\cos4\theta_p+1221 \pi  m_f\nn\\
%%%%%%%%%%%%
&-&7092 T\bigg\}\log\frac{8 p^2}{m_D^2} + 12\bigg\{\frac{128 m_f \cos ^2\theta_p (2-\cos2 \theta_p)}{3 \left(\cosh \left(\frac{m_f}{T}\right)+1\right)} + 2  (52 T-9 \pi  m_f)\cos 2 \theta_p+  (\pi  m_f-6 T)\cos 4 \theta_p \nn\\
&-& 15 \pi  m_f+94 T-16 m_f^2 T \left(16f_k (\cos2 \theta_p-9) \cos ^2\theta_p+g_k (20 \cos2 \theta_p - \cos4\theta_p + 13)\right)\bigg\}\log^2\frac{8 p^2}{m_D^2}\Bigg]
\eea
\subsubsection{Counterterm}
The appropriate counterterms should make the integral in Eq.~\eqref{F_gqp} ultraviolet convergent, and it should not introduce any infrared divergences. Our choice for the subtraction counterterm for $c$ mode is
\bea
&&\hspace{-0.5cm}\omega_c^{CT} 
= \left[p^2+\frac{m_D^2}{2}\right.\nn\\
&+&\left.\sum_f\frac{q_fB^2 g^2 \left[\(\pi m_f - 4T+32 m_f^2 T g_k\) \(190 \cos2\theta_p + 125 \cos4\theta_p -27\) -8192m_f^2 T g_k \right]}{6144 \pi ^2 m_f^2 T}\right.\nn\\
&-&\left.\sum_f\frac{g^2q_fB^2 \(10 \cos2\theta_p+5 \cos4\theta_p-23\) \left(32m_f^2 T g_k+\pi m_f - 4T\right)}{1024 m_f^2\, T\, \pi^2}\log\left\{\frac{8(p^2+m_D^2/2)}{m_D^2}\right\}\right]^{1/2}\nn\\
&-&\frac{\left(m_D^2+\sum_f\frac{q_fB^2 g^2 \left[\(\pi m_f - 4T+32 m_f^2 T g_k\) \(160 \cos2\theta_p + 155 \cos4\theta_p -63\) - 3072m_f^2 T g_k \right]}{1536 \pi ^2 m_f^2 }\right)^2}{16\left(p^2+\frac{m_D^2}{2}\right)^{3/2}}
\left[\left(1-\sum_f \frac{g^2 (q_fB)^2 }{3072 \pi ^2 }\right.\right.\nn\\
&\times&\frac{32 m_f^2 T g_k (349-90 \cos 2 \theta_p + 45 \cos4 \theta_p) + 9 (\pi m_f-4 T)(53 - 10 \cos 2 \theta_p + 5 \cos4 \theta_p ) }{m_f^2 m_D^2 T}-\sum_f\frac{ g^2 (q_fB)^2 }{512}\nn\\
&\times&\left.\left.\frac{(10 \cos2 \theta_p + 5 \cos4 \theta_p - 23) \left(32 m_f^2 T g_k+\pi  m_f - 4 T\right)}{ \pi ^2 m_f^2 m_D^2 T}\log\frac{8p^2+4m_D^2}{m_D^2} \right)\log\frac{8p^2+4m_D^2}{m_D^2}-2\right],\ \label{omega_c_CT}
\eea
and for the $d$ mode is
\bea
&&\hspace{-0.5cm}\omega_d^{CT} =  \Bigg[p^2+\frac{m_D^2}{2}\nn\\
&+&\sum_f\frac{g^2 (q_fB)^2}{3072 \pi ^2 m_f^2 T} \Bigg\{\frac{128 m_f (13-11 \cos2\theta_p) \cot^2\theta_p}{3\left(\cosh \left(m_f/T\right)+1\right)}+ (356 T-71 \pi  m_f)\cos2 \theta_p + 60T + 27 \pi m_f\nn\\
&+&8 m_f^2 T \csc^2\theta_p \left\{ \frac{16}{m_f^2}-\frac{7 \pi}{m_fT}  +  32 f_k (53-9 \cos2 \theta_p) \cos^2\theta_p - g_k (308 \cos2\theta_p - 53 \cos4 \theta_p + 193)
\right\} \nn\\
&+&12\left(\frac{32 m_f (\cos2\theta_p-2) \cot ^2\theta_p}{3 \left(\cosh \frac{m_f}{T}+1\right)}+(\pi m_f-6 T) \left(\cos2\theta_p+8 \csc ^2\theta_p\right)+46 T -8\pi m_f \right.\nn\\
&-&16m_f^2T\Big\{4 f_k (9-\cos2\theta_p) \cot ^2\theta_p+ g_k \left(9-\cos2\theta_p -8 \csc ^2\theta_p\right)\Big\} \bigg)\log\frac{8p^2+4m_D^2}{m_D^2}\Bigg\}\Bigg]^{1/2}\nn\\
%%%%%%%%%%%
&-&\frac{1}{16\left(p^2+\frac{m_D^2}{2}\right)^{3/2}}\left\{m_D^2-\sum_f\frac{g^2 (q_fB)^2 \csc ^2\theta_p}{9216 \pi ^2 m_f^2 T}\left(\frac{64 m_f \cos ^2\theta_p (94 \cos2\theta_p-89)}{\cosh \left(m_f/T\right)+1}+ 405 \pi m_f-1764 T\right.\right.\nn\\
&+&6 \cos2\theta_p (101\pi  m_f-392 T)-3 \cos4\theta_p (101 \pi  m_f-476 T)-1536 m_f^2 T f_k (85-18 \cos2 \theta_p) \cos ^2\theta_p  \nn\\
&+&96 m_f^2 T g_k (248 \cos2\theta_p - 83 \cos4\theta_p +163)\bigg)\Bigg\}^2\left[\left(1+\sum_f \frac{g^2 (q_fB)^2 }{512 \pi ^2 m_b^2 m_D^2 T }\Bigg\{\frac{192m_f \cot ^2\theta_p}{\cosh \left(\frac{m_f}{T}\right)+1} + 179 m_f \pi\right.\right.\nn\\
&-&1108 T+(21 \pi  m_f-76 T)\cos2\theta_p +4(256 T-39 \pi m_f) \csc ^2\theta_p -256m_f^2 T f_k(\cos2\theta_p-23) \cot ^2\theta_p\nn\\
&-&8 m_f^2 T g_k(84 \cos2\theta_p + 23 \cos4\theta_p + 53) \csc ^2\theta_p\Bigg\}+\sum_f\frac{ g^2 (q_fB)^2 }{384\pi ^2 m_b^2 m_D^2 T}\Bigg\{\frac{32 m_f (\cos2\theta_p-2) \cot ^2\theta_p}{\cosh \left(m_f/T\right)+1}\nn\\
%%%%%%%%
&+&3 \left(46 T-8 \pi m_f + (\pi m_f - 6T) \left(\cos2\theta_p+8 \csc^2\theta_p\right)\right)-192 m_f^2 T f_k (9-\cos 2\theta_p) \cot^2\theta_p\nn\\
&-&\left.48 m_f^2 T g_k \left(9-\cos 2\theta_p - 8 \csc ^2\theta_p\right)\Bigg\}\log\frac{8p^2+4m_D^2}{m_D^2} \right)\log\frac{8p^2+4m_D^2}{m_D^2}-2\Bigg].\label{omega_d_CT}
\eea
Using the counterterms in Eqs.~\eqref{omega_c_CT}~and~\eqref{omega_d_CT}, the total renormalized quasiparticle contribution of the gluon free energy from Eq.~\eqref{F_gqp} can be written as
\bea
F_{g,QP}^{\rm ren}&=& d_A\int\frac{p^2dp\sin\theta_p d\theta_p d\phi}{(2\pi)^3} \bigg[T\ln\frac{1-e^{-\beta \omega_b}}{1-e^{-\beta p}}+\frac{\omega_b-p}{2}+T\ln\big(1-e^{-\beta \omega_c}\big)+T\ln\big(1-e^{-\beta \omega_d}\big)\nn\\
&&+\frac{\omega_c-\omega_c^{CT}}{2} +\frac{\omega_d - \omega_d^{CT}}{2}\bigg].\label{FgQPren}
\eea

\subsection{Landau damping part}
\subsubsection{$b$ mode}
The Landau damping part for the $b$ mode in the gluon free energy is
\bea
F^1_{g,LD}&=&\frac{1}{\pi}\int_{\bf p}\int_0^p \,dp_0\, \phi_b\bigg[\frac{1}{e^{\beta p_0}-1}+\frac{1}{2}\bigg],
\label{Fg1LD}
\eea
where the angle $\phi_b$ is given by
\bea
\phi_b&=& \tan^{-1} \frac{\text{Im}(P^2-b)}{\text{Re}(P^2-b)}\nn\\
&=&\tan ^{-1} \frac{\pi  m_D^2 p_0}{2 p^3+m_D^2 (2 p-L p_0)} - \frac{2 \pi  m_D^2 \delta m_D^2  p p_0}{\left\{2 p^3+m_D^2 (2 p-L p_0)\right\}^2+\pi ^2 m_D^4 p_0^2}\nn\\
&-& \sum_f\frac{p^3 p_0(q_fB)^2 }{768 \pi  \left\{\left(2 p m_D^2+2 p^3-p_0 L m_D^2\right)^2+\pi ^2 p_0^2 m_D^4\right\}} 
%%%%%%
\left[\frac{19 \pi }{m_f T}-\frac{9 \pi  p_0^2}{m_f p^2 T} \right.\nn\\
&-&\left.\frac{8}{m_f^2}-\frac{384 p_0^2 (f_k + g_k)}{p^2}+\frac{24 p_0^2}{m_f^2 p^2} + \frac{m_D^2 }{p^2}\left\{128 \left(17 f_k+9 g_k \right)+\frac{19 \pi }{ m_f^2 T}-\frac{8 T}{m_f^2 T}\right\} \right.\nn\\
&+&128 (17 f_k+9 g_k)+\left.3 \cos2 \theta_p \left(128 (f_k+g_k)-\frac{8}{m_f^2}+\frac{3 \pi }{m_f T}\right) \left(\frac{m_D^2}{p^2}-\frac{3 p_0^2}{p^2}+1\right)\right],
\eea
where $L=\log\frac{p+p_0}{p-p_0}$. We must choose a subtraction term to remove the ultravilolet divergence in $b$ mode without introducing any infrared divergences. Our choice for the subtraction term for the $b$ mode is
\bea
&&\mathcal{F}_{g,\rm{LD}}^{1,\rm{sub}}\nn\\
&=&\frac{1}{\pi} \int\limits_{\mathbf{p}} \int\limits_{0}^{p} \frac{dp_0}{2}\Bigg[\frac{ \pi m_{D}^{2} p_0}{2 p^{3}} \frac{2}{e^{\beta p_0}-1}+\frac{\pi m_{D}^{2} p_0}{2 p\left(p^{2}+ m_{D}^{2}\right)} + \frac{ \pi m_{D}^{4} p_0^{2}\, L}{4 p^{2}\left(p^{2}+ m_{D}^{2}\right)^{2}} -\frac{\pi\,  \delta m_D^2 m_D^2 \, p_0}{2 p \left(p^2+m_D^2\right)^2}\nn\\
&-&\sum_f\frac{\(q_fB\)^2p_0}{768m_f^2\pi T}\left\{\frac{ 5 \pi  m_f+8 T+128 m_f^2 T (7 f_k+3 g_k)-3 \cos2 \theta_p \left(3 \pi  m_f-8 T+128 m_f^2 T (f_k+g_k)\right)}{ p^3 }\right.\nn\\
&+& \left.\frac{3P^2   \left\{3 \pi  m_f-8 T+128 m_f^2 T (f_k+g_k)\right\}(1+3 \cos2 \theta_p)}{2 p^5 T}\right\}\frac{1}{e^{\beta p_0}-1} -\sum_f \frac{g^2(q_fB)^2p_0}{1536 m_f^2 \pi T}\nn\\
&\times&\left\{\frac{ 128 m_f^2 T (7 f_k + 3 g_k)+5 \pi  m_f+8 T - 3 \cos (2 \theta_p) \left(128 m_f^2 T (f_k+g_k)+3 \pi  m_f-8 T\right)}{p \left(m_D^2+p^2\right)}\right.\nn\\
&-&\left.\frac{3 P^2 (3 \cos2 \theta_p +1) \left(128 m_f^2 T (f_k+g_k)+3 \pi  m_f-8 T\right)}{2 p \left(p^2+m_D^2\right)^2}\right\}
-\sum_f\frac{g^2 m_D^2\(q_fB\)^2p_0}{3072 \pi m_f^2 T}\nn\\
&\times&\left\{\frac{8 T+29 \pi  m_f + 128 m_f^2 T (31 f_k+15 g_k) - 3 \cos2\theta_p \left(3 \pi  m_f-8 T+128 m_f^2 T (f_k+g_k)\right)}{p \left(p^2+m_D^2\right)^2}\right.\nn\\
&+&\frac{4 p_0}{p}\frac{ 3 \cos2 \theta_p \left(128 m_f^2 T (f_k+g_k)+3 \pi  m_f-8 T\right)-8 T \left(16 m_f^2 (7 f_k+3 g_k)+1\right)-5 \pi  m_f}{\left(p^2+m_D^2\right)^2} \nn\\
&\times& \left(1-\frac{p_0 L}{2 p}\right) -\left.\frac{6 P^2 p_0 (1 + 3 \cos2 \theta_p ) \left(128 m_f^2 T (f_k+g_k)+3 \pi  m_f-8 T\right)}{p \left(p^2+m_D^2\right)^3}\frac{p_0L }{2 p}\right\}\Bigg].\label{sub_b}
\eea

\subsubsection{$c$ mode}
The Landau damping part for the $c$ mode in the gluon free energy is
\bea
F^2_{g,LD}&=&-\frac{1}{\pi}\int_{\bf p}\int_0^p \,d\omega \phi_c\bigg[\frac{1}{e^{\beta \omega}-1}+\frac{1}{2}\bigg],
\eea
where the angle $\phi_c$ is given by
\bea
\phi_c&=& \tan^{-1} \frac{\text{Im}(P^2-c)}{\text{Re}(P^2-c)}\nn\\
&=&\tan ^{-1}\frac{\pi m_D^2 p_0 P^2}{4 p^3P^2 - 2m_D^2 p p_0^2 + m_D^2 p_0 P^2 L}
+\sum_f \frac{16 (q_fB)^2 g^2 g_k m_D^2 p^3 p_0P^2}{3 \pi  \left\{\left(2 m_D^2 p p_0^2 - 4 p^3 P^2 - m_D^2 p_0 P^2 L \right)^2+\pi ^2 m_D^4 p_0^2 P^4\right\}}\nn\\
%%%
&+& \sum_f \frac{g^2p_0p (q_fB)^2 \left(32 g_k m_f^2 T+\pi m_f - 4 T\right)}{128 \pi  m_f^2 T \left\{\left(2 m_D^2 p p_0^2 - 4 p^3 P^2  - m_D^2 p_0 P^2 L \right)^2+\pi ^2 m_D^4 p_0^2 P^4\right\}}\Bigg[\left\{\frac{m_D^2p_0}{p}\(\frac{2p_0}{p}- \frac{P^2 L}{p^2}\)-4  P^2\right\}\nn\\
&\times&\Bigg\{\frac{15}{4} (1-\cos \theta_p \cot \theta_p) \left(P^4 \cos2\theta_p +8 P^2 p_0^2 \cos ^2\theta_p - \cos^4\theta_p \left(p^4-10 p^2 p_0^2+\frac{35 p_0^4}{3}\right)\right)+11 p^4-7 p^2 p_0^2\nn\\
&+&5 p_0^2 \cos 2 \theta_p \left(3 P^2+2 p_0^2\right)-10 p_0^4 \csc ^2\theta_p +30 p_0^4\Bigg\}
-\frac{m_D^2P^2}{6p^2}\Bigg\{p^2 p_0^2 (97-40 \cos2 \theta_p -25 \cos 4 \theta_p ) \csc ^2\theta_p\nn\\
&-&6 L p p_0 \left(11 p^2+6 p_0^2 \csc ^2\theta_p+5 p_0^2\right)+15 p^4 (1-\cos \theta_p \cot \theta_p) \left(3  \sin ^4\theta_p-\frac{p_0^2}{p^2} \left(\frac{35 \cos ^4\theta_p}{3}-10 \cos ^2\theta_p+1\right)\right)\nn\\
&+&24\left\{\frac{15}{8} P^4 \sin ^4\theta_p (\cos \theta_p \cot\theta_p-1)+3 P^2p_0^2 \left(\csc^2\theta_p+\frac{5}{4} \left(\sin ^22\theta_p+2 \cos 2\theta_p-2 \sin 2\theta_p \cos^3\theta_p\right)+2\right)\right.\nn\\
&+&\left.5p_0^4 \left(\frac{1}{5}-\cos ^4\theta_p+\cos2\theta_p-\frac{8 \cot ^2\theta_p}{5}+\cos ^5\theta_p \cot \theta_p\right)\right\} \left(1-\frac{Lp_0}{2 p}\right) \Bigg\}  \Bigg].
\eea
To remove the ultraviolet divergence without introducing any infrared divergences, our choice for the subtraction term for the $c$ mode is
\bea
&&\mathcal{F}_{g,\rm{LD}}^{2, \rm{sub}}\nn\\
&=&-\frac{1}{\pi} \int\limits_{\mathbf{p}} \int\limits_{0}^{p} \frac{dp_0}{2}\Bigg[\frac{ \pi m_D^{2}p_0}{4 p^3} \frac{2}{e^{\beta p_0}-1}- \frac{\pi m_D^2p_0 P^2}{4 p^{3}\left(-P^2+\frac{m_D^2}{2}\right)}-\frac{\pi m_D^4p_0P^4}{8p^5\left(-P^2+\frac{m_D^2}{2}\right)^2}\left(  \frac{p_0}{2p} L-1\right)\nn\\
&+& \sum_f\frac{5g^2(q_fB)^2 \left[32 g_k m_f^2 T+ \pi m_f-4T\right]}{512 p^3 \pi  m_f^2 T \sin^2\theta_p} \frac{2}{e^{\beta p_0}-1}\Bigg\{\frac{\frac{7}{5}\sin^2\theta_p- \cos ^4\theta_p+ \sin\theta_p \cos ^6\theta_p+ \sin ^4\theta_p \cos ^2\theta_p}{p \(p^2-p_0^2+\frac{m_D^2}{2}\)}\nn\\
%%%%%%%%
&+&\frac{52+338 \cos2 \theta_p+300 \cos 4 \theta_p-50 \cos6 \theta_p-5 \sin\theta_p-45 \sin3 \theta_p-65 \sin5 \theta_p-25 \sin7 \theta_p}{320 p^3}\nn\\
&+&\frac{P^2  (60+102 \cos2 \theta_p + 420 \cos4 \theta_p - 70 \cos6 \theta_p +17 \sin \theta_p-3 \sin3 \theta_p-55 \sin5 \theta_p - 35 \sin7 \theta_p)}{512 p^5}
\Bigg\} \nn\\
&+& \sum_f\frac{5g^2 (q_fB)^2 p_0 \left[32g_k m_f^2T + \pi m_f -4 T\right]}{256 \pi  m_f^2  T \sin^2\theta_p}\left\{\frac{\frac{7}{5}\sin^2\theta_p- \cos ^4\theta_p+ \sin\theta_p \cos ^6\theta_p+ \sin ^4\theta_p \cos ^2\theta_p}{p \(p^2-p_0^2  + \frac{m_D^2}{2}\)}\right.\nn\\
%%%%%%%%
&-&\frac{52+338 \cos2 \theta_p+300 \cos 4 \theta_p-50 \cos6 \theta_p-5 \sin\theta_p-45 \sin3 \theta_p-65 \sin5 \theta_p-25 \sin7 \theta_p}{320 p^3\(p^2-p_0^2  + \frac{m_D^2}{2}\)}P^2\nn\\
&+&\frac{P^6  (60+102 \cos2 \theta_p + 420 \cos4 \theta_p - 70 \cos6 \theta_p +17 \sin \theta_p-3 \sin3 \theta_p-55 \sin5 \theta_p - 35 \sin7 \theta_p)}{512 p^5\(p^2-p_0^2  + \frac{m_D^2}{2}\)^2}
\Bigg\}\nn\\
&-&
%%%%%%
\sum_f\frac{g^2(q_fB)^2m_D^2 g_kp_0P^2}{3 \pi  p^3 \left(p^2-p_0^2+\frac{m_D^2}{2}\right)^2} +\sum_f \frac{5g^2(q_fB)^2m_D^2 p_0\left[32 g_k m_f^2 T+\pi m_f-4 T\right]\csc^2\theta_p}{32768 \pi  m_f^2 T }\nn\\
&\times&\Bigg\{- \frac{\frac{394}{5} \cos2\theta_p + 12 \cos4\theta_p - 2 \cos6 \theta_p -5 \sin\theta_p - 9 \sin3\theta_p - 5 \sin5 \theta_p -  \sin7 \theta_p -\frac{124}{5}}{p^3 \(-P^2+\frac{m_D^2}{2}\)^2}\frac{ P^2 p_0 L}{p}  \nn\\
&-&\frac{89 \sin\theta_p + 213 \sin3 \theta_p + 161 \sin5 \theta_p + 37 \sin7 \theta_p - \frac{8274}{5} \cos2 \theta_p - 444 \cos4 \theta_p + 74 \cos6 \theta_p + \frac{1164}{5}}{12 p^3 \(-P^2+\frac{m_D^2}{2}\)^2}P^2\nn\\
&-&\frac{41 \sin\theta_p + 213 \sin3 \theta_p + 257 \sin5 \theta_p + 85 \sin7 \theta_p -\frac{7602}{5} \cos2 \theta_p - 1020 \cos 4 \theta_p + 170 \cos6 \theta_p - \frac{948}{5}}{12 p^5\(-P^2+\frac{m_D^2}{2}\)^2}P^4\nn\\
&+&\frac{ 17 \sin\theta_p - 3 \sin3 \theta_p - 55 \sin5 \theta_p - 35 \sin7 \theta_p + 102 \cos2 \theta_p + 420 \cos4 \theta_p - 70 \cos6 \theta_p + 60}{4 p^7\(-P^2+\frac{m_D^2}{2}\)^3} P^8 \left[\frac{L p_0}{2 p}-1\right]\nn\\
&+&\frac{\sin\theta_p + 9 \sin3 \theta_p + 13 \sin5 \theta_p + 5 \sin7 \theta_p - \frac{338}{5} \cos2 \theta_p - 60 \cos4 \theta_p + 10 \cos6 \theta_p - \frac{52}{5}}{p^5\(-P^2+\frac{m_D^2}{2}\)^2}\frac{P^4L p_0 }{p}\Bigg\} \Bigg].\label{sub_c}
\eea

\subsubsection{$d$ mode}
The Landau damping part for the $d$ mode in the gluon free energy is
\bea
F^3_{g,LD}&=&-\frac{1}{\pi}\int_{\bf p}\int_0^p \,d\omega \phi_d\bigg[\frac{1}{e^{\beta \omega}-1}+\frac{1}{2}\bigg],
\eea
where the angle $\phi_d$ is given by
\bea
\phi_d&=& \tan^{-1} \frac{\text{Im}(P^2-d)}{\text{Re}(P^2-d)}\nn\\
&&\hspace{-.6cm}=\tan ^{-1}\frac{\pi m_D^2 p_0 P^2}{4 p^3P^2 - 2m_D^2 p p_0^2 + m_D^2 p_0 P^2 L}
-\sum_f \frac{g^2 p_0 p (q_fB)^2 \csc ^2\theta_p}{768 \pi^2 m_f^2  P^2T \left\{\left(\frac{2 m_D^2 p p_0^2}{P^2}- Lm_D^2 p_0-4 p^3\right)^2+\pi ^2 m_D^4 p_0^2\right\}}\nn\\
&\times&\Bigg[6 \pi  \left(32 g_k m_f^2 T+\pi  m_f-4 T\right) \left(m_D^2 p_0^2+p^4+3 p_0^2 \left(2 p^2-5 p_0^2\right)\right)+\left(\frac{2 m_D^2 p_0^2}{p^2P^2}- \frac{Lm_D^2 p_0}{p^3}-4 \right)\cos ^2\theta_p \nn\\
&\times&\left\{\frac{128 \pi  m_f p_0^2 \left(\cos ^2\theta_p \left(3 p^2-5 p_0^2\right)+3 p_0^2\right)}{\cosh \left(\frac{m_f}{T}\right)+1} - 3 \pi ^2 m_f \left(3 p^4-28 p^2 p_0^2+69 p_0^4\right) + 12 \pi  T \left(3 p^4-10 p^2 p_0^2+71 p_0^4\right)\right.\nn\\
&-&32 m_f^2 T p_0^4 \left\{8 f_k \left(\frac{56p^3}{p_0^3}-\frac{27 \pi  p^2}{ p_0^2}-3 \pi \right)+3 \pi  g_k \left(\frac{3 p^4}{p_0^4}- \frac{46 p^2}{ p_0^2} + 67 \right)\right\} - \frac{3\pi p_0^2  \cos ^2\theta_p}{2}\bigg\{ 4 T \left(9 +\frac{151 P^2}{p^2	}\right)  \nn\\ 
&-&\pi  m_f  \left(1+\frac{139 P^2}{p^2}\right)- \frac{15 p^2}{p_0^2} \left(32 g_k m_f^2 T+\pi  m_f-4 T\right) + 32 m_f^2 Tg_k \left(\frac{16f_k}{g_k } \left(\frac{3 P^2}{p^2}+2\right)+7-\frac{127 P^2}{p^2}\right) \bigg\}\Bigg\}\nn\\
&-&128 m_f   \pi  m_D^2 p_0^2 \cos ^2\theta_p\left\{\frac{8+\frac{6 P^2}{p^2}-\frac{3 L p_0^3}{p^3}+\cos^2\theta_p \left(\frac{5 L p_0^3}{p^3}-\frac{3 L p_0}{p}-\frac{10 p_0^2}{p^2}+\frac{8}{3}\right)}{ \cosh \frac{m_f}{T}+1} + \frac{3 \pi}{64}  \left(29 - \frac{69p_0^2}{p^2}\right)\right.\nn\\
&-&\frac{T }{16 m_f}\left(87-\frac{213p_0^2}{p^2}\right) + \frac{m_f T}{2}  \left(224 f_k+71 g_k +\frac{24 f_k p_0^2}{p^2}  - \frac{201 g_k p_0^2}{p^2} \right)+\frac{3 \pi  L}{128}\left[\frac{69 p_0^3}{p^3}-\frac{28 p_0^2}{p^2}+\frac{3 p}{p_0}\right.\nn\\
&-&\left.\frac{4T}{\pi m_f}\left(\frac{71 p_0^3}{p^3}-\frac{10 p_0}{p}+\frac{3 p}{p_0}\right) - \frac{p_0 32 g_k m_f T }{\pi p} \left(\frac{8f_k}{g_k}\left(\frac{p_0^2}{p^2}+9\right)-\frac{3 p^2}{p_0^2}-\frac{67 p_0^2}{p^2}+46\right)\right]-\frac{\pi  \cos^2\theta_p}{128} \nn\\
&\times& \left[\frac{32 m_fp_0^2 T}{\pi  p^2}\left(14f_k+\frac{275 g_k p^2}{p_0^2}-381 g_k\right)-\frac{4 T}{\pi  m_f}\left(275-\frac{453 p_0^2}{p^2}\right)-\frac{417 p_0^2}{p^2}+275 + \frac{3 L}{2}\left\{\frac{15 p}{p_0}-\frac{138p_0}{p}\right.\right.\nn\\
&+&\left.\frac{139 p_0^3}{p^3}+\frac{32g_k m_f T }{\pi }\left(\frac{16 f_k }{g_k}\left(\frac{p}{p_0}-\frac{3 p_0^3}{p^3}\right) +\frac{127 p_0^3}{p^3}-\frac{134 p_0}{p}+\frac{15 p}{p_0}\right)-\frac{4 T \left(\frac{151 p_0^3}{p^3}-\frac{142 p_0}{p}+\frac{15 p}{p_0}\right)}{\pi  m_f}\right\}\Bigg]\Bigg\}\Bigg].\nn\\
\eea
Our choice for the subtraction term for the $d$ mode to remove the divergence is
\bea
&&\mathcal{F}_{g,\rm{LD}}^{3,\,\rm{sub}}\nn\\
&=&-\frac{1}{\pi} \int\limits_{\mathbf{p}} \int\limits_{0}^{p} \frac{dp_0}{2}\Bigg[\frac{ \pi m_D^{2}p_0}{4 p^3} \frac{2}{e^{\beta p_0}-1}- \frac{\pi m_D^2p_0 P^2}{4 p^{3}\left(-P^2+\frac{m_D^2}{2}\right)}-\frac{\pi m_D^4p_0P^4}{8p^5\left(-P^2+\frac{m_D^2}{2}\right)^2}\left(\frac{p_0}{2p} L-1\right)\nn\\
&+&\sum_f\frac{2g^2(q_fB)^2}{e^{\beta p_0}-1}\left\{\frac{14 p_0^2 f_k \cot ^2\theta_p}{3 \pi ^2 p^2 \left(p^2-p_0^2+\frac{ m_D^2}{2}\right)}- \frac{3p_0 \csc^2\theta_p}{768\pi m_f^2 p T \left(p^2-p_0^2+\frac{ m_D^2}{2}\right)}\left[\frac{32 m_f(2-\cos2\theta_p) \cos^2\theta_p}{\cosh \frac{m_f}{T}+1}\right.  \right.\nn\\
&+ & \cos ^2\theta_p \left(640m_f^2 T f_k - 192 m_f^2 T g_k-11 \pi m_f + 64 T\right) + 2 \cos^4\theta_p \left(16 m_f^2 T (g_k-4 f_k)+\pi m_f - 6 T\right) \nn\\
&+&32 m_f^2 T g_k + \pi  m_f -4 T\Bigg]+ \frac{3 p_0\csc^2\theta_p}{1536\pi m_f^2 p^3 T}\left(\frac{32 m_f (5-7 \cos 2 \theta_p) \cos^2\theta_p}{3\left(\cosh \left(\frac{m_f}{T}\right)+1\right)}+6 \left(32 m_f^2 T g_k + \pi m_f - 4 T\right)\right.\nn\\
&+&5 \cos^4\theta_p \left(7 \pi  m_f -32 T -64 m_f^2 T (2 f_k - 3 g_k)\right)+11\cos^2\theta_p \left(24T -5 \pi m_f + 128 m_f^2 T (f_k - g_k) \right)\Bigg)\nn\\
&+&\frac{3 p_0 \csc^2\theta_p P^2}{6144 \pi m_f^2 p^5 T}\left(\frac{128 m_f (1-5 \cos2 \theta_p) \cos^2\theta_p}{3 \left(\cosh \frac{m_f}{T} + 1\right)}+ \cos^2\theta_p \left(64 m_f^2 T (8f_k - 67 g_k) - 138 \pi m_f + 568 T\right)\right.\nn\\
&+&  15 \left(32 m_f^2 T g_k+\pi  m_f-4 T\right) + \cos^4\theta_p \left(139 \pi m_f-604 T-32m_f^2 T (48 f_k-127 g_k)\right)\Bigg)\Bigg\}\nn\\
%%%%%   CT2%%%%%%%%%
&+&\sum_f g^2(q_fB)^2\left\{\frac{14 p_0^2 f_k \cot ^2\theta_p}{3 \pi ^2 p^2 \left(-P^2+\frac{m_D^2}{2} \right)}-\frac{3p_0 \csc^2\theta_p}{768\pi m_f^2 p T \left(p^2-p_0^2+\frac{ m_D^2}{2}\right)}\left[\frac{32 m_f(2-\cos2\theta_p) \cot^2\theta_p}{\cosh \frac{m_f}{T}+1}\right.  \right.\nn\\
&+ & \cos ^2\theta_p \left(640m_f^2 T f_k - 192 m_f^2 T g_k-11 \pi m_f + 64 T\right) + 2 \cos^4\theta_p \left(16 m_f^2 T (g_k-4 f_k)+\pi m_f - 6 T\right) \nn\\
&+&32 m_f^2 T g_k + \pi  m_f -4 T\Bigg]+\frac{3 p_0 P^2\csc^2\theta_p}{1536\pi m_f^2 p^3 T\(P^2-\frac{m_D^2}{2}\)}\Bigg[\frac{32 m_f (5-7 \cos 2 \theta_p) \cos^2\theta_p}{3\left(\cosh \left(\frac{m_f}{T}\right)+1\right)}+6 \left(32 m_f^2 T g_k + \pi m_f\right)\nn\\
&-&24T+5 \cos^4\theta_p \left(7 \pi  m_f -32 T -64 m_f^2 T (2 f_k - 3 g_k)\right)+11\cos^2\theta_p \left(24T -5 \pi m_f + 128 m_f^2 T (f_k - g_k) \right)\Bigg]\nn\\
&+&\frac{3 p_0 \csc^2\theta_p P^4}{6144 \pi m_f^2 p^5 \(P^2-\frac{m_D^2}{2}\) T}\left(\frac{128 m_f (1-5 \cos2 \theta_p) \cos^2\theta_p}{3 \left(\cosh \frac{m_f}{T} + 1\right)}+ \cos^2\theta_p \left(64 m_f^2 T (8f_k - 67 g_k) - 138 \pi m_f \right)\right.\nn\\
&-&  568 T\cos^2\theta_p +15 \left(32 m_f^2 T g_k+\pi  m_f-4 T\right) + \cos^4\theta_p \left(139 \pi m_f-604 T-32m_f^2 T (48 f_k-127 g_k)\right)\Bigg)\Bigg\}\nn\\
%%%%%%%CT2 , g^2 mD^2 %%%%%%
&+&\sum_f g^2 m_D^2 (q_fB)^2\left[\frac{(2-\cos2\theta_p) \cot^2\theta_p P^2L}{48 \pi  m_f p^2 T \left(\cosh\frac{m_f}{T} + 1\right) \left(P^2-\frac{m_D^2}{2}\right)^2}  + \frac{7P^2\cot^2\theta_p }{3 \pi ^2 p^2 \left(P^2-\frac{m_D^2}{2}\right)^2} \left\{f_k +\frac{3\pi L}{1792}\left(\frac{29 }{m_f^2}\right.\right.\right.\nn\\
&-&\left.\left.\left.\frac{5 \pi }{m_f T}+288 f_k - 88 g_k - \left(\frac{3}{m_f^2}-\frac{\pi }{2m_f T}-8 (g_k-4f_k)\right) \cos2 \theta_p + \left(16 g_k-\frac{2}{m_f^2}+\frac{\pi }{2 m_f T}\right) \sec^2\theta_p \right)\right\}\right.\nn\\
&-&\frac{p_0 P^2}{384 \pi  p^3 \left(P^2-\frac{m_D^2}{2}\right)^2}\left\{\frac{4 \left(25-17 \cos 2\theta_p\right) \cot ^2\theta_p}{3 m_f T \left(\cosh \frac{m_f}{T} + 1\right)} +\frac{219 \pi }{32 m_f T}-\frac{261}{8 m_f^2} -\frac{448f_k L}{\pi}-616f_k+209g_k\right.\nn\\
&+& \left. \frac{5 \cos2\theta_p}{T}\left(24 f_k - 13 g_k+\frac{25}{8 m_f^2}-\frac{19 \pi }{32 m_f}\right) -  16 \csc^2\theta_p \left(\frac{31 \pi ^2}{64 \pi  m_f T} -\frac{5}{2 m_f^2}-\frac{28 f_k L}{\pi }-46 f_k+13 g_k\right)\right\}\nn\\
&+&\frac{P^2 L}{16 \pi p^4 \left(P^2-\frac{m_D^2}{2}\right)}\left\{\frac{112 f_k \cot ^2\theta_p}{3 \pi  L} +\frac{3 \cos ^2\theta_p}{m_f T \left(\cosh \frac{m_f}{T}+1\right)} -\left(8 g_k-  7 f_k  - \frac{23}{16 m_f^2} + \frac{39 \pi }{128 m_f T} \right)\cos2\theta_p\right.\nn\\
&-&\left.  21 f_k + 4 g_k - \frac{29}{16 m_f^2} + \frac{37 \pi }{128 m_f T} - 4 \csc ^2\theta_p \left(2 g_k-7 f_k -\frac{11}{16 m_f^2}+\frac{15 \pi }{128 m_f T}\right)\right\} -\frac{7 p_0 P^4}{768 \pi  p^5 \left(P^2-\frac{m_D^2}{2} \right)^2}\nn\\
&\times&\left\{\frac{8 \left(31-47 \cos (2 \theta_p\right) \cot ^2\theta_p}{21 m_f T \left(\cosh \frac{m_f}{T}+1\right)}  -\frac{309 \pi }{224 m_f T}+\frac{393}{56 m_f^2} + \frac{11\cos 2\theta_p}{7}  \left(48 f_k-77 g_k+\frac{101}{8 m_f^2}-\frac{89 \pi }{32 m_f T}\right)\right.\nn\\
&-&\left.\frac{128 f_k L}{\pi }-80 f_k -23 g_k  + \frac{16}{7} \csc ^2\theta_p \left(\frac{56 f_k L}{\pi }+68 f_k-23 g_k +\frac{25}{8 m_f^2}-\frac{11 \pi }{16 m_f T}\right) \right\} + \frac{P^6L}{256 \pi  p^6 \left(P^2-\frac{m_D^2}{2} \right)^2}\nn\\
&\times&\left\{\frac{8 (5-7 \cos2\theta_p) \cot^2\theta_p}{3 m_f T \left(\cosh\frac{m_f}{T} +1\right)} - \frac{121}{32} \left(32 g_k-\frac{4}{m_f^2}+\frac{\pi }{m_f T}\right) - \cos2\theta_p \left(247 g_k-128 f_k-\frac{311}{8 m_f^2}+\frac{279 \pi }{32 m_f T}\right) \right\}\nn\\
&+&\frac{p_0 (1-5 \cos2\theta_p) \cot^2\theta_p P^8 \left(1-\frac{p_0L}{2 p}\right)}{48 \pi  m_f p^7 T \left(\cosh \frac{m_f}{T}+1\right) \left(P^2-\frac{m_D^2}{2}\right)^3}
%%%%
 - \frac{p_0 \csc ^2\theta_p P^8}{64 \pi  p^7 \left(P^2-\frac{m_D^2}{2}\right)^3}\left\{2 \cos ^2\theta_p \left(8f_k - 67g_k + \frac{71}{8m_f^2}-\frac{69 \pi }{32 m_f T}\right) \right.\nn\\
 &+& \left.15 g_k -\frac{15}{8 m_f^2}+\frac{15 \pi }{32 m_f T} + \cos ^4\theta_p \left(127g_k-48 f_k -\frac{151}{8 m_f^2}+\frac{139 \pi }{32 m_f T}\right)\right\}  \nn\\
 &-&\frac{P^{10}\csc ^2\theta_p L}{128 \pi  p^8 \left(P^2-\frac{m_D^2}{2}\right)^3}\left\{\frac{15}{8 m_F^2} -15g_k-\frac{15 \pi }{32m_fT} - \left(16f_k-134 g_k+\frac{71}{4 m_f^2}-\frac{69 \pi }{16 m_fT}\right)\cos ^2\theta_p \right.\nn\\
 &+&\left.\left.\cos ^4\theta_p \left(48 f_k - 127 g_k+\frac{151}{8 m_f^2}-\frac{139 \pi }{32 m_fT}\right)\right\}\right].\label{sub_d}
\eea
Using the subtraction terms for $b, c$, and $d$ modes as given in Eqs.~\eqref{sub_b}, \eqref{sub_c} and~\eqref{sub_d} respectively,  the total renormalized LD contribution of the gluon free energy is given by
\bea
\mathcal{F}_{g,\rm{Ld}}^{\rm ren}=d_A\(\mathcal{F}_{g,\rm{Ld}}^1- \mathcal{F}_{g,\rm{Ld}}^{1,\,\rm{sub}} + \mathcal{F}_{g,\rm{Ld}}^2- \mathcal{F}_{g,\rm{Ld}}^{2,\rm{sub}} + \mathcal{F}_{g,\rm{Ld}}^3- \mathcal{F}_{g,\rm{Ld}}^{3,\rm{sub}}\).
\label{FgLDren}
\eea
Therefore, the total renormalized pressure in terms of quasiparticle poles and the Landau-damping term come from Eqs.~\eqref{FqQPren},~\eqref{FqLDren},~\eqref{FgQPren},~\eqref{FgLDren} and become
\bea
P^{QP+LD} = - \left[F_{q,\,\rm QP}^{\rm ren} + F_{q,\,\rm LD}^{\rm ren}+ F_{g,\,\rm QP}^{\rm ren} + F_{g,\,\rm LD}^{\rm ren} \right].
\eea

\bibliographystyle{JHEP}

\end{document}